\documentclass[useAMS,usegraphicx]{mn2e}
\usepackage{amsmath,amssymb}
\usepackage{color,graphicx,colortbl,url,multirow}
\usepackage[utf8]{inputenc}
\usepackage{rotating}
\usepackage{caption}
\usepackage[longnamesfirst]{natbib}
\usepackage{subfig}
\usepackage{multirow}
\bibpunct[, ]{(}{)}{;}{a}{}{,}

\shortcites{Braine1993,Compiegne2010,Cortese2011,Dunne2000,Heiderman2010,Helfer2003,Ivison2011,James2002,deJong1984,Lada2012,McGaugh1991,Narayanan2008,Narayanan2011,Narayanan2012,Pierini2001,Press2007,Yao2003,Mao2010,Sanders2003,Soifer1989,Zwaan1997,deVaucouleurs1991,Magdis2011b,Mannucci2010,Sakamoto1995,Schweitzer2006,Totani2011,Vergani2007,Coccato2011,Condon1998,Paturel2003,Galametz2011}

\arrayrulecolor{black}

\newcommand{\degree}{\ensuremath{^\circ}}
\newcommand{\mum}{\ensuremath{\,\mu\text{m}}}
\renewcommand{\thefootnote}{\fnsymbol{footnote}}

\newcommand{\hatlas}{\mbox{H-ATLAS}}
\newcommand{\Herschel}{\textit{Herschel}}
\newcommand{\IRAS}{\textit{IRAS}}
\newcommand{\WISE}{\textit{WISE}}
\newcommand{\hour}{\ensuremath{^\text{h}}}

\newcommand{\hi}{\text{H{\sc i}}}
\newcommand{\htwo}{\ensuremath{\text{H}_{2}}}

\newcommand{\Jykms}{\mbox{\ensuremath{{\rm \,Jy\,km\,s}^{-1}}}}
\newcommand{\Kkmspc}{\mbox{\ensuremath{{\rm \,K\,km\,s}^{-1}\,{\rm pc}^2}}}
\newcommand{\Kkms}{\mbox{\ensuremath{{\rm \,K\,km\,s}^{-1}}}}
\newcommand{\kms}{\mbox{\ensuremath{{\rm \,km\,s}^{-1}}}}
\newcommand{\cott}{\mbox{\text{CO(3--2)}}}
\newcommand{\coto}{\mbox{\text{CO(2--1)}}}
\newcommand{\cooz}{\mbox{\text{CO(1--0)}}}
\newcommand{\subcott}{\ensuremath{_{(3-2)}}}
\newcommand{\subcoto}{\ensuremath{_{(2-1)}}}
\newcommand{\subcooz}{\ensuremath{_{(1-0)}}}

\newcommand{\chisq}{\ensuremath{\chi^2}}
\newcommand{\Lsun}{\ensuremath{\,\text{L}_{\odot}}}
\newcommand{\Msun}{\ensuremath{\,\text{M}_{\odot}}}
\newcommand{\Lfir}{\ensuremath{L_{\text{FIR}}}}
\newcommand{\Ltir}{\ensuremath{L_{\text{TIR}}}}

\newcommand{\Mhtwo}{\ensuremath{M_{\text{H}_2}}}
\newcommand{\Mdust}{\ensuremath{M_{\text{dust}}}}
\newcommand{\Mstar}{\ensuremath{M_{\star}}}
\newcommand{\ie}{\mbox{\text{i.e.}}}
\newcommand{\eg}{\mbox{\text{e.g.}}}

\newcommand{\etal}{et~al.}
\title[\hatlas: CO in local submm galaxies]{\Herschel\footnotemark[1]-ATLAS: correlations between dust and gas in local submm-selected galaxies}
\author[N.\ Bourne et al.]
       {N.\ Bourne,$\!^1$\footnotemark[2]\ 
	L.\ Dunne,$\!^2$  
	G.\,J.\ Bendo,$\!^3$  
	M.\,W.\,L.\ Smith,$\!^4$
	C.\,J.\,R.\ Clark,$\!^4$\newauthor
	D.\,J.\,B.\ Smith,$\!^5$
	E.\,E.\ Rigby,$\!^6$
	M.\ Baes,$\!^7$
	L.\,L.\ Leeuw,$\!^8$
	S.\,J.\ Maddox,$\!^2$\newauthor
	M.\,A.\ Thompson,$\!^5$
	M.\,N.\ Bremer,$\!^9$
	A.\ Cooray,$\!^{10}$
	A.\ Dariush,$\!^{11}$
	G.\ de Zotti,$\!^{12,13}$\newauthor
	S.\ Dye,$\!^1$
	S.\ Eales,$\!^4$
	R.\ Hopwood,$\!^{14,15}$
	E.\ Ibar,$\!^{16}$
	R.\,J.\ Ivison,$\!^{17}$
	M.\,J.\ Jarvis,$\!^{18,19}$\newauthor
	M.\,J.\ Micha{\l}owski,$\!^{7,20}$
	K.\ Rowlands,$\!^1$
	E.\ Valiante $\!^4$
	\vspace*{1mm}\\
	$^1$School of Physics \& Astronomy, University of Nottingham,
	University Park, Nottingham NG7 2RD, UK\\
	$^2$Department of Physics \& Astronomy, University of Canterbury, Christchurch, New Zealand\\ 
	$^3$Jodrell Bank Centre for Astrophysics, University of Manchester, Alan Turing Building, Manchester M13 9PL, UK\\
	$^4$School of Physics \& Astronomy, Cardiff University, Queen's Buildings, The Parade, Cardiff CF24 3AA, UK\\
	$^5$Centre for Astrophysics Research, STRI, University of Hertfordshire, College Lane, Hatfield AL10 9AB, UK\\
	$^6$Leiden Observatory, J.H. Oort Building, P.O. Box 9513, NL-2300 RA Leiden, The Netherlands\\
	$^7$Sterrenkundig Observatorium, Universiteit Gent, Krijgslaan 281 S9, B-9000 Gent, Belgium\\
	$^8$College of Graduate Studies, University of South Africa, P.\,O. Box 392, Unisa, 003, South Africa\\
	$^9$H.\,H.\,Wills Physics Laboratory, University of Bristol, Tyndall Avenue, Bristol BS8 1TL, UK\\
	$^{10}$Department of Physics and Astronomy, University of California, Irvine, CA 92697, USA\\
	$^{11}$Institute of Astronomy, University of Cambridge, Madingley Road, Cambridge CB3 0HA, UK\\
	$^{12}$INAF-Osservatorio Astronomico di Padova, Vicolo Osservatorio 5, I-35122 Padova, Italy\\
	$^{13}$SISSA, Via Bonomea 265, I-34136 Trieste, Italy\\
	$^{14}$Physics Department, Imperial College London, South Kensington Campus, London SW7 2AZ, UK \\
	$^{15}$Department of Physical Sciences, The Open University, Milton Keynes MK7 6AA, UK\\
	$^{16}$Instituto de Astrof\'isica, Facultad de F\'isica, Pontificia Universidad Cat\'olica de Chile, Casilla 306, Santiago 22, Chile\\
	$^{17}$UK Astronomy Technology Centre, The Royal Observatory, Blackford Hill, Edinburgh EH9 3HJ, UK\\
	$^{18}$Astrophysics, Department of Physics, Keble Road, Oxford OX1 3RH, UK\\
	$^{19}$Department of Physics, University of the Western Cape, Private Bag X17, Bellville 7535, South Africa\\
	$^{20}$SUPA, Institute for Astronomy, University of Edinburgh, Royal Observatory, Edinburgh, EH9 3HJ, UK}
\begin{document}
\maketitle
\begin{abstract}
We present an analysis of CO molecular gas tracers in a sample of 500\mum-selected \Herschel-ATLAS galaxies at $z<0.05$ ($cz<14990$\kms). Using $22-500\mum$ photometry from \WISE, \IRAS\ and \Herschel, with \hi\ data from the literature, we investigate correlations between warm and cold dust, and tracers of the gas in different phases.
The correlation between global \cott\ line fluxes and FIR--submm fluxes weakens with increasing IR wavelength ($\lambda\gtrsim60\mum$), as a result of colder dust being less strongly associated with dense gas.
Conversely, \coto\ and \hi\ line fluxes both appear to be better correlated with longer wavelengths, suggesting that cold dust is more strongly associated with diffuse atomic and molecular gas phases, consistent with it being at least partially heated by radiation from old stellar populations.
The increased scatter at long wavelengths implies that sub-millimetre fluxes are a poorer tracer of SFR.
Fluxes at 22 and 60\mum\ are also better correlated with diffuse gas tracers than dense \cott, probably due to very-small-grain emission in the diffuse interstellar medium, which is not correlated with SFR.
The FIR/CO luminosity ratio and the dust mass/CO luminosity ratio both decrease with increasing luminosity, as a result of either correlations between mass and metallicity (changing CO/\htwo) or between CO luminosity and excitation [changing \cott/\cooz].
\end{abstract}


\section{Introduction}

\renewcommand{\thefootnote}{\fnsymbol{footnote}}
\footnotetext[1]{\Herschel\ is an ESA space observatory with science instruments provided by European-led Principal Investigator consortia and with important participation from NASA.}
\footnotetext[2]{E-mail: nbourne22@gmail.com}
\renewcommand{\thefootnote}{\arabic{footnote}}

The cold interstellar medium (ISM) is a vital building block of galaxies, and building up an observational understanding of its properties and processes presents one of the current challenges for astronomy.
Dense molecular gas, which fuels star formation, has traditionally been observed in galaxies using the emission lines of the CO molecule, but this tracer is fraught with uncertainties due to varying metallicity and optical thickness \citep[\eg][]{Dickman1986,Genzel2011}. These problems can only be surmounted by the calibration of the CO tracer against independent molecular gas tracers, such as the ISM cooling lines of carbon \citep{Malhotra2001,Pierini2001}. A more widely available tracer is the dust mass, which is correlated with the \htwo\ mass assuming a dust-to-gas ratio that is dependent on metallicity \citep[\eg][]{Cox1986,Issa1990,Dunne2000,James2002}. 
However, the relationship between metallicity and dust-to-gas ratio is a matter of active research in studies of local spiral galaxies \citep{Draine2007,Munoz-Mateos2009,Bendo2009,Magrini2011,Leroy2011a,Leroy2013,Smith2012a} and dwarfs \citep{Lisenfeld1998,Walter2007}.

The revolution in wide-field imaging at far-infrared (FIR) to sub-millimetre (submm) wavelengths, provided by telescopes such as \Herschel\ \citep{Pilbratt2010}, means that we now have unprecedented coverage of the spectral energy distribution (SED) of dust in galaxies at both low and high redshifts. We must learn how to interpret these observations in terms of the ISM mass and star-formation activity within galaxies. 
Emission from dust in star-forming galaxies (SFGs) is dominated by large grains in thermal equilibrium, comprising a number of different phases including a cold ($\sim20$\,K), diffuse component and a warmer ($\sim50$\,K) component associated with star-forming regions \citep{deJong1984,Dunne2001}. Emission from these warm dust grains is directly correlated with the star-formation rate (SFR) since it is heated by emission from young, massive stars \citep{Kennicutt1998a,Kennicutt2012}. It is less clear whether the emission from cold dust is directly tied to the SFR, since it is likely to be partially heated by emission from the older stellar population \citep[\eg][]{Helou1986,Walterbos1996,Bendo2010,Bendo2011,Boquien2011,Totani2011,Groves2012,Smith2012a}.

The Schmidt-Kennicutt law \citep[SK law;][]{Schmidt1959,Kennicutt1998} describes a universal correlation in the surface densities of gas (usually \htwo) and SFR in galactic disks, with a non-linear slope and a significant amount of scatter.
The slope depends on the emission line used to trace the gas, since higher-energy transitions occur in denser and hotter gas \citep{Gao2004a,Krumholz2007,Narayanan2008}, and it is this dense gas that directly feeds star formation. Observations show that starbursts and high-redshift submm galaxies (SMGs) deviate from the SK law of disk galaxies, either as a result of sampling bias favouring the upper end of the intrinsic scatter, or due to a difference in the CO-to-\htwo\ conversion factor in extreme environments \citep{Daddi2010,Genzel2010,Narayanan2011,Krumholz2012,Feldmann2012}. 
It has been suggested, however, that this observation results from of an excitation bias in using lines with differing excitation energies \citep{Ivison2011}. 
It also depends upon the assumption that IR luminosities trace the SFR consistently at high and low luminosities, but an increased contribution from dust heated by older stars could affect the apparent SFRs of lower-luminosity galaxies in relation to high-luminosity starbursts and SMGs.

It is important to understand the gas--SFR relationship in SFGs of all luminosities, and to explore the biases resulting from different selection functions. 
In this paper we compare measurements at FIR wavelengths ($\lambda<200\mum$), which are sensitive to warmer dust, and submm wavelengths ($\lambda>200\mum$), which trace the more massive cold dust component \citep{Dunne2001}.
Far-infrared selection favours high-SFR galaxies but does not necessarily select galaxies with massive ISM content. This is likely to be biased towards galaxies in the upper extreme of the SK scatter, characterised by starbursts with high star-formation efficiencies and short gas depletion times. 
We instead sample nearby galaxies with high dust masses by selecting in the submm, which is an unbiased tracer of the cold ISM. 
Continuum emission in the submm is a strong tracer of the dust mass and is connected to the total gas mass via the dust/gas ratio, and to the SFR via the SK law. 
A blind submm survey is therefore much less sensitive to the star-formation properties of galaxies than samples selected at shorter FIR wavelengths \citep[\eg][]{Yao2003,Gao2004a}, and provides a more direct selection of gas-rich galaxies than targeted surveys based on characteristics such as visual morphology \citep[\eg][]{Young1995,Helfer2003,Mao2010}.

Little is known about galaxies selected in this way since before \Herschel\ it was impossible to conduct blind sky surveys at $\lambda>100\mum$ that sample a sufficiently large volume at low redshift. The largest local submm survey prior to \Herschel\ was the SCUBA Local Universe Galaxy Survey (SLUGS), which used the SCUBA camera to survey galaxies selected from \IRAS\ \citep{Dunne2000} and optical \citep{Vlahakis2005} samples. \citet{Yao2003} measured correlations between nuclear CO emission in the SLUGS 60\mum\ sample and the FIR luminosity (\Lfir) and dust mass (\Mdust), using both \cooz\ as a \textit{total} molecular gas tracer and \cott\ as a \textit{dense} gas tracer. 
Similar studies were carried out by \citet{Gao2004a} sampling IR/CO-selected galaxies; and \citet{Mao2010} sampling a range of spirals, starbursts and AGN.
Recently, \citet{Wilson2012} studied correlations between molecular gas and the FIR in \hi-selected galaxies as part of the JCMT Nearby Galaxies Legacy Survey (NGLS; \citealp{Wilson2009}), while \citet{Corbelli2012} targeted galaxies in the Virgo cluster, showing that molecular gas is more closely correlated with the submm than with shorter wavelengths. 
\citet{Clemens2013} studied a sample of 234 local galaxies selected at 545\,GHz (550\mum) from the \textit{Planck} early release, and demonstrated that dust mass (derived from FIR-mm SEDs) and atomic gas are strongly correlated, while the relationship between dust mass and SFR depends on dust temperature. 

We study a sample selected at 500\mum\ from the \Herschel\ Astrophysical Terahertz Large Area Survey \citep[\hatlas;][]{Eales2010a}. \hatlas\ covers the largest sky area of any extragalactic submm survey, hence it probes the largest volume of the local Universe in five photometric bands from 100 to 500\mum. It is ideal for producing a sample of nearby galaxies selected by their cold dust content. 
This unique sample can be used to explore the relative correlations between the FIR/submm wavebands and the various gas phases, including diffuse atomic \hi\ and molecular \htwo\ traced by \coto\ and \cott. 
In this paper we describe the sample selection and data acquisition in Section~\ref{sec:data}, before analysing the results in Section~\ref{sec:analysis}. The implications of the results are discussed in Section~\ref{sec:disc} and conclusions are summarised in Section~\ref{sec:conc}.

\section{Data}
\label{sec:data}
\subsection{The sample}

The sample was selected from the three equatorial fields of \hatlas\ Phase~1, which are centred at R.A. of approximately 9\hour, 12\hour\ and 14.5\hour, and Dec. $\sim0\degree$, and cover approximately $160$\,deg$^2$ in total. 
The \hatlas\ survey provides PACS \citep{Poglitsch2010} imaging at 100 and 160\mum\ and SPIRE \citep{Griffin2010} imaging at 250, 350 and 500\mum, as described by \citet{Ibar2010} and \citet{Pascale2010}. Source extraction and photometry were conducted using the MADX algorithm (S.\,J.~Maddox \etal\ in prep.) and will be described in an upcoming paper (E.~Valiante \etal\ in prep.); the procedure is also described by \citet{Rigby2010}.
Optical counterparts in the Sloan Digital Sky Survey Data Release 7 \citep[SDSS;][]{Abazajian2009} have been identified using the technique described by \citet{Smith2011a}, including redshifts from the Galaxy and Mass Assembly survey \citep[GAMA;][]{Driver2010}, and will be described in an upcoming paper (N.~Bourne \etal\ in prep.).

The 17 brightest 500\mum\ sources with redshifts $z<0.05$, and fluxes $S_{500}>250$\,mJy, were selected from the catalogues (see Table~\ref{tab:phottable}). In addition to these, the brightest source from the \hatlas\ Science Demonstration Phase (SDP~1), and the two brightest early-type galaxies from SDP (SDP~4 and SDP~15, morphologically classified by \citealp{Rowlands2011}) were included for comparison.
Lenses can potentially contaminate a sample selected at this wavelength \citep[\eg][]{Negrello2010}, but the blue submm colours in our sample indicate that none are lenses. Only one galaxy (NGC~5713) has a prior \cott\ measurement in the literature; this was a single-beam pointing targeting the nucleus \citep{Yao2003}. The galaxy was retained in the sample because the total extended CO flux had not been measured. 

\subsection{Photometric data}
\label{sec:hatlasdata}
We used \hatlas\ photometry from the catalogues described above.
In addition to MADX point-source fluxes, extended sources were measured using elliptical apertures with semimajor axis given by the sum in quadrature of the optical isophotal radius and 1.6 times the full-width at half-maximum (FWHM) of the beam \citep[c.f.][]{Rigby2010}. We adopted systematic errors of seven per cent on SPIRE fluxes and 10 per cent on PACS fluxes, following the guidance in the Observing Manuals.\footnote{\url{http://herschel.esac.esa.int/Docs/SPIRE/html/spire_om.html}; \url{http://herschel.esac.esa.int/Docs/PACS/html/pacs_om.html}}

Flux from extended sources in the PACS maps can be lost due to the high-pass filter (HPF) used in the naive-projection map-making \citep[see][for further details]{Ibar2010}. Bright sources were masked in the HPF, however many of the sources in the sample have low surface brightness in the PACS maps and hence were not adequately masked. We therefore rejected the 100\mum\ data for the sample, as well as the 160\mum\ fluxes of NGC~5705 and UGC~09215, after inspecting the HPF masks in comparison with the extent of 250\mum\ emission.
The sample is covered by FIR all-sky surveys with the \textit{Infrared Astronomical Satellite} \citep[\IRAS;][]{Neugebauer1984} and \textit{AKARI}
\citep{Murakami2007}.
Point-source catalogues from \textit{AKARI} are available at 9 and 18\mum\ \citep{Ishihara2010a}, and at 65, 90, 140 and 160\mum\ \citep{Yamamura2009a}. 
However, since the \textit{AKARI} bands have PSF sizes ranging from 5.6 to 61~arcsec (FWHM), these measurements will not be representative of the total flux of these galaxies (which are typically $\sim1$~arcmin across).
We therefore obtained additional FIR photometry from \IRAS\ at 60 and 100\mum; there were an insufficient number of detections at 12 and 25\mum\ for the purposes of this study. 
The 60 and 100\mum\ bands have PSFs with 1.44 and 2.94~arcmin FWHM respectively \citep{Soifer1989}, which means that several of the galaxies in the sample may be partially resolved by \IRAS.
Hence the point-source fluxes in the Faint Source Catalogue \citep[FSC;][]{Moshir1992} are potentially underestimated, and only four of the galaxies have extended flux measurements in the Revised Bright Galaxy Sample \citep[RBGS;][]{Sanders2003}. We therefore measured extended-source fluxes in the raw \IRAS\ scans using the Scan Processing and Integration Tool (Scanpi)\footnote{Scanpi provided by the NASA/IPAC Infrared Science Archive: \url{http://scanpiops.ipac.caltech.edu:9000/applications/Scanpi/}}, following the procedures outlined by \citet{Sanders2003}. Results were inspected visually and with reference to the FSC and RBGS. For unresolved sources, the agreement between Scanpi and FSC was within 15 per cent, which is equal to the systematic errors quoted by \citet{Soifer1989}. We therefore adopted this fraction as the systematic error on 
all \IRAS\ fluxes.
\label{sec:scanpi}

We extended the photometric coverage to 22\mum\ using aperture fluxes from the \textit{Wide-Field Infrared Survey Explorer} \citep[\WISE;][]{Wright2010}, obtained from the NASA/IPAC Infrared Science Archive. The \WISE\ elliptical apertures are based on the apertures defined in the 2MASS Extended Source Catalog \citep{Skrutskie2006}, accounting for the PSF of \WISE\ (FWHM 12~arcsec at 22\mum). Comparing these apertures with the size of the sources at 250\mum\ (which has comparable resolution; FWHM 18~arcsec), we found that the apertures of all but five of the sample were sufficient to enclose all of the emission. We discarded the 22\mum\ data for NGC~5496, NGC~5705 and NGC~5584, which had apertures that missed substantial amounts of the 250\mum\ emission, as well as NGC~5713 and UGC~09215, which had no aperture measurements.
Following the \WISE\ Explanatory Supplement,\footnote{\url{http://wise2.ipac.caltech.edu/docs/release/allsky/expsup/}} we applied a correction factor of 0.90 to the 22\mum\ fluxes for a rising (red) SED, and we assumed a calibration uncertainty of six per cent.
The photometry measured for the sample are listed in Table~\ref{tab:phottable}.

\begin{sidewaystable*}
\caption[The CO sample and photometry]{The CO sample, sorted by 500\mum\ flux, with the \hatlas\ Phase~1 IDs, NED identifiers and J2000 coordinates, B-band $D_{25}$ isophote (arcmin) from HyperLEDA, and FIR/submm photometry in Jy.
}
{\small
\setlength{\tabcolsep}{4pt}
\begin{center}
\begin{tabular}{l l c c l l c c c c c c c}
\hline
\hatlas\ ID       & NED ID                &    R.A. &     Dec.       & Redshift  & $D_{25}$ & $S_{22}$           &    $S_{60}$            &      $S_{100}$         &      $S_{160}$       &      $S_{250}$           &      $S_{350}$         &        $S_{500}$           \\
\hline
 J120023.7-010553 & NGC~4030              & 12:00:24 & $-$01:06:00   & 0.004887  &  3.8  &   $1.582 \pm  0.096 $ & $ 20.88   \pm  3.13 $  &  $45.87  \pm  6.88 $  &  $ 60.53  \pm 6.05 $  &  $ 33.79  \pm   2.37  $  &  $ 14.54 \pm   1.02 $  & $   5.00  \pm  0.35  $   \\
 J144455.9+015719 & NGC~5746              & 14:44:56 & $+$01:57:18   & 0.005751  &  7.2  &   $0.397 \pm  0.025 $ & $  2.31   \pm  0.35 $  &  $10.37  \pm  1.56 $  &  $ 24.45  \pm 2.45 $  &  $ 17.81  \pm   1.25  $  &  $  8.82 \pm   0.63 $  & $   3.49  \pm  0.26  $   \\
 J144011.1-001725 & NGC~5713              & 14:40:12 & $-$00:17:20   & 0.006334  &  2.5  &          --           & $ 23.74   \pm  3.56 $  &  $36.55  \pm  5.49 $  &  $ 36.48  \pm 3.65 $  &  $ 15.64   \pm  1.10  $  &  $  6.34 \pm   0.45 $  & $   2.05  \pm  0.15  $   \\
 J143740.9+021729 & NGC~5690              & 14:37:41 & $+$02:17:27   & 0.005847  &  3.2  &   $0.527 \pm  0.032 $ & $  5.99   \pm  0.90 $  &  $16.07  \pm  2.41 $  &  $ 20.85  \pm 2.09 $  &  $ 11.31   \pm  0.79  $  &  $  5.16 \pm   0.37 $  & $   1.96  \pm  0.15  $   \\
 J144056.2-001906 & NGC~5719              & 14:40:56 & $-$00:19:06   & 0.005781  &  3.1  &   $0.569 \pm  0.034 $ & $  8.76  \pm   1.31 $  &  $17.68  \pm  2.66 $  &  $ 19.04  \pm 1.90 $  &  $ 10.00   \pm  0.71  $  &  $  4.70 \pm   0.35 $  & $   1.68  \pm  0.14  $   \\
 J142223.4-002313 & NGC~5584              & 14:22:24 & $-$00:23:16   & 0.005464  &  3.1  &          --           & $  2.32    \pm 0.35 $  &  $ 5.28  \pm  0.83 $  &  $ 8.29   \pm 0.84 $  &  $  5.69    \pm 0.40  $  &  $  3.18 \pm   0.23 $  & $   1.30  \pm  0.10  $   \\
 J144424.3+014046 & NGC~5740              & 14:44:24 & $+$01:40:47   & 0.005243  &  2.7  &   $0.261 \pm  0.016 $ & $  2.98   \pm  0.45 $  &  $ 6.68  \pm  1.03 $  &  $ 7.77   \pm 0.07 $  &  $  4.91   \pm  0.35  $  &  $  2.34 \pm   0.17 $  & $   0.89  \pm  0.07  $   \\
 J143949.5-004305 & NGC~5705              & 14:39:50 & $-$00:43:07   & 0.005864  &  1.3  &          --           & $  0.50    \pm 0.10 $  &  $ 1.42  \pm  0.33 $  &         --            &  $  1.74    \pm 0.13  $  &  $  1.20 \pm   0.10 $  & $   0.61  \pm  0.06  $   \\
 J142327.2+014335 & UGC~09215             & 14:23:27 & $+$01:43:35   & 0.004660  &  2.1  &          --           & $  1.51    \pm 0.23 $  &  $ 2.95  \pm  0.46 $  &         --            &  $  2.13    \pm 0.16  $  &  $  1.32 \pm   0.10 $  & $   0.53  \pm  0.05  $   \\
 J141137.7-010928 & NGC~5496              & 14:11:38 & $-$01:09:33   & 0.005140  &  2.7  &          --           & $  0.89    \pm 0.14 $  &  $ 2.08  \pm  0.37 $  &  $ 2.86   \pm 0.29 $  &  $  2.55    \pm 0.19  $  &  $  1.50 \pm   0.13 $  & $   0.71  \pm  0.07  $   \\
 J144611.2-001324 & NGC~5750              & 14:46:11 & $-$00:13:23   & 0.005627  &  2.7  &   $0.052 \pm  0.004 $ & $  0.62  \pm   0.11 $  &  $ 2.40  \pm  0.38 $  &  $ 3.85  \pm  0.07 $  &  $  2.59  \pm   0.18  $  &  $  1.26 \pm   0.09 $  & $   0.46  \pm  0.04  $   \\
 J143753.3-002354 & NGC~5691              & 14:37:53 & $-$00:23:56   & 0.006238  &  1.9  &   $0.239 \pm  0.014 $ & $  3.72   \pm  0.56 $  &  $ 6.61  \pm  1.02 $  &  $ 6.35   \pm 0.06 $  &  $  2.92   \pm  0.21  $  &  $  1.25 \pm   0.09 $  & $   0.46  \pm  0.04  $   \\
 J115705.9+010732 & CGCG~013-010          & 11:57:06 & $+$01:07:32   & 0.039503  & 0.81  &   $0.104 \pm  0.006 $ & $  3.14   \pm  0.47 $  &  $ 7.12  \pm  1.08 $  &  $ 6.38   \pm 0.64 $  &  $  2.87   \pm  0.21  $  &  $  1.15 \pm   0.09 $  & $   0.40  \pm  0.04  $   \\
 J114923.8-010501 & NGC~3907B             & 11:49:24 & $-$01:05:02   & 0.020774  &  1.5  &   $0.060 \pm  0.004 $ & $  0.70  \pm   0.11 $  &  $ 2.18  \pm  0.35 $  &  $ 2.67  \pm  0.27 $  &  $  1.80  \pm   0.13  $  &  $  0.81 \pm   0.07 $  & $   0.29  \pm  0.03  $   \\
 J141215.6-003759 & CGCG~018-010          & 14:12:16 & $-$00:37:59   & 0.025596  & 0.75  &   $0.070 \pm  0.004 $ & $  1.30  \pm   0.20 $  &  $ 3.29  \pm  0.53 $  &  $ 3.80  \pm  0.05 $  &  $  1.99  \pm   0.14  $  &  $  0.82 \pm   0.06 $  & $   0.29  \pm  0.03  $   \\
 J140808.5-014208 & NGC~5478              & 14:08:09 & $-$01:42:08   & 0.025147  & 0.97  &   $0.053 \pm  0.003 $ & $  0.69  \pm   0.13 $  &  $ 1.80  \pm  0.39 $  &  $ 2.65  \pm  0.27 $  &  $  1.76  \pm   0.13  $  &  $  0.76 \pm   0.06 $  & $   0.26  \pm  0.03  $   \\
\multicolumn{1}{c}{--} & NGC~2861          & 09:23:37 & $+$02:08:11   & 0.016965  &  1.3  &   $0.056 \pm  0.004 $ & $  0.76  \pm   0.12 $  &  $ 2.13  \pm  0.38 $  &  $ 3.26  \pm  0.33 $  &  $  1.90  \pm   0.13  $  &  $  0.73 \pm   0.05 $  & $   0.25  \pm  0.02  $   \\
\hline
 J090401.1+012729 & 2MASX-  & \multirow{2}{*}{09:04:01} & \multirow{2}{*}{$+$01:27:29} & \multirow{2}{*}{0.053439} & \multirow{2}{*}{0.54}                                     &   \multirow{2}{*}{$0.075 \pm  0.005 $} & \multirow{2}{*}{$  2.76 \pm    0.42 $}  &  \multirow{2}{*}{$ 4.36  \pm  0.67 $}  &  \multirow{2}{*}{$ 3.84 \pm  0.05  $}  &  \multirow{2}{*}{$  1.61 \pm   0.12   $}  &  \multirow{2}{*}{$  0.61 \pm   0.04 $}  & \multirow{2}{*}{$   0.19  \pm  0.02  $}   \\
 (SDP~1)          & \,J090401+012729 & &  \\
 J090917.0-010959 & \multirow{2}{*}{CGCG~006-008}       & \multirow{2}{*}{09:09:17} & \multirow{2}{*}{$-$01:09:59} & \multirow{2}{*}{0.027489} & \multirow{2}{*}{0.95}         &   \multirow{2}{*}{$0.028 \pm  0.002 $} & \multirow{2}{*}{$  0.54  \pm   0.09 $}  &  \multirow{2}{*}{$ 1.51  \pm  0.27 $}  &  \multirow{2}{*}{$ 1.34 \pm  0.05  $}  &  \multirow{2}{*}{$  0.69  \pm  0.05   $}  &  \multirow{2}{*}{$  0.28 \pm   0.03 $}  & \multirow{2}{*}{$   0.10  \pm  0.02  $}   \\
 (SDP~4)          & & &  \\
 J091205.8+002655 & 2MASX- & \multirow{2}{*}{09:12:06}  & \multirow{2}{*}{$+$00:26:56}  & \multirow{2}{*}{0.054493} & \multirow{2}{*}{0.31}                                    &   \multirow{2}{*}{$0.068 \pm  0.004 $} & \multirow{2}{*}{$  0.80  \pm   0.13 $}  &  \multirow{2}{*}{$ 1.19  \pm  0.24 $}  &  \multirow{2}{*}{$ 0.90  \pm 0.03  $}  &  \multirow{2}{*}{$  0.36  \pm  0.03   $}  &  \multirow{2}{*}{$  0.14 \pm   0.01 $}  & \multirow{2}{*}{$   0.06  \pm  0.01  $}   \\
 (SDP~15) & \,J091205+002656	&  \\
\hline
\end{tabular}
\end{center}
}
{\footnotesize Notes:
Fluxes in Jy are from WISE (22\mum), IRAS (60, 100\mum), PACS (160\mum) and SPIRE (250, 350, 500\mum) as described in the text. 
Galaxies down to, and including, NGC~2861 constitute a complete sample of $z<0.05$ galaxies with $S_{500}>250$\,mJy in the Phase~1 fields. 
NGC~2861 is on the edge of the 12\hour\ field and does not have an \hatlas\ ID because it is outside of the mask used for source extraction. 
SPIRE coverage this close to the edge of the map is shallower than elsewhere but is sufficient for imaging this bright source.}
\label{tab:phottable}
\end{sidewaystable*}

\subsection{CO data}
\label{sec:codata}
We observed galaxies in the sample with the RxA and HARP instruments on the 15m James Clerk Maxwell Telescope (JCMT), between 2011~February and August, using jiggle and grid mapping modes in order to collect emission from the full optical/submm extent of each source. RxA was used to observe the \coto\ line at 230.5\,GHz (rest frame), with a half-power beam-width (HPBW) of 21~arcsec, while HARP was used to observe \cott\ at 345.8\,GHz, with a HPBW of 14~arcsec. 
Galaxies were observed for typical integration times of around 20min, obtaining rms noise levels between 40 and 500\,mK ($T_A^\star$), which were predicted to obtain $5\sigma$ detections of the line in the integrated cube.
Two of the largest galaxies (NGC~5746 and NGC~5705) were not observed due to the prohibitively long integration time required to cover them, while a further two (NGC~4030 and UGC~09215) were not observed since they were part of the \Herschel\ Reference Survey (HRS; \citealp{Boselli2010}) CO sample. A global \cott\ measurement for NGC~4030 was later obtained from the HRS data set (M.\,W.\,L.~Smith et~al. in prep.) and is included in the analysis in Section~\ref{sec:analysis}; the measurement of UGC~09215 was unavailable.
The rest of the sample were all observed with either HARP, RxA or both instruments. 
Data were reduced (including first-order baseline-fitting, de-spiking and masking of timelines affected by varying baselines and high noise) and converted into cubes using the \textsc{starlink} software library,\footnote{\url{http://starlink.jach.hawaii.edu/starlink}} and following the \mbox{ACSIS} cookbook\footnote{\url{http://www.jach.hawaii.edu/JCMT/spectral_line/data_reduction/acsisdr/}}.

We measured the global flux from the entire emitting region of each source by integrating the data cubes along the spectral axis between the expected line edges (i.e. across the {full-width at zero intensity}, FWZI), and in the spatial pixels within an elliptical aperture.
Initial estimates of the line widths were based on the \hi\ 21\,cm line measured in the HIPASS cubes (see Section~\ref{sec:hidata}) and on ancillary CO data where available from the literature. 
These estimates were also informed by additional \cooz\ line widths measured in spectra obtained from a companion observing program on the Nobeyama Radio Observatory (NRO; PI: L.~Leeuw). These data will be described in an upcoming publication by L.~Leeuw \etal\ (in prep.). We subtracted second-order baselines, smoothed the spectra to 60\kms\ resolution, and measured the FWZI of the lines. Line widths are presented in Table~\ref{tab:aperdata}. The NRO data are single-beam (15\arcsec\ HPBW) pointings of the central regions of the galaxies, hence the intensity enclosed may not be the total emission from the galaxy, and it is likely that the \coto\ and \cott\ lines integrated over the entire galaxy are broader. 

The axis ratio and orientation of the elliptical apertures were based on profile fits to the 250\mum\ images (which have a similar beam size to the CO maps).
Since the distribution of (dense) molecular gas does not directly trace that of either the dust or the stars, the size of the emitting region cannot necessarily be inferred from the extent of the submm or optical images. We therefore used a range of aperture sizes to plot a curve of growth for each source, \ie\ the cumulative aperture flux as a function of semimajor axis ($a$), in order to determine the optimal size of aperture in which to integrate the flux. 
Having done this, the line widths were revised where necessary using the FWZI of the integrated spectrum within the aperture, and curves-of-growth were re-measured using the new line widths until consistent results were obtained.
Errors were then calculated by integrating within the chosen aperture on a variance map, in which the variance of each pixel is calculated from the wings of the baseline-subtracted spectrum, masking out the expected velocity range of any emission, and adding in quadrature the uncertainty on the baseline subtraction. This latter was estimated from the covariance matrix of the polynomial coefficients from the baseline fit. 
Where no detection was obtained at $2.5\sigma$ confidence, we calculated $3\sigma$ upper limits from the error in an aperture equal to the optical size, making the assumption that the distribution of molecular gas is unlikely to extend beyond the optical disk. 

\begin{figure*}
\raisebox{1cm}{\includegraphics[scale=0.253]{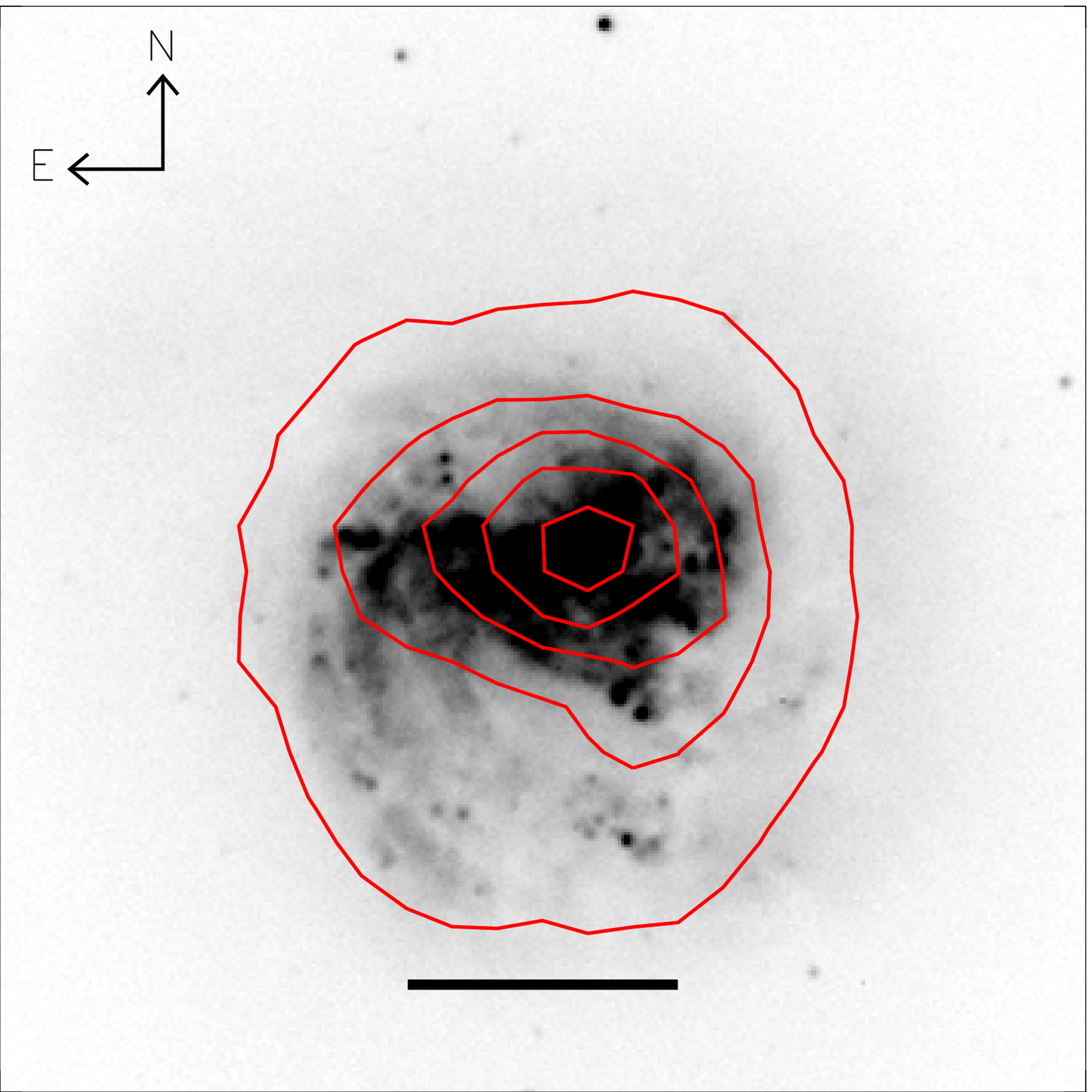}}
\includegraphics[scale=0.55,clip=true,trim=0.5cm 0 0.5cm 0]{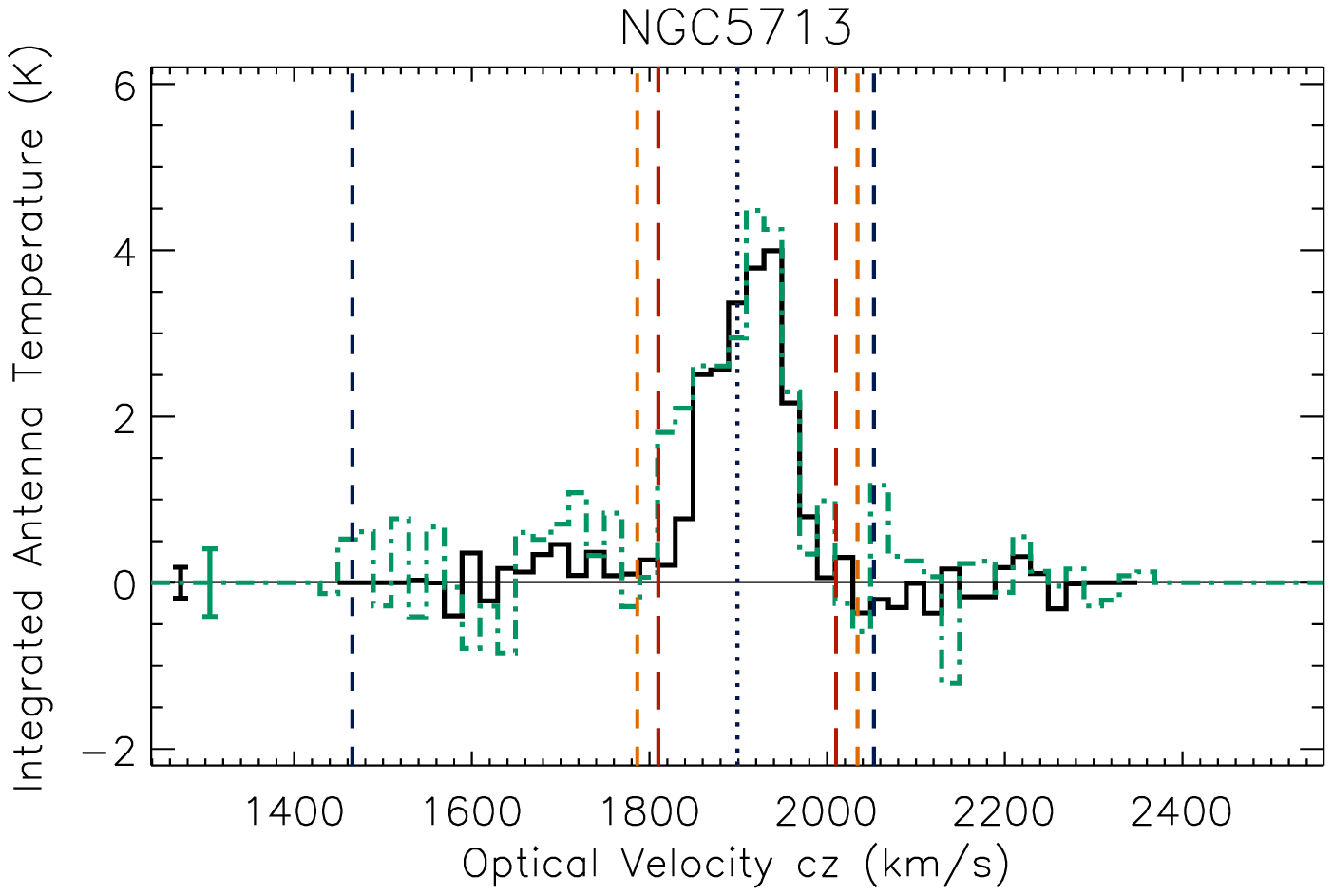}
\includegraphics[scale=0.55,clip=true,trim=1cm 0 0.5cm 0]{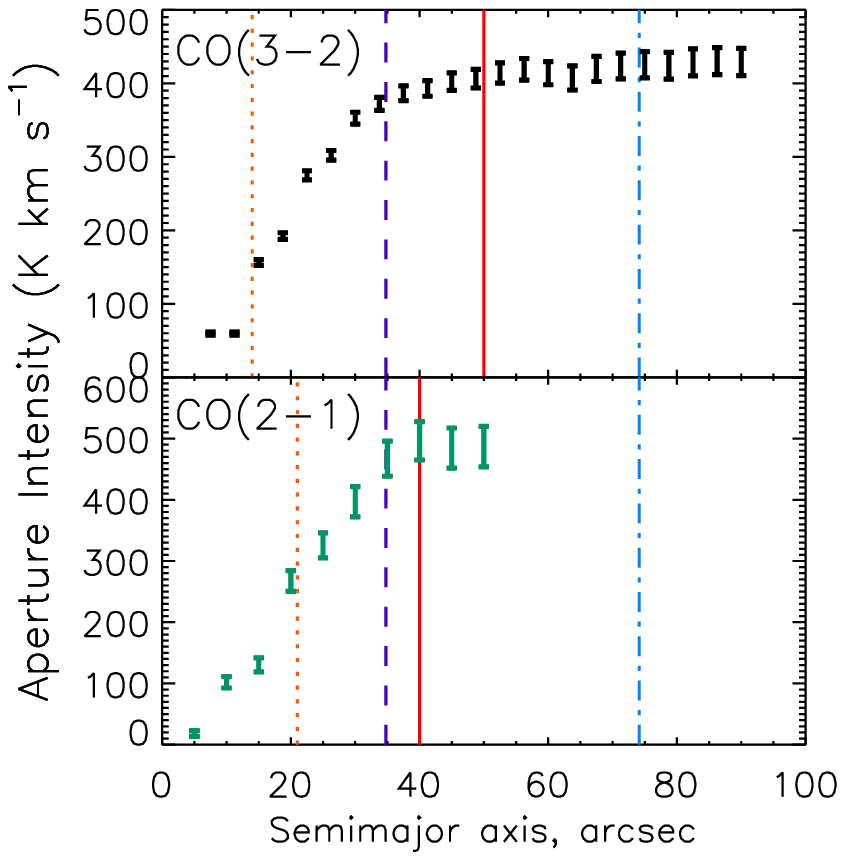}
\\
\raisebox{1cm}{\includegraphics[scale=0.253]{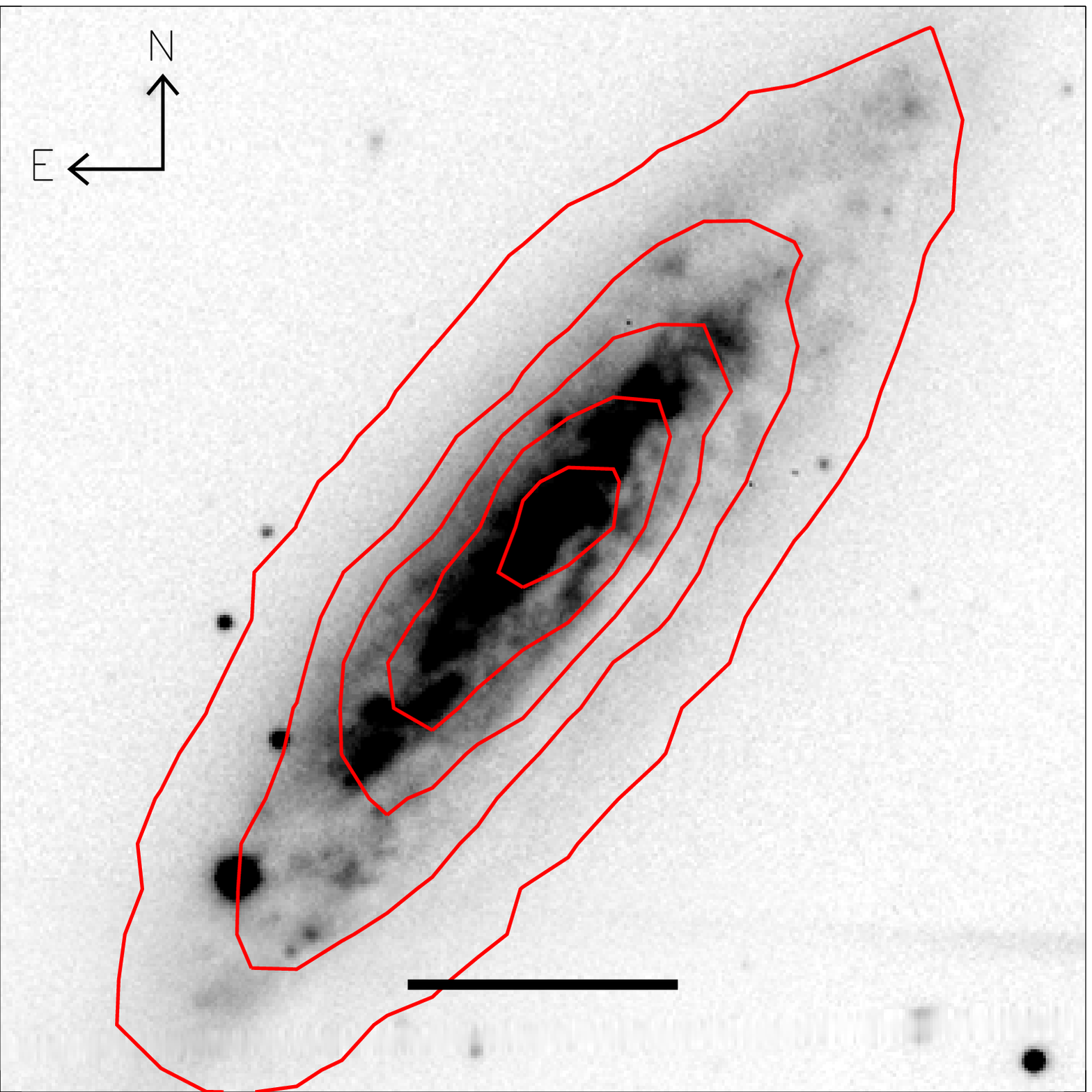}}
\includegraphics[scale=0.55,clip=true,trim=0.5cm 0 0.5cm 0]{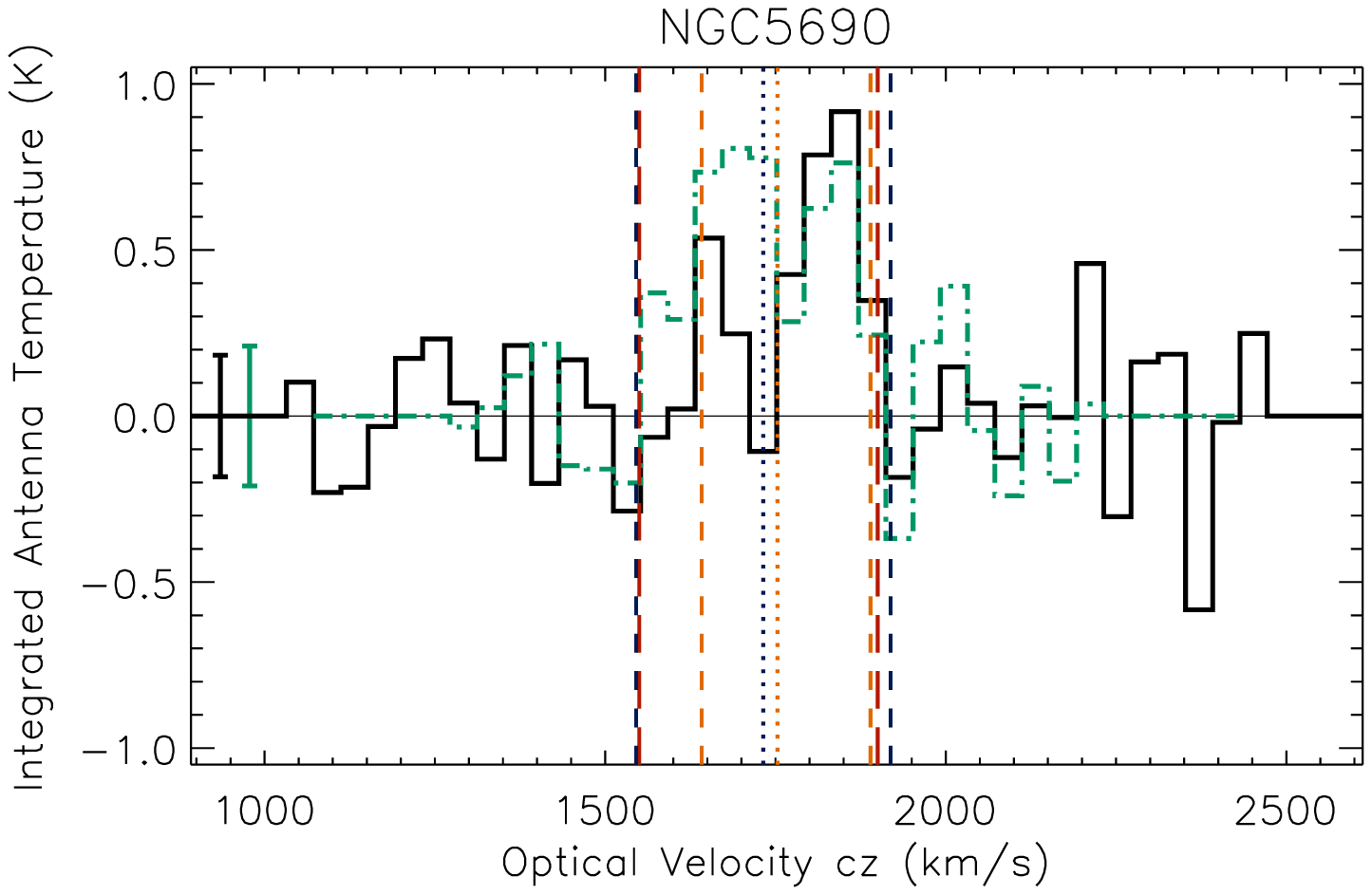}
\includegraphics[scale=0.55,clip=true,trim=1cm 0 0.5cm 0]{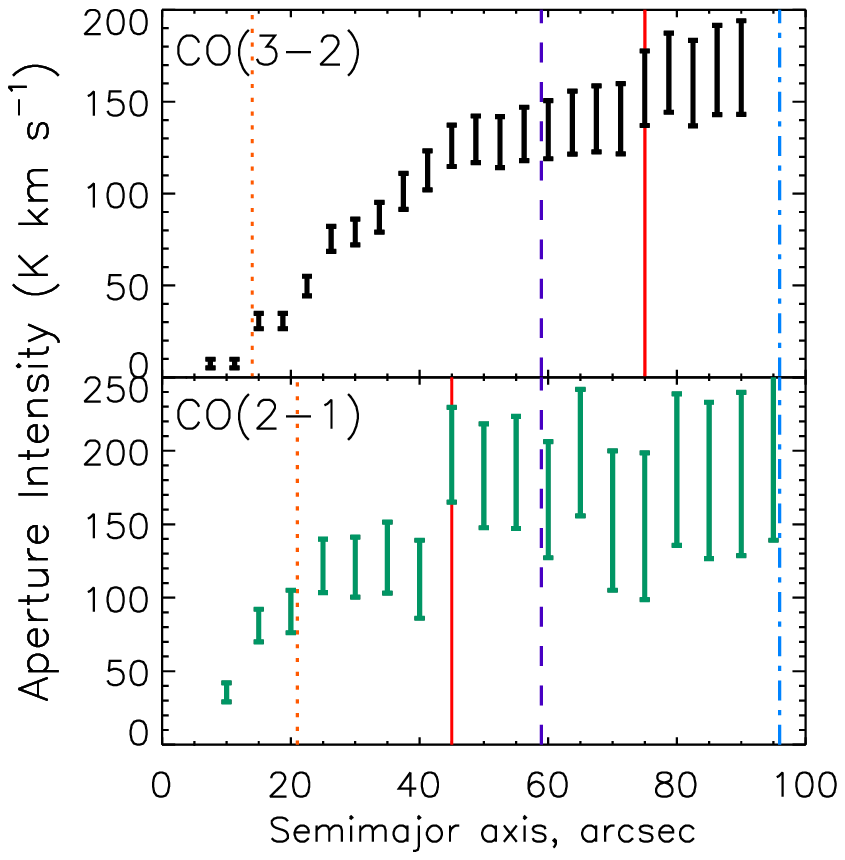}
\\
\raisebox{1cm}{\includegraphics[scale=0.253]{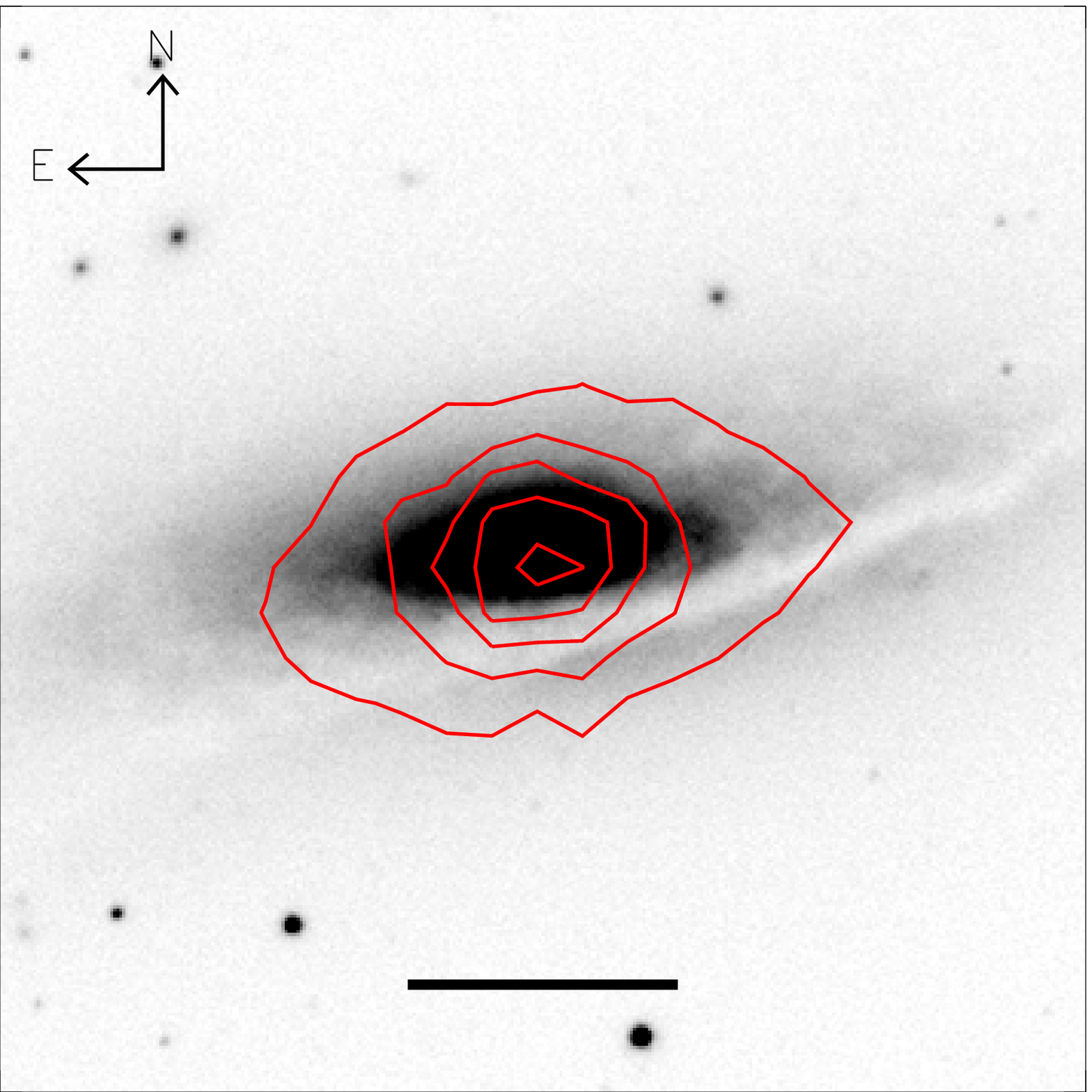}}
\includegraphics[scale=0.55,clip=true,trim=0.5cm 0 0.5cm 0]{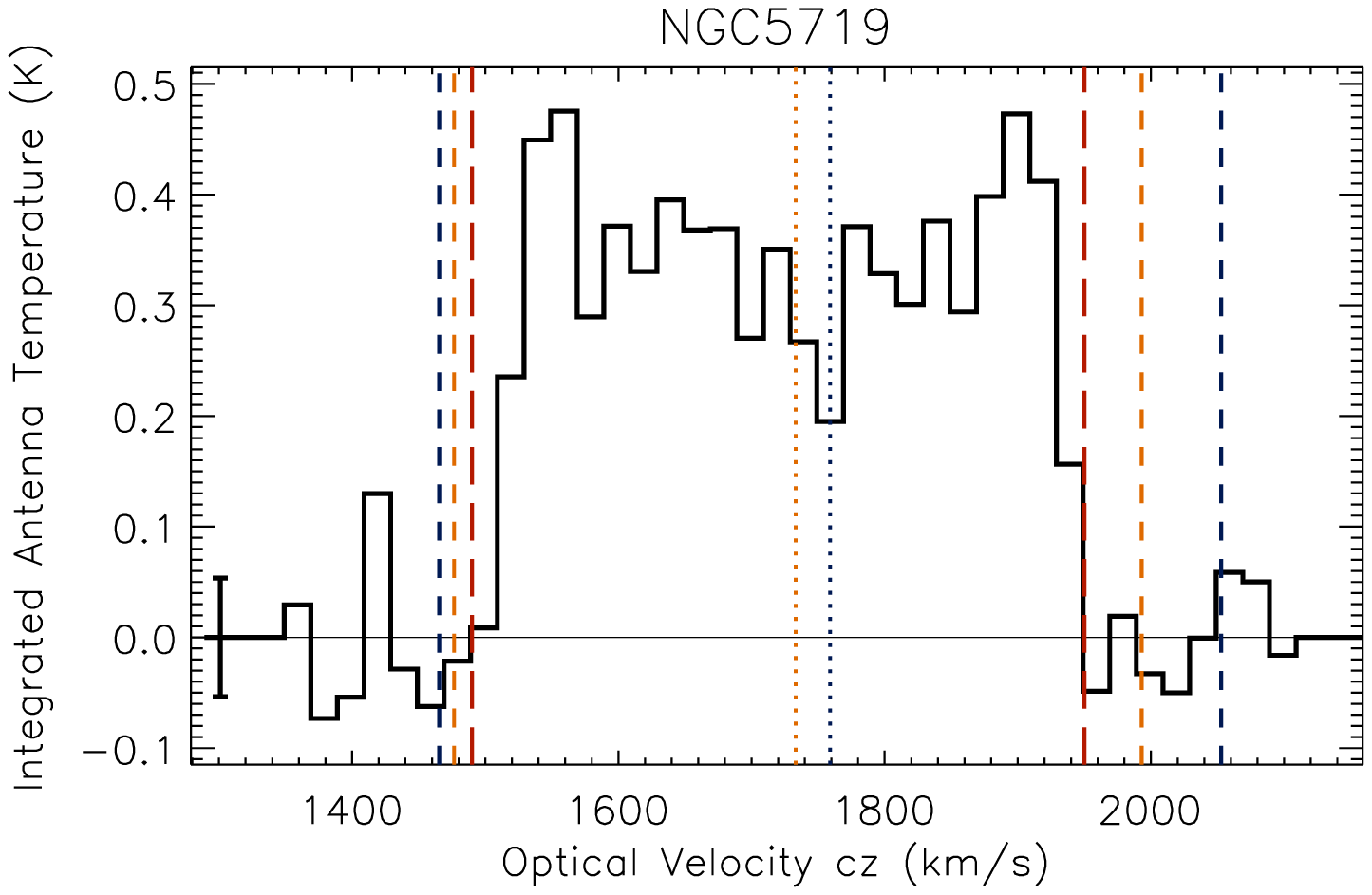}
\includegraphics[scale=0.55,clip=true,trim=1cm 0 0.5cm 0]{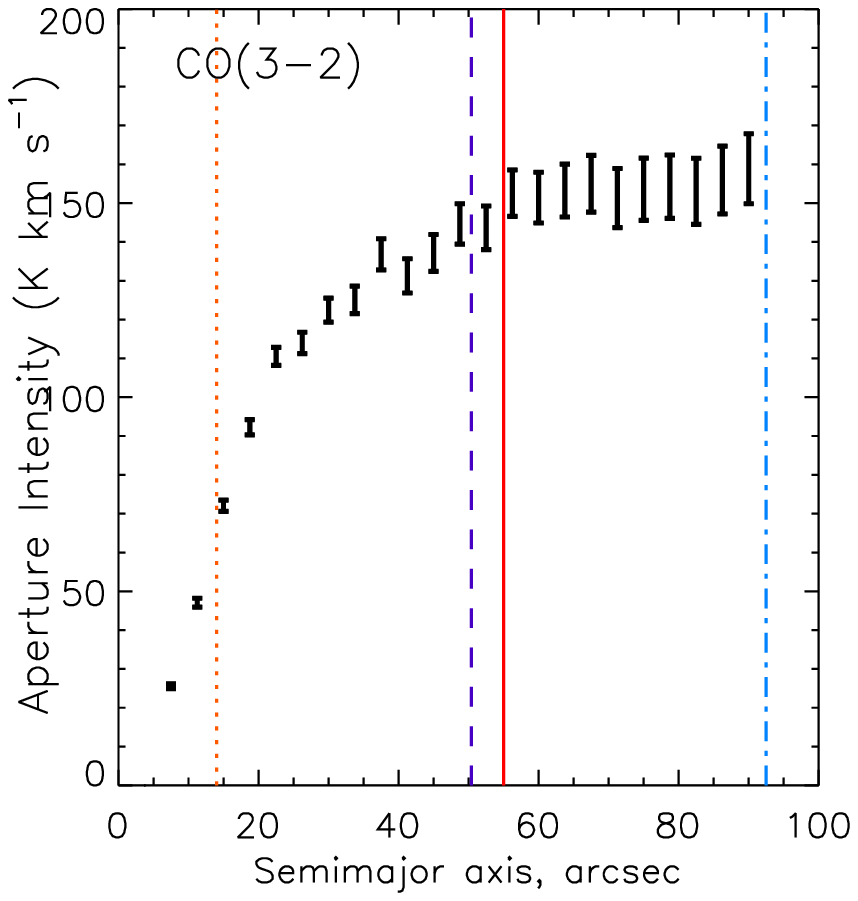}
\\
\raisebox{1cm}{\includegraphics[scale=0.253]{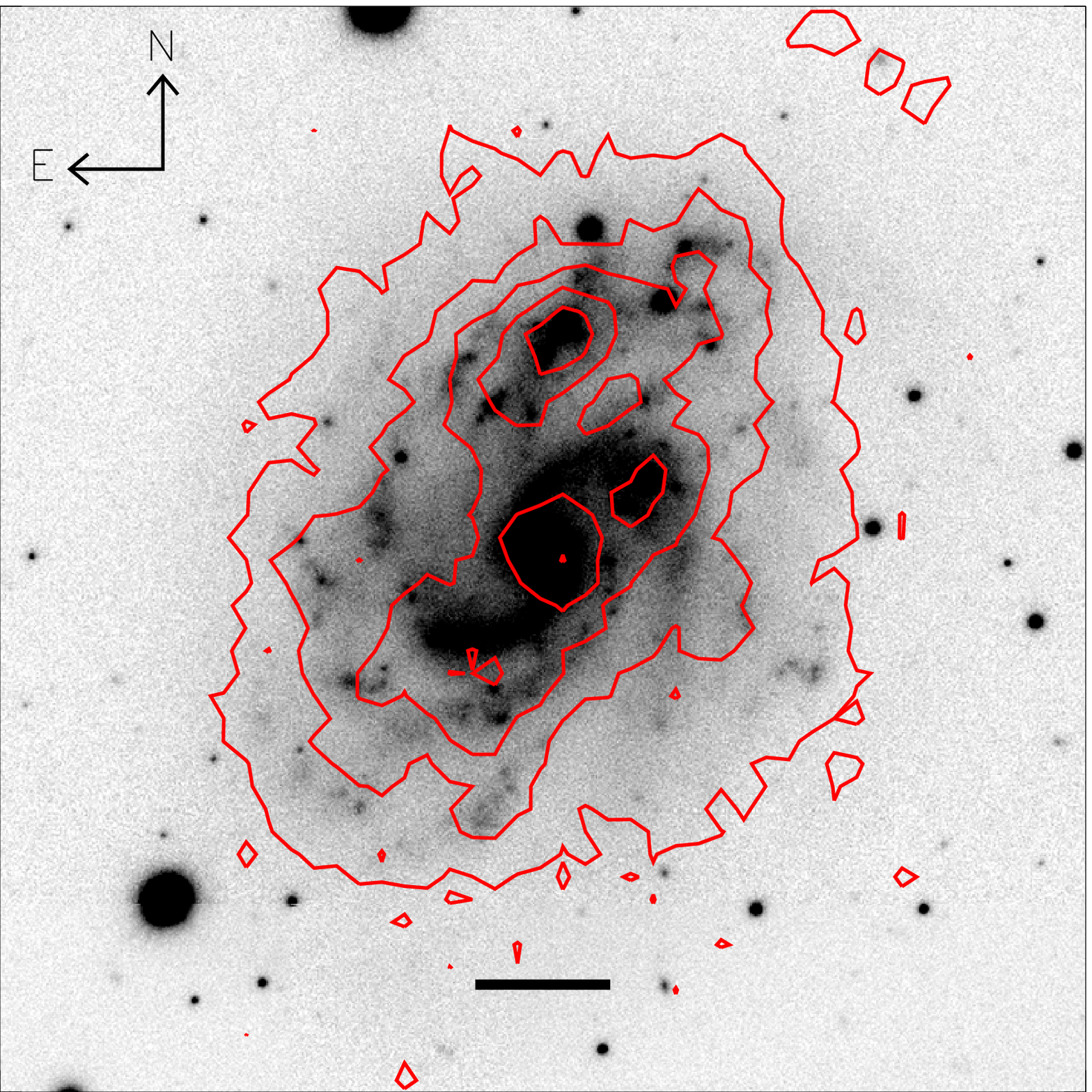}}
\includegraphics[scale=0.55,clip=true,trim=0.5cm 0 0.5cm 0]{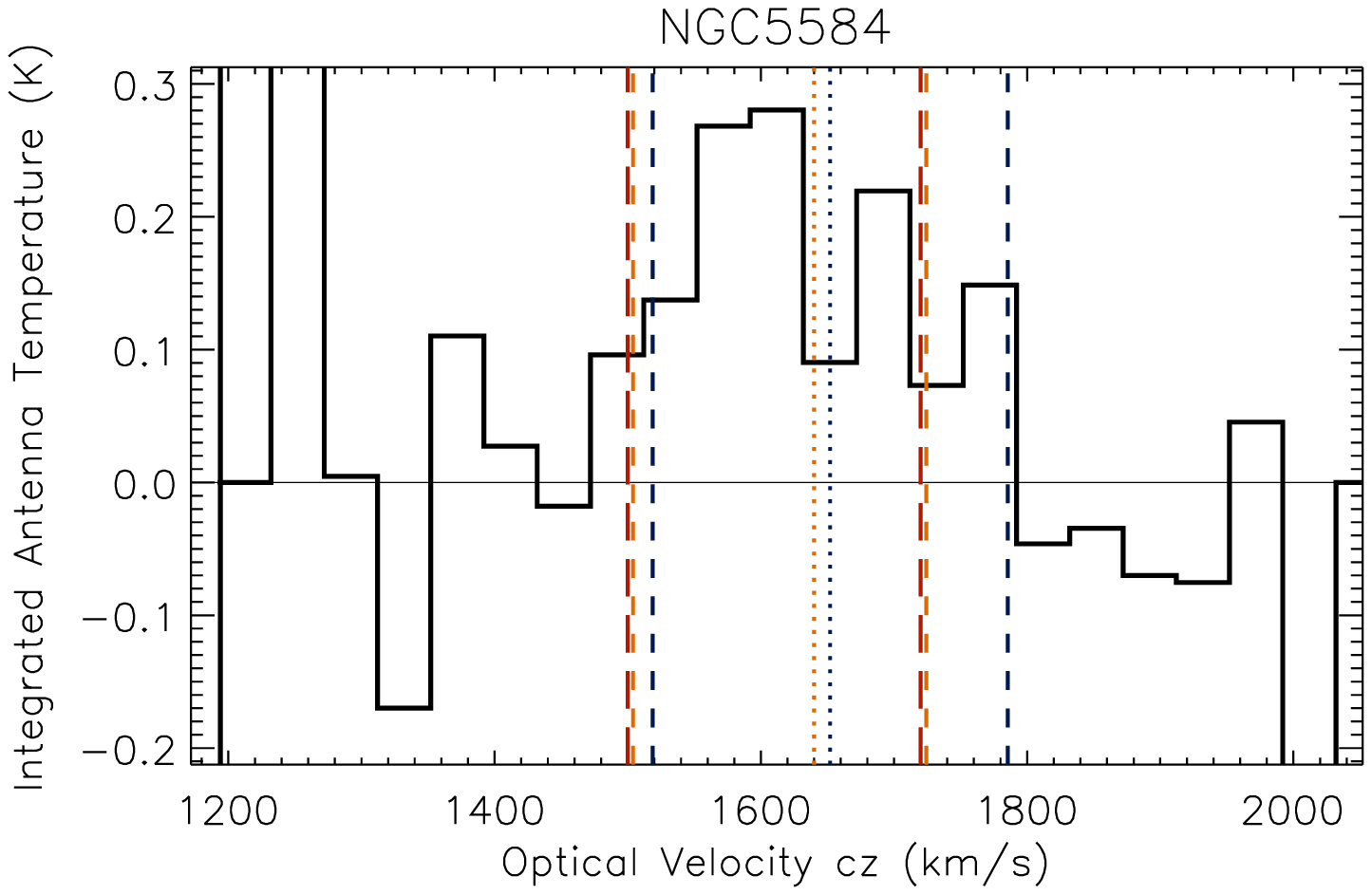}
\includegraphics[scale=0.55,clip=true,trim=1cm 0 0.5cm 0]{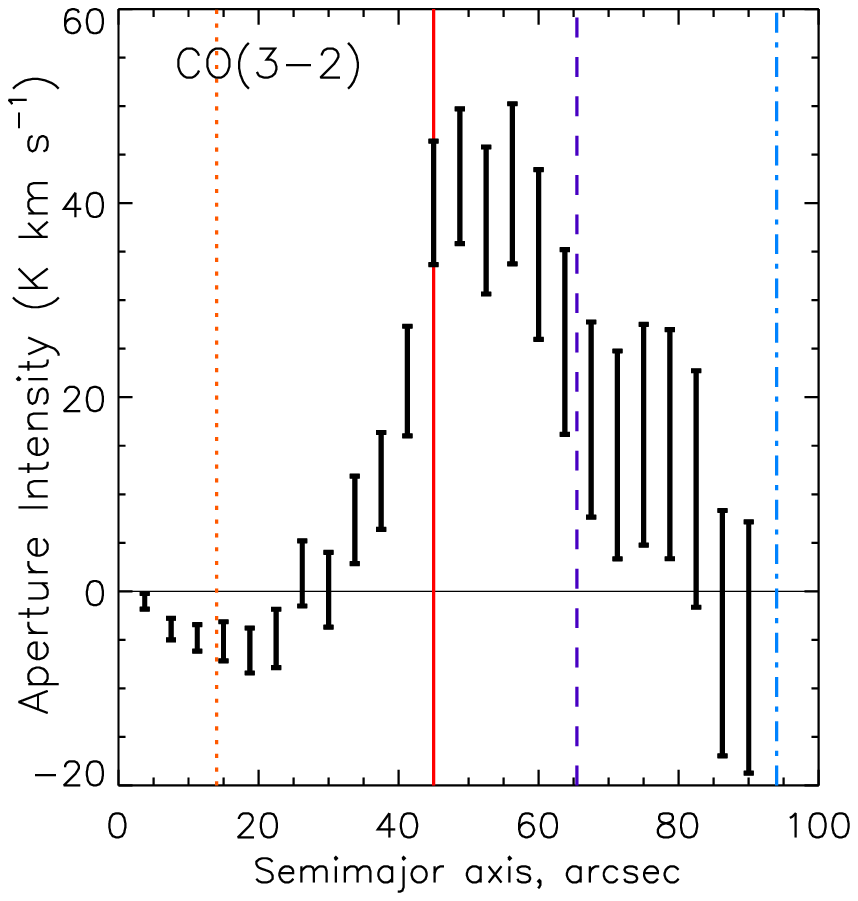}
\caption{\textbf{Left:} SDSS $r$-band images (2\arcmin$\times$2\arcmin) with 250\mum\ contours at the 10$^\text{th}$, 30$^\text{th}$, 50$^\text{th}$, 70$^\text{th}$, \&\ 90$^\text{th}$ percentiles. The horizontal black bar in each image is 30\arcsec\ in length. \textbf{Middle:} aperture-integrated spectra in HARP (black) and RxA (green); blue lines = \hi\ line width from HIPASS; orange = CO line width from literature; red = CO width integrated over for moment maps. 
\textbf{Right:} curves of growth showing integrated intensity as a function of aperture semimajor axis; blue dot-dashed line = semimajor axis of B-band $d_{25}$ isophote; orange dotted line = FWHM of HARP/RxA beam; violet dashed line = 250\mum\ HWHM along major axis; red solid line = aperture chosen to enclose all CO emission.
SDP~4 and SDP~15 were observed as single pointings so there is no curve of growth.
The anomalous curve-of-growth of NGC~5584 is discussed in Section~\ref{sec:codata}.
}
\label{fig:spectra}
\end{figure*}

Curves of growth and spectra for the galaxies that were observed are shown in Figure~\ref{fig:spectra}.
Some of the cubes produced fluctuating curves of growth as a result of poor baseline subtraction. The worst of these is NGC~5584, which has low surface brightness and is extended across the full HARP footprint. The HARP cube appears to contain emission coincident with the spiral arms and within the velocity range of literature CO and \hi\ detections, but the cumulative flux falls in larger apertures due to residual baselines that could not be adequately subtracted with a first or second order fit. As a result it is impossible to be sure whether the flux measured is over- or underestimated, hence this measurement was excluded from the analysis. The curves of growth for CGCG~018-077 and SDP~1 show similar fluctuations outside of the error bars due to residual baselines at high radii, but these are restricted to pixels outside of the optical and submm emission regions, and visual inspection of the cubes indicates that the baselines in the centre have been adequately subtracted so that the aperture measurements 
used are reliable.

The CO measurements are summarised in Table~\ref{tab:aperdata}, and are compared to CO line widths from the literature and the \hi\ widths measured in Section~\ref{sec:hidata}. Literature CO line widths and those from the NRO data set are often smaller than those in the HARP and RxA cubes because they are single-beam measurements containing only the nucleus. \hi\ line widths can differ in general because they trace a different gas phase (which is generally more extended), although in the case of NGC~5713 and NGC~5719 the \hi\ lines are blended.

\begin{figure*}
\raisebox{1cm}{\includegraphics[scale=0.253]{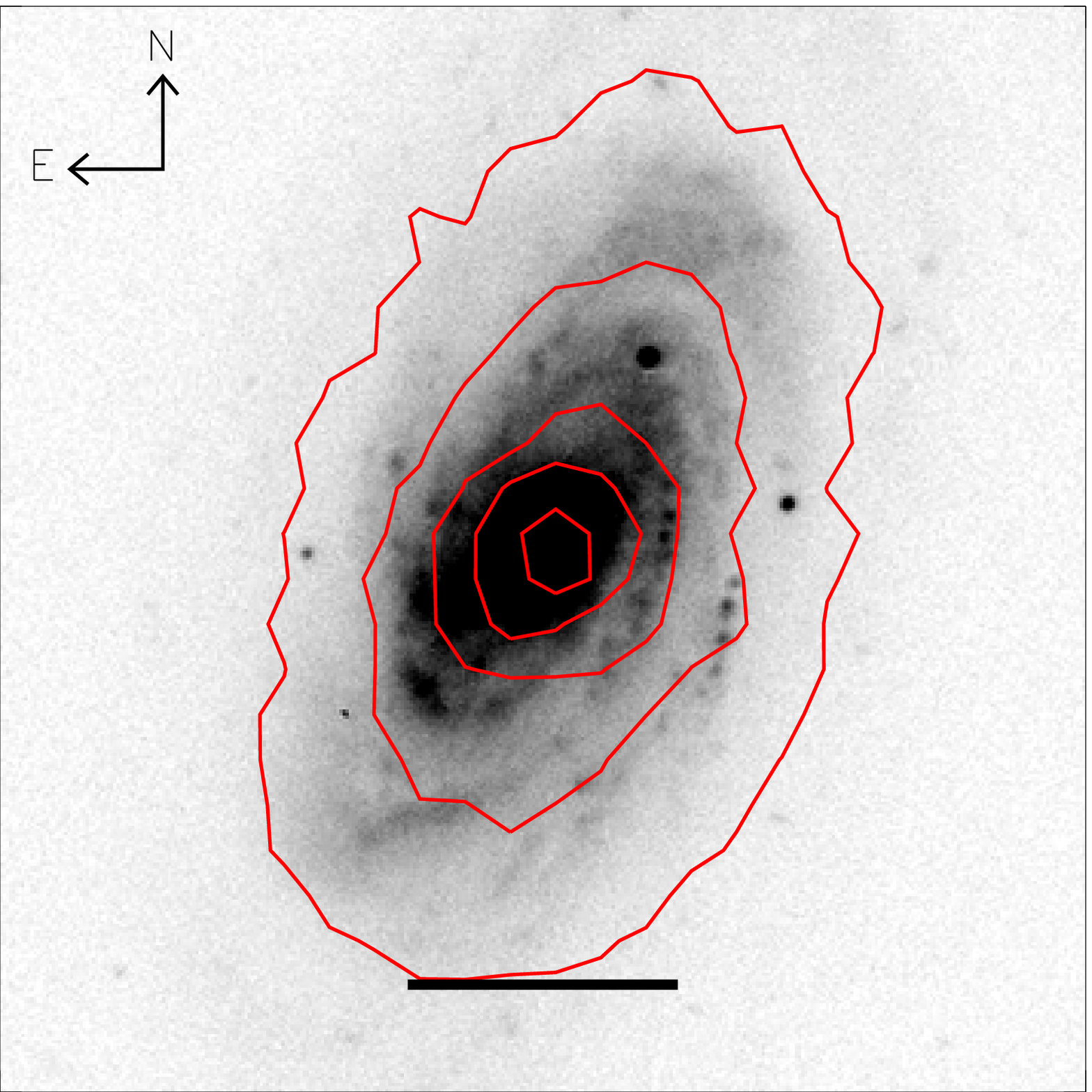}}
\includegraphics[scale=0.55,clip=true,trim=0.5cm 0 0.5cm 0]{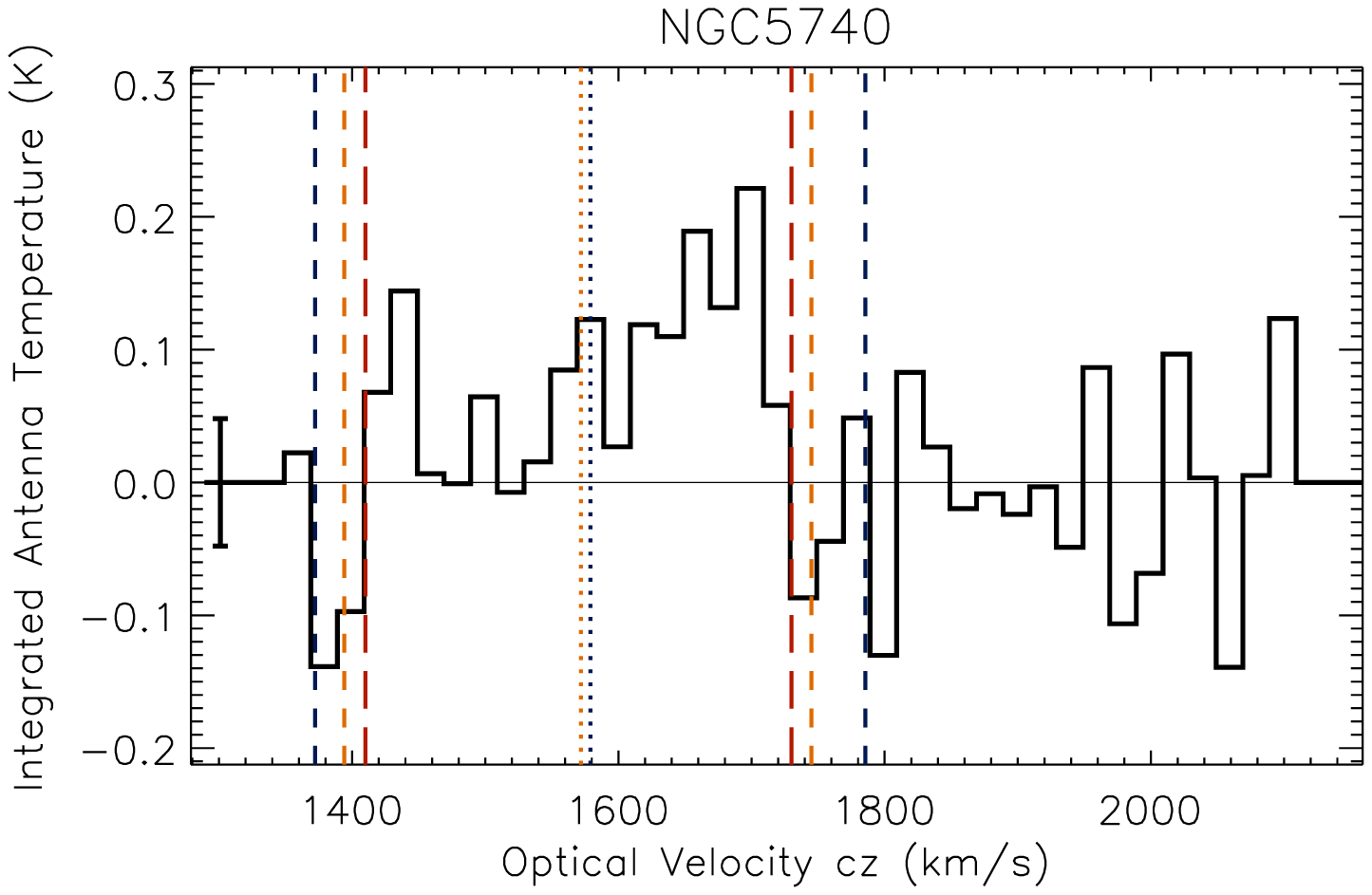}
\includegraphics[scale=0.55,clip=true,trim=1cm 0 0.5cm 0]{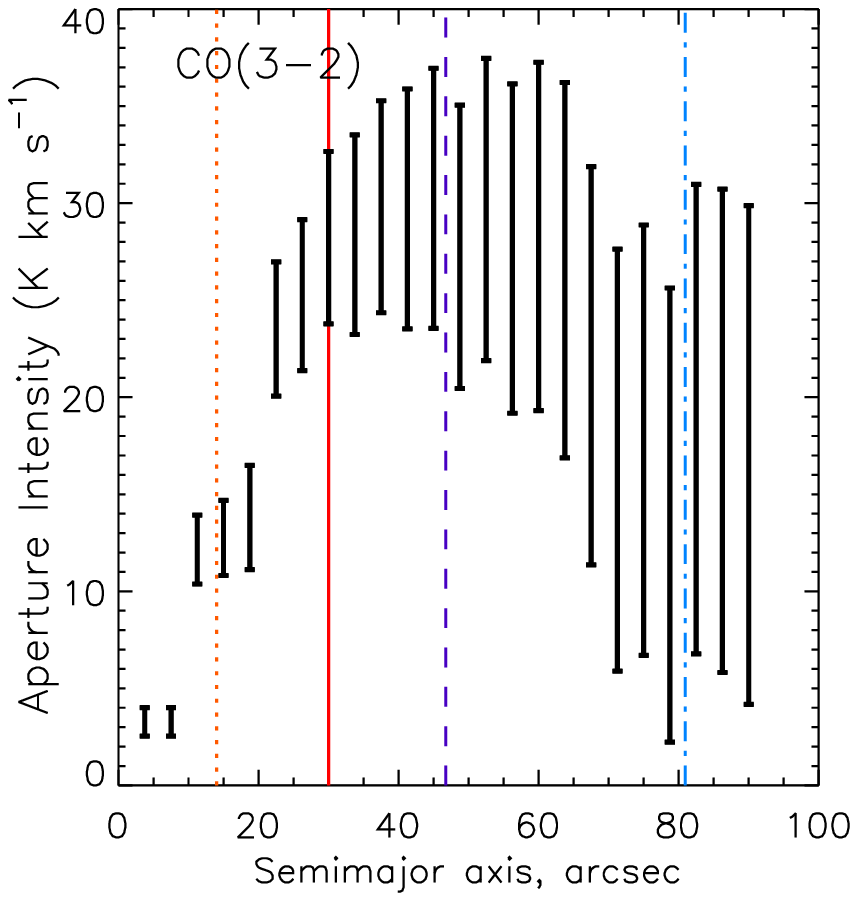}
\\
\raisebox{1cm}{\includegraphics[scale=0.253]{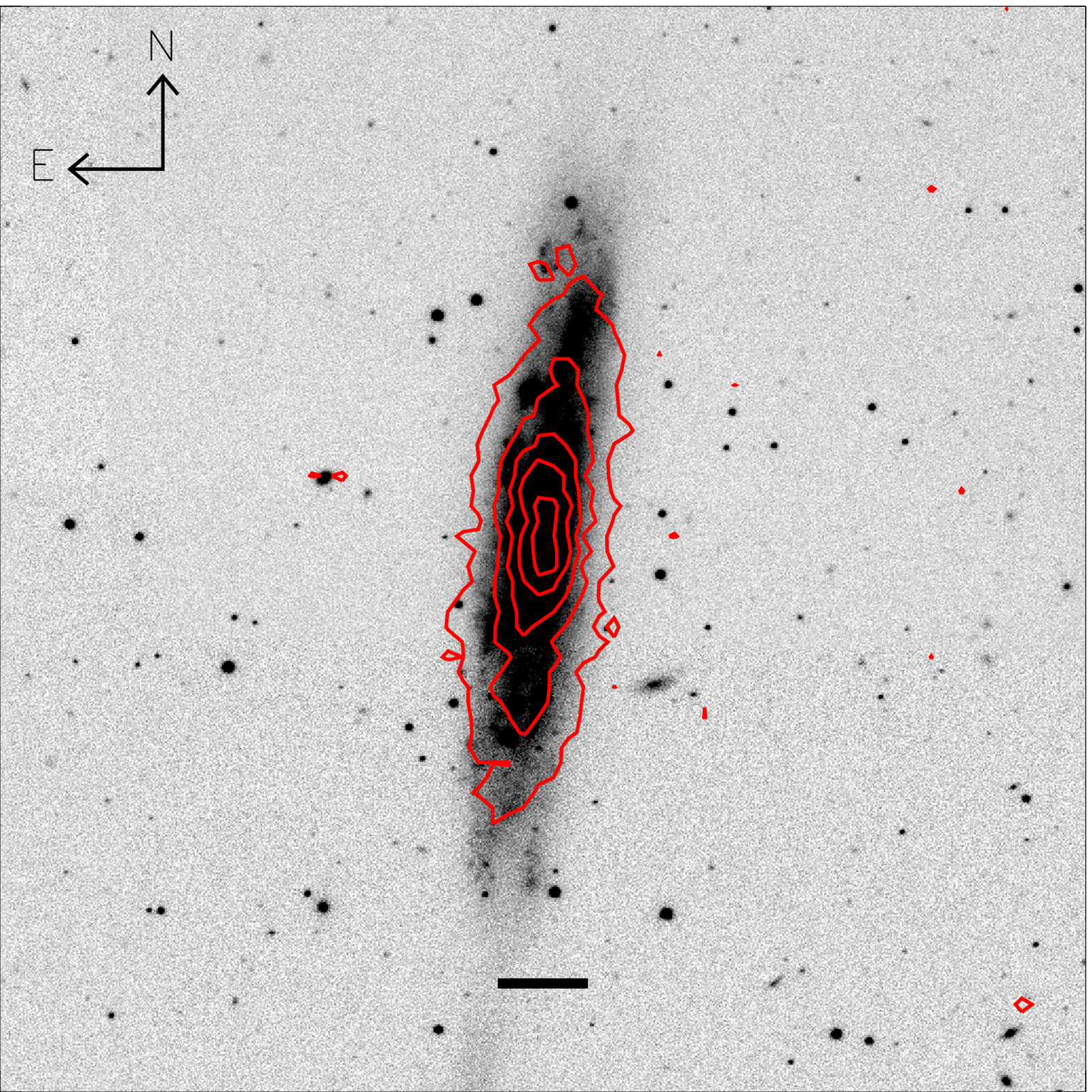}}
\includegraphics[scale=0.55,clip=true,trim=0.5cm 0 0.5cm 0]{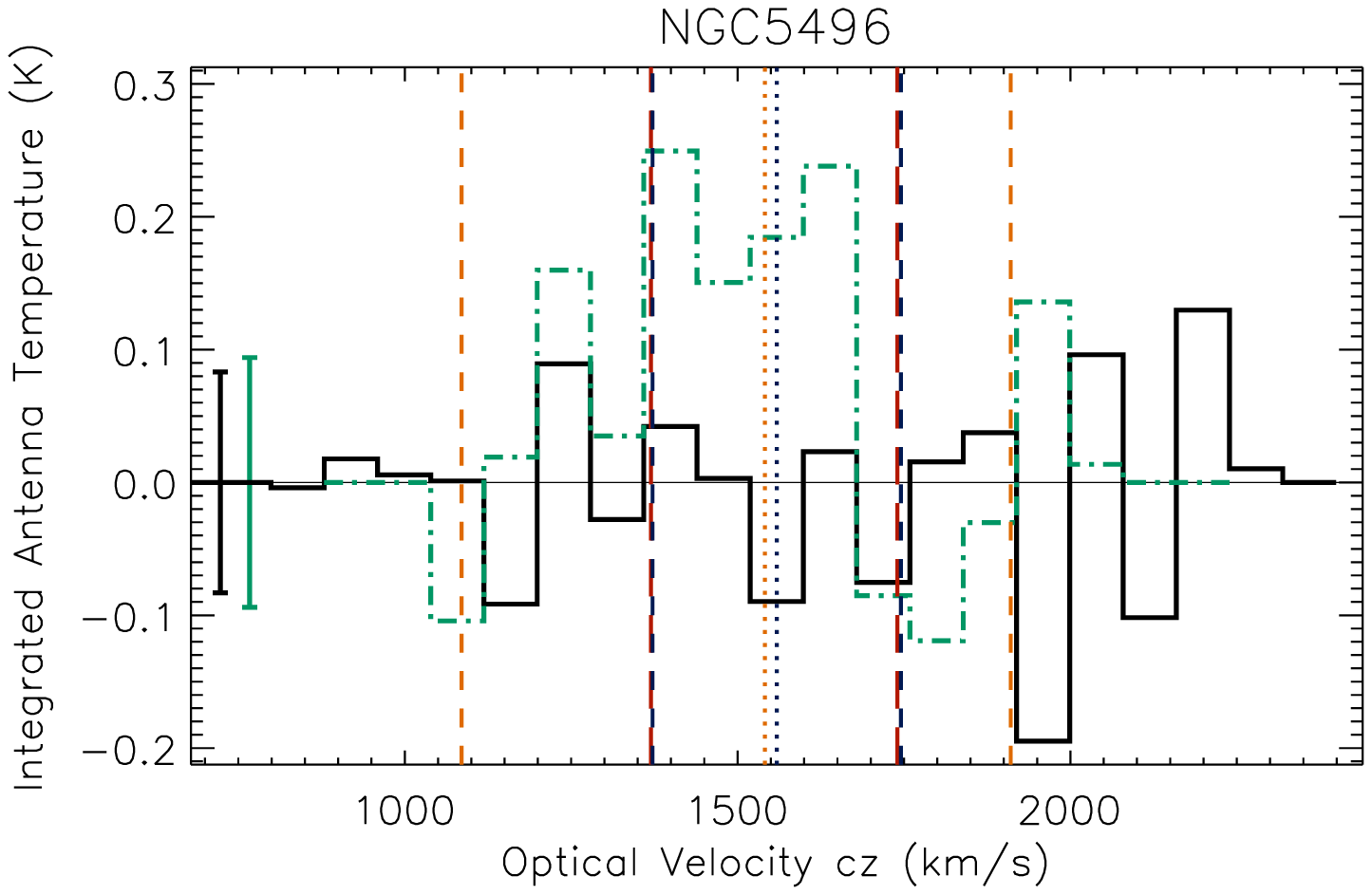}
\includegraphics[scale=0.55,clip=true,trim=1cm 0 0.5cm 0]{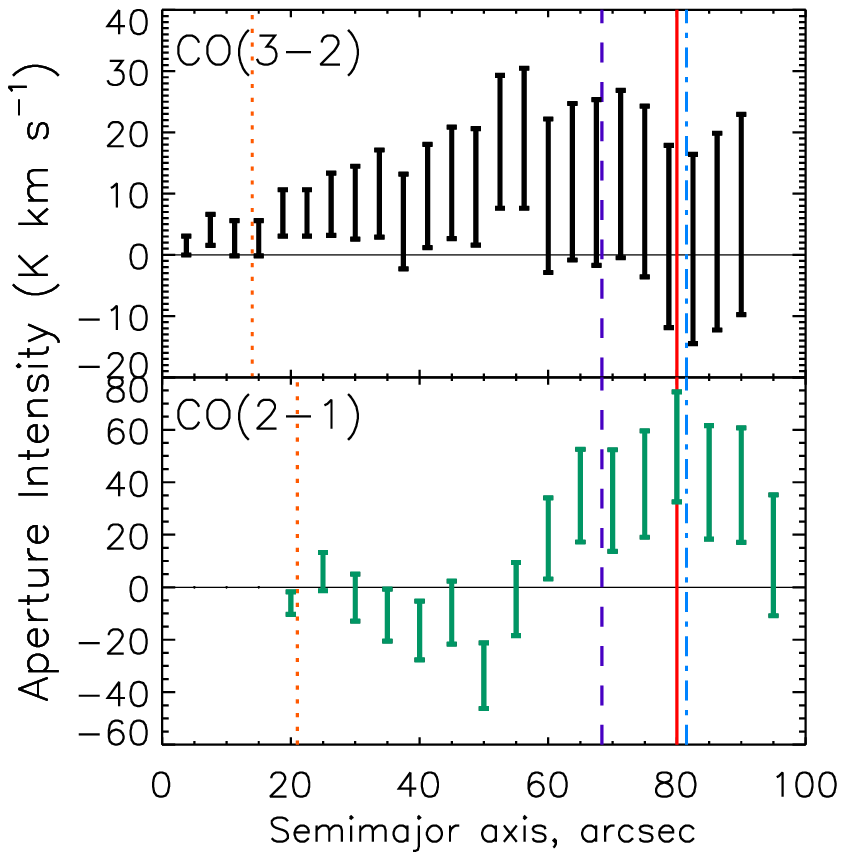}
\\
\raisebox{1cm}{\includegraphics[scale=0.253]{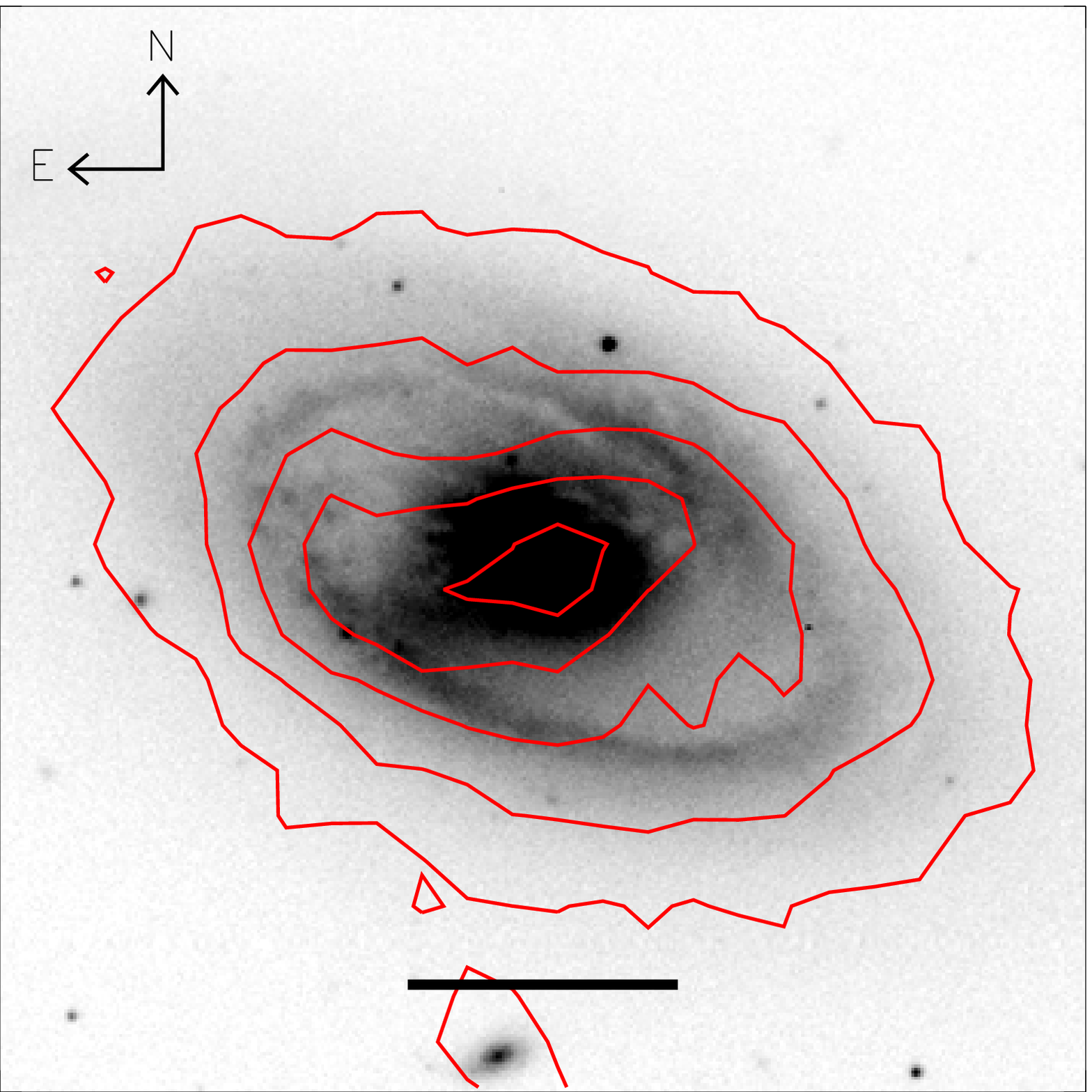}}
\includegraphics[scale=0.55,clip=true,trim=0.5cm 0 0.5cm 0]{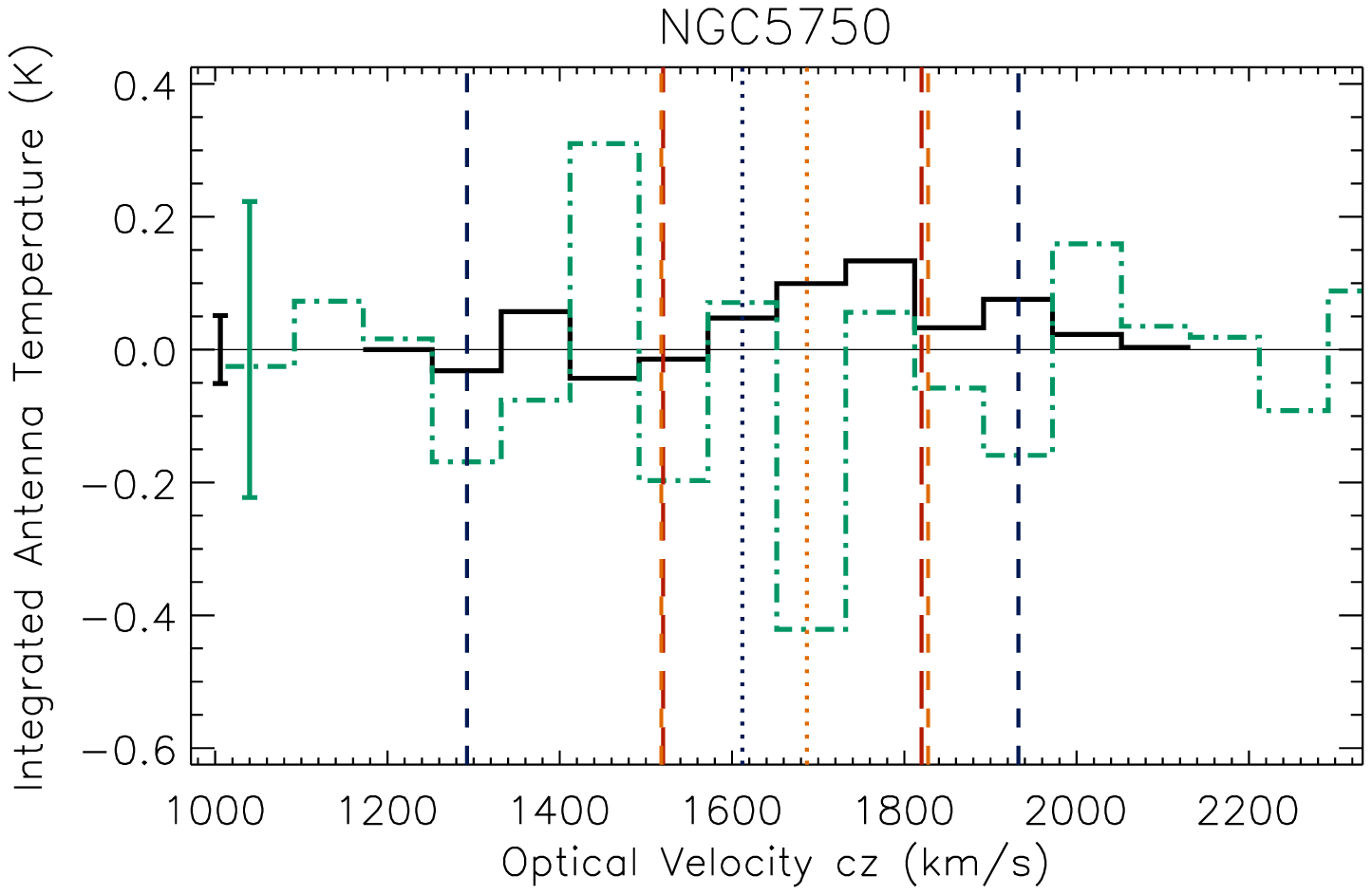}
\includegraphics[scale=0.55,clip=true,trim=1cm 0 0.5cm 0]{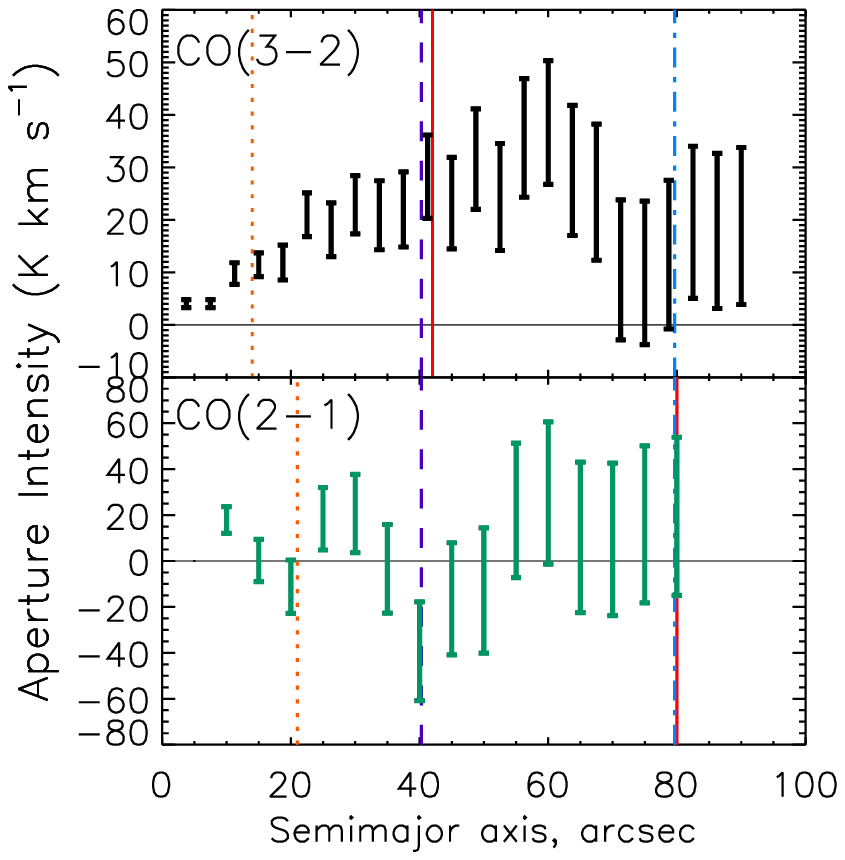}
\\
\raisebox{1cm}{\includegraphics[scale=0.253]{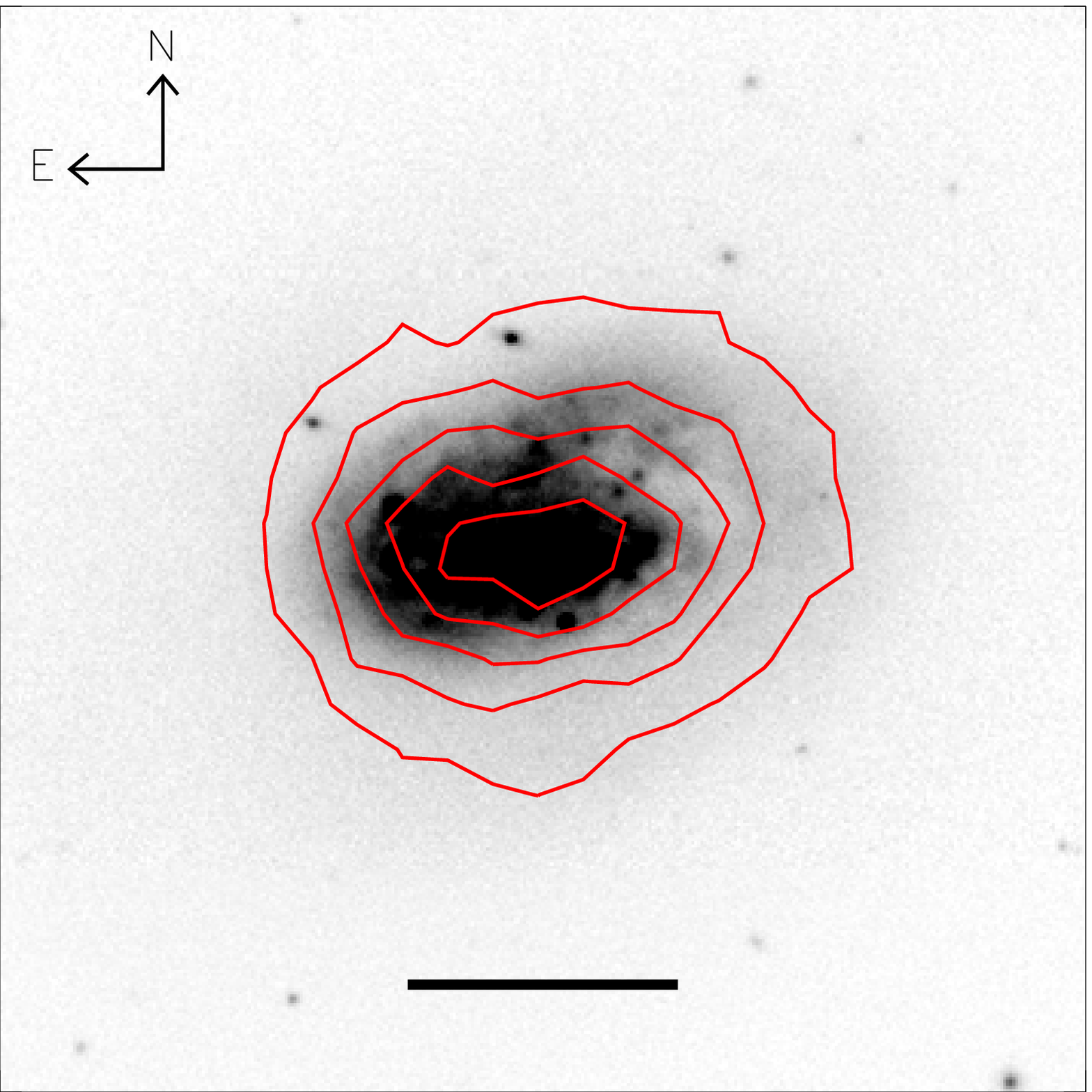}}
\includegraphics[scale=0.55,clip=true,trim=0.5cm 0 0.5cm 0]{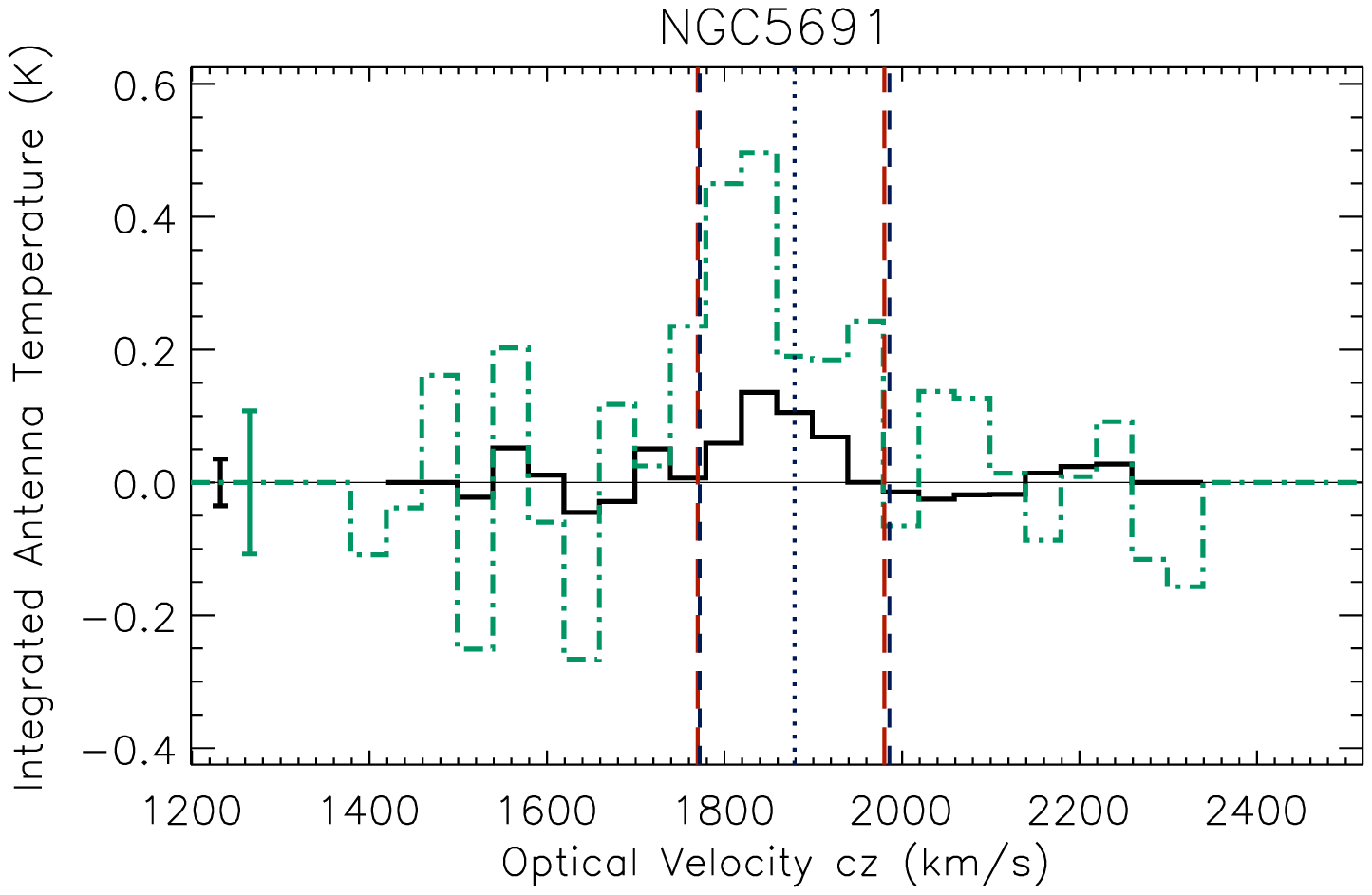}
\includegraphics[scale=0.55,clip=true,trim=1cm 0 0.5cm 0]{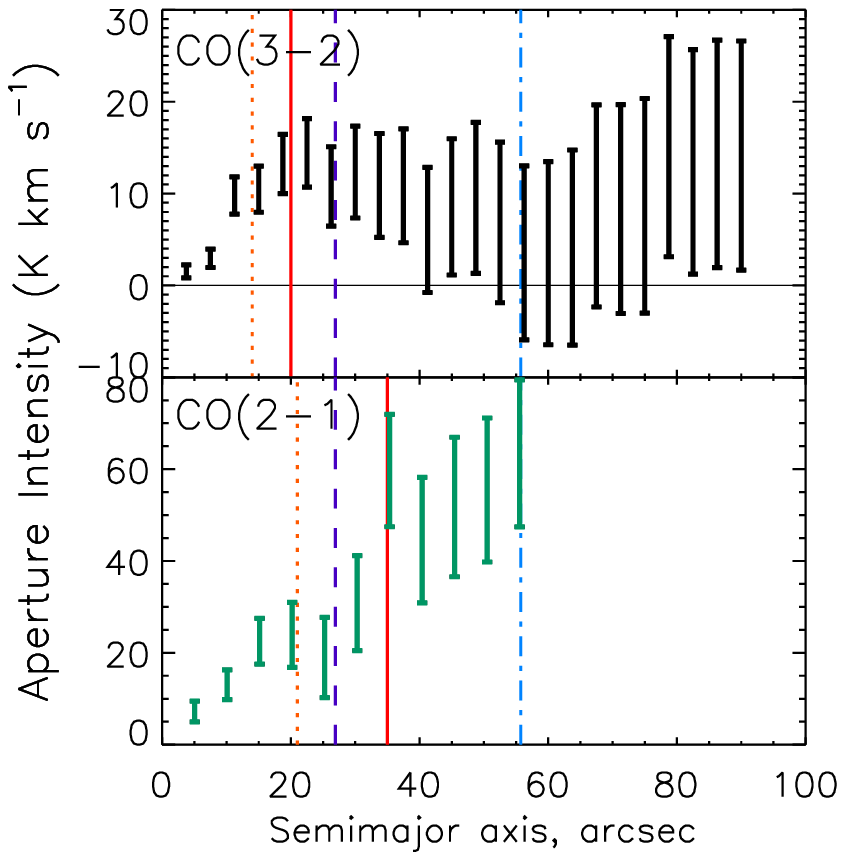}
\\
\centering {\bf Figure~1.} (continued)
\label{fig:spectra2}
\end{figure*}
\begin{figure*}
\raisebox{1cm}{\includegraphics[scale=0.253]{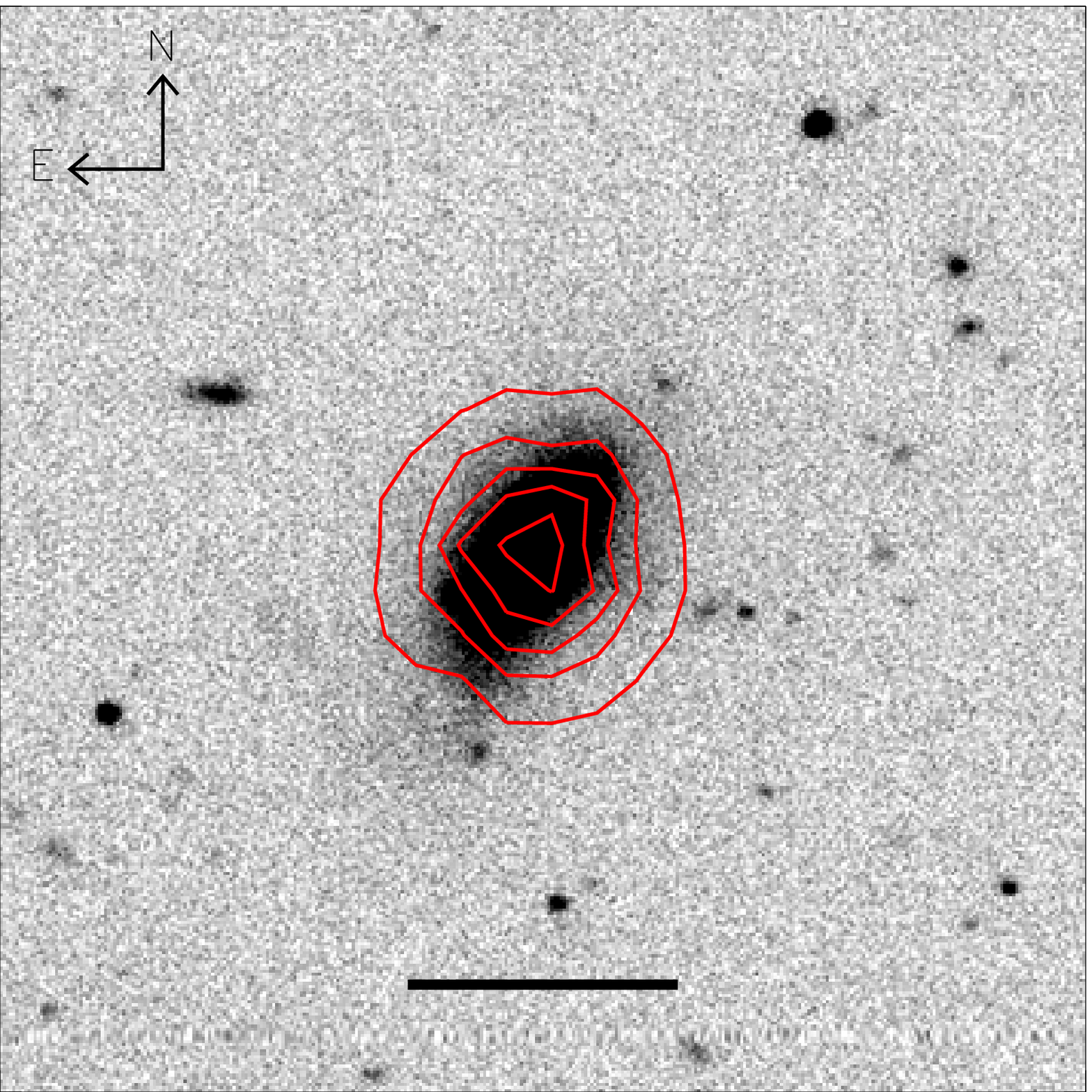}}
\includegraphics[scale=0.55,clip=true,trim=0.5cm 0 0.5cm 0]{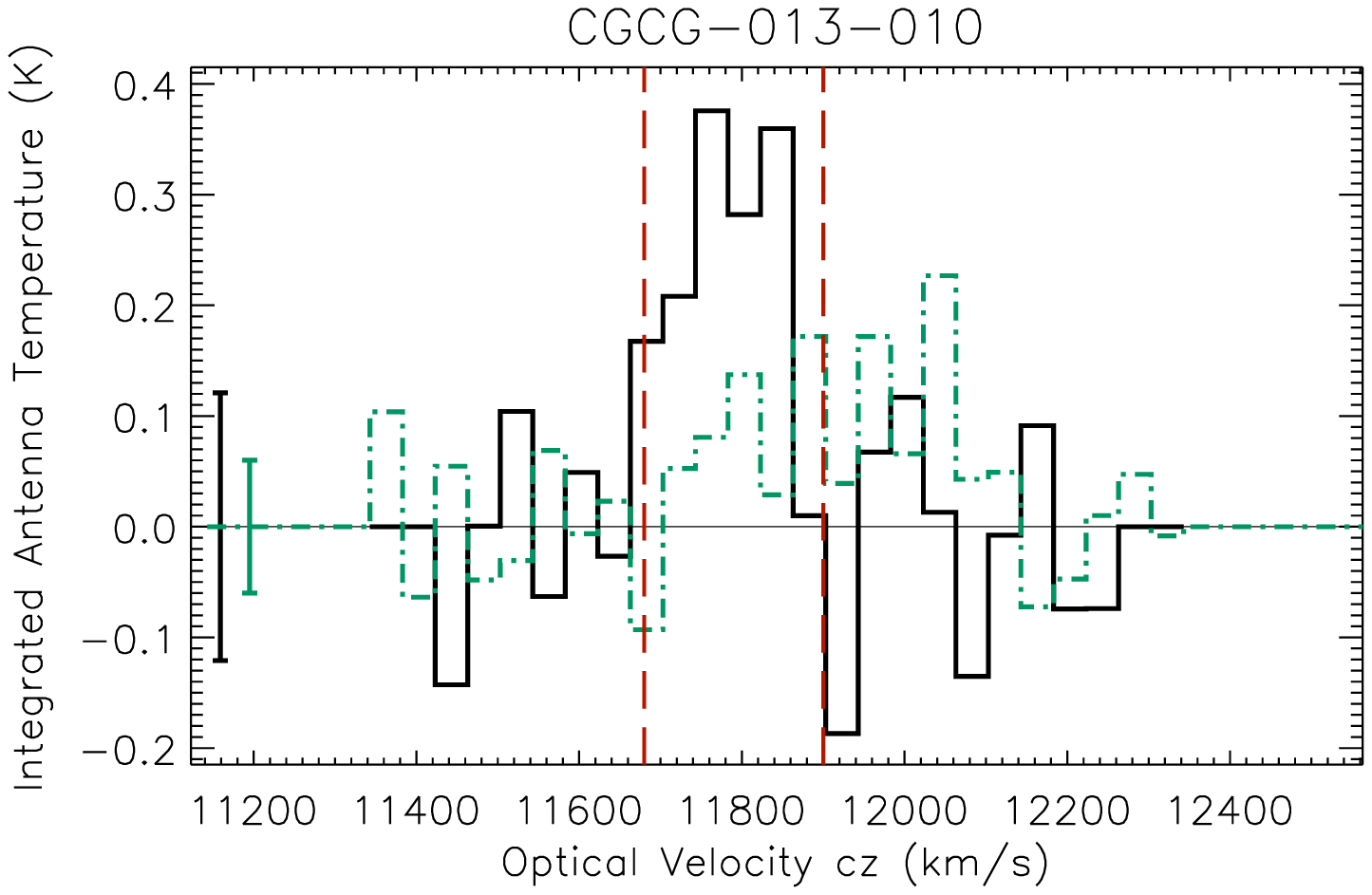}
\includegraphics[scale=0.55,clip=true,trim=1cm 0 0.5cm 0]{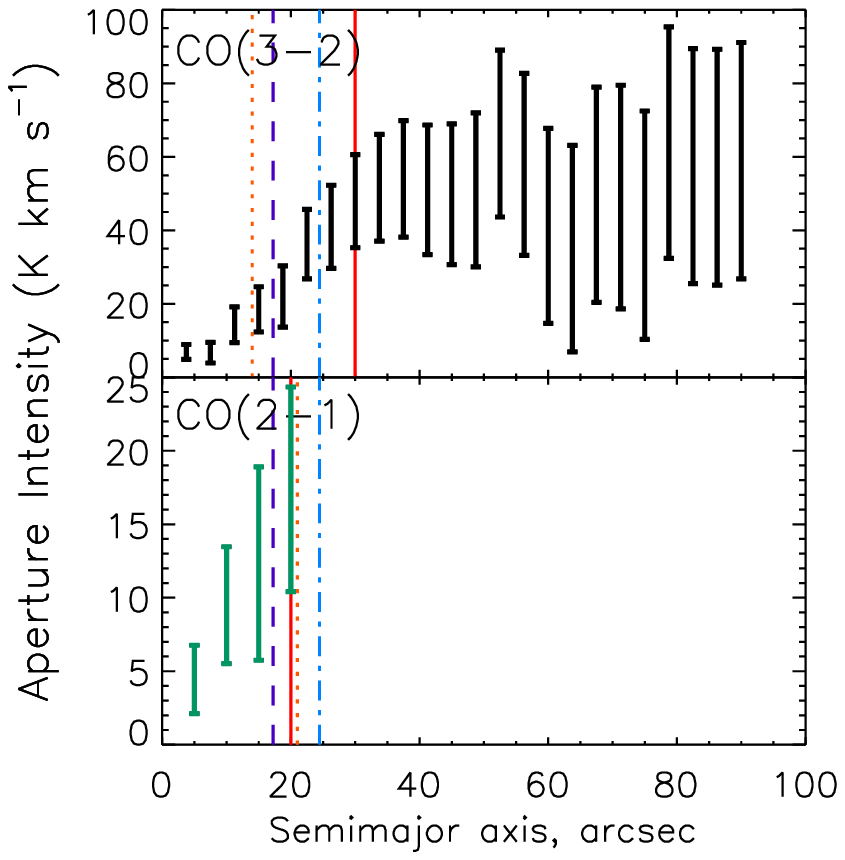}
\\
\raisebox{1cm}{\includegraphics[scale=0.253]{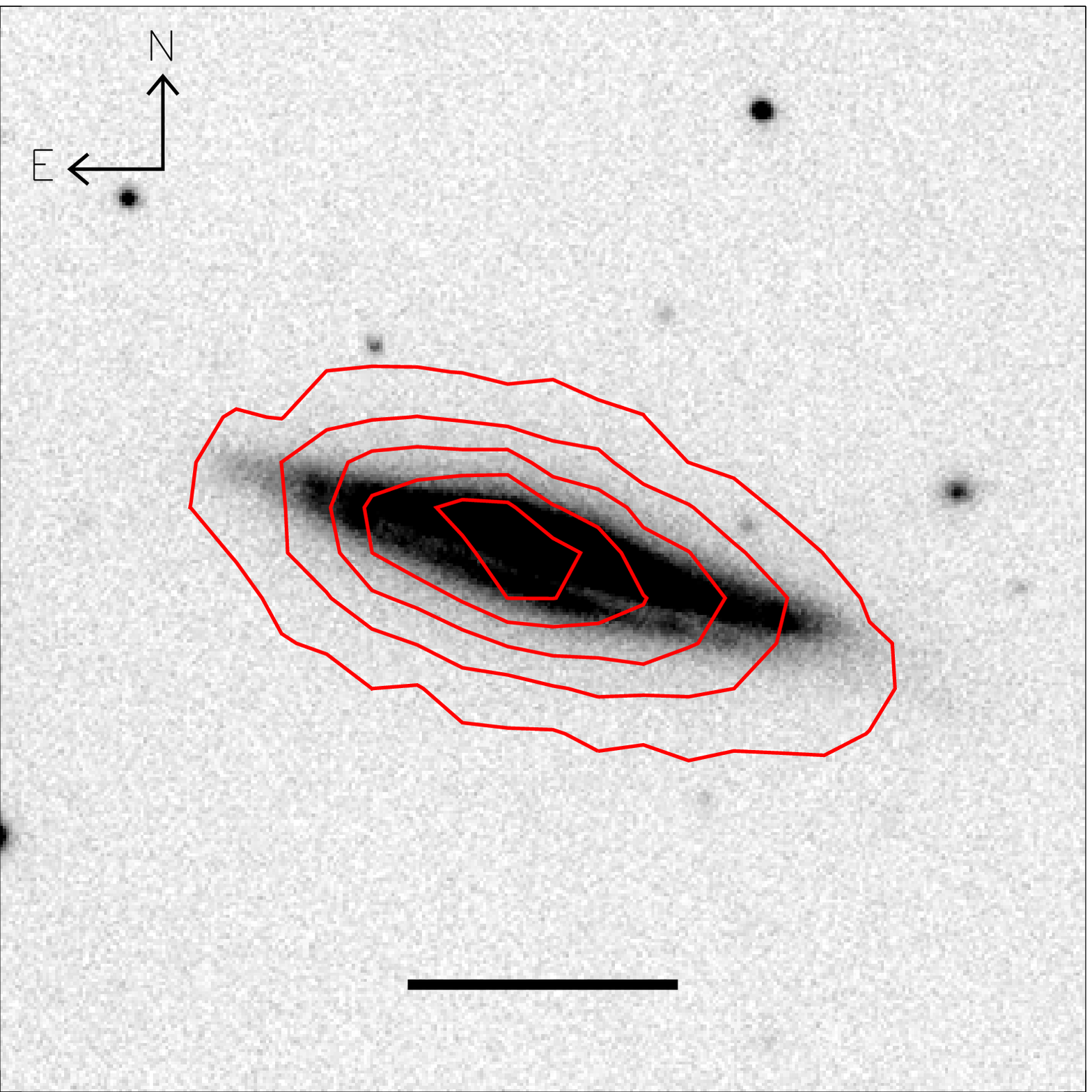}}
\includegraphics[scale=0.55,clip=true,trim=0.5cm 0 0.5cm 0]{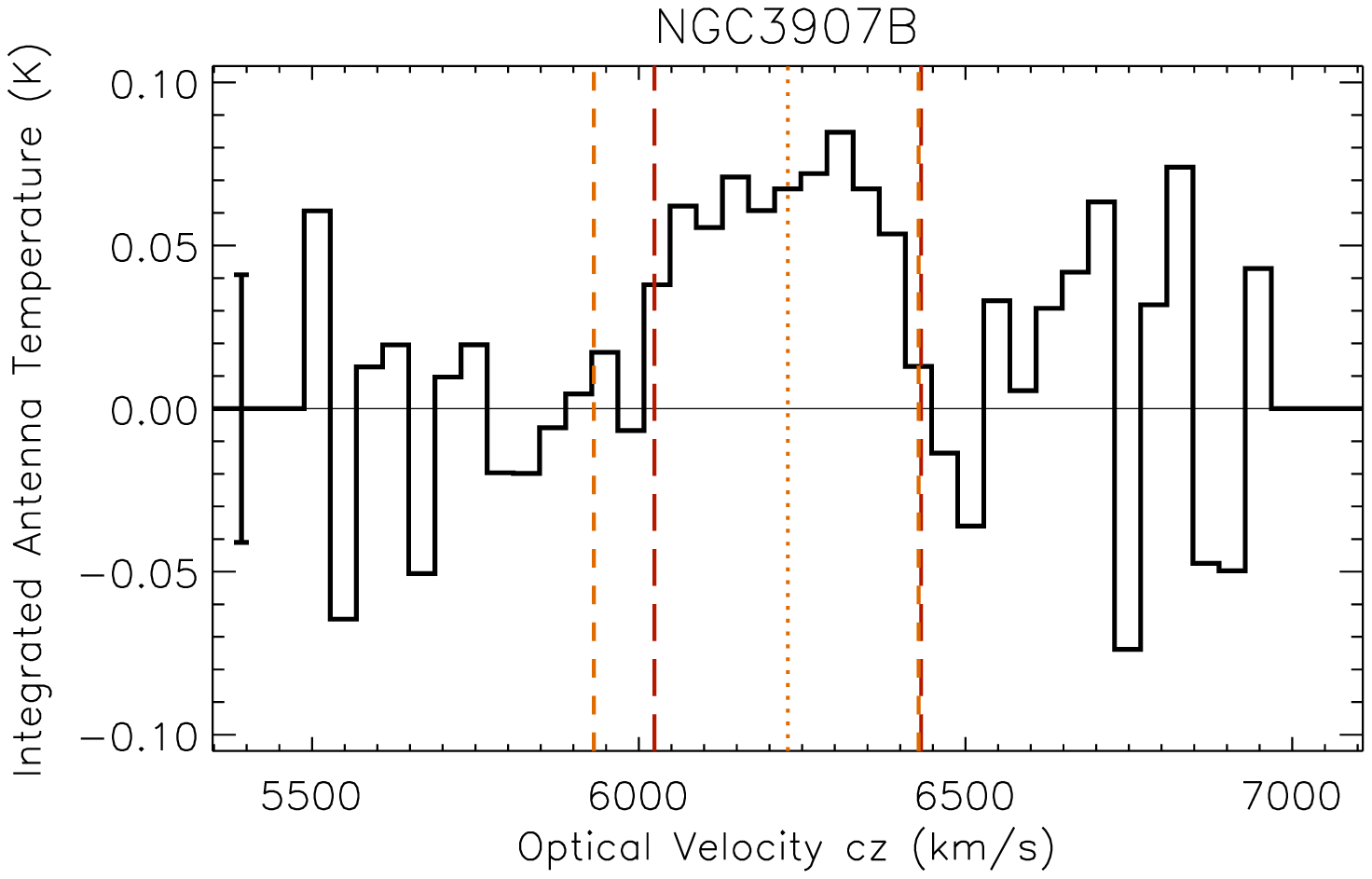}
\includegraphics[scale=0.55,clip=true,trim=1cm 0 0.5cm 0]{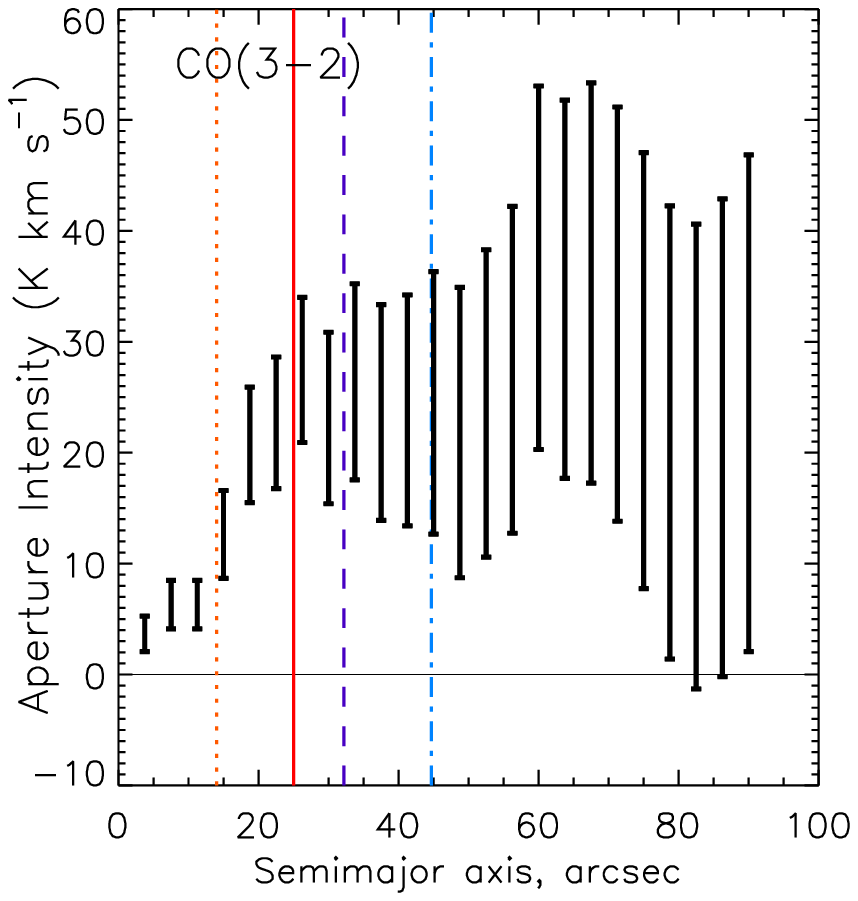}
\\
\raisebox{1cm}{\includegraphics[scale=0.253]{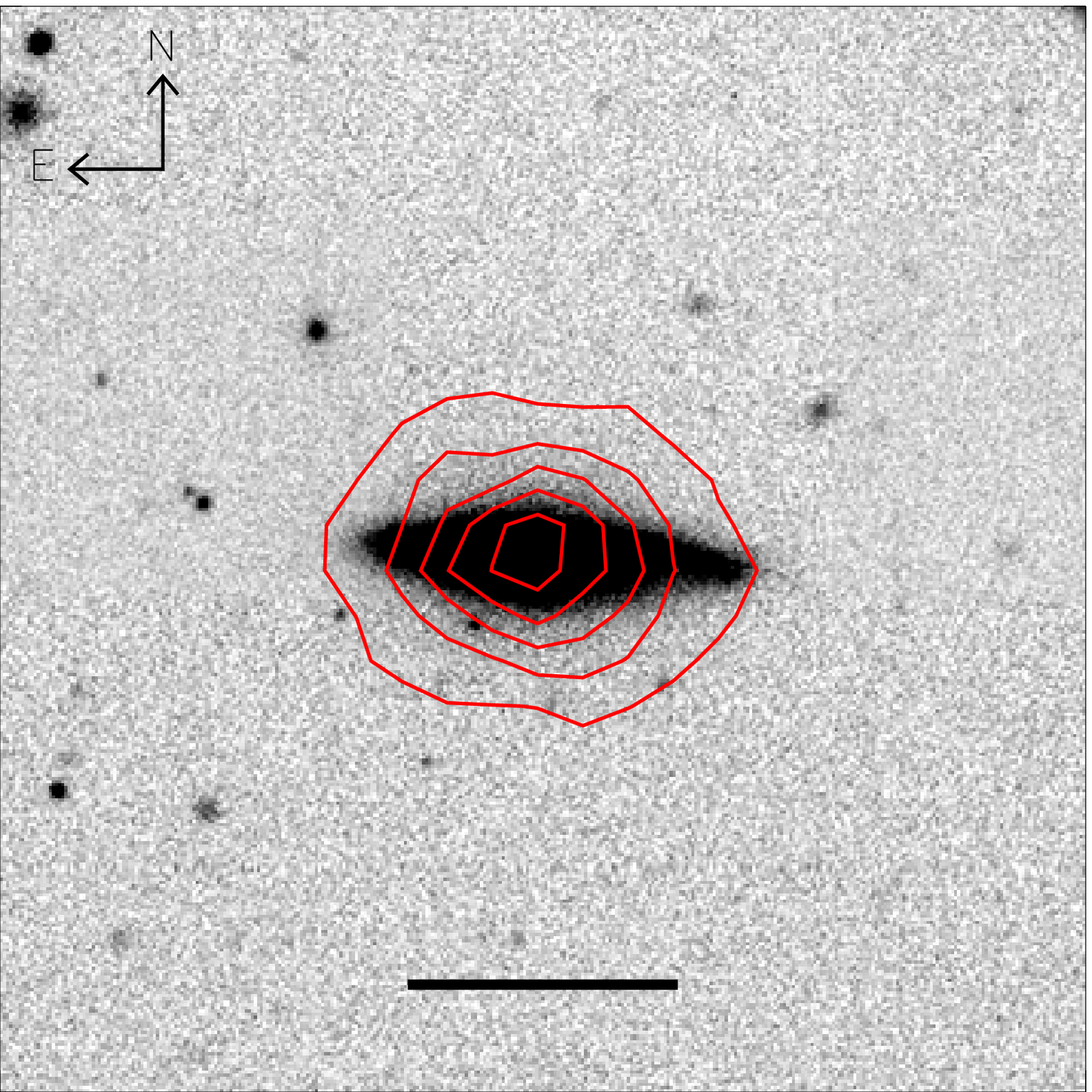}}
\includegraphics[scale=0.55,clip=true,trim=0.5cm 0 0.5cm 0]{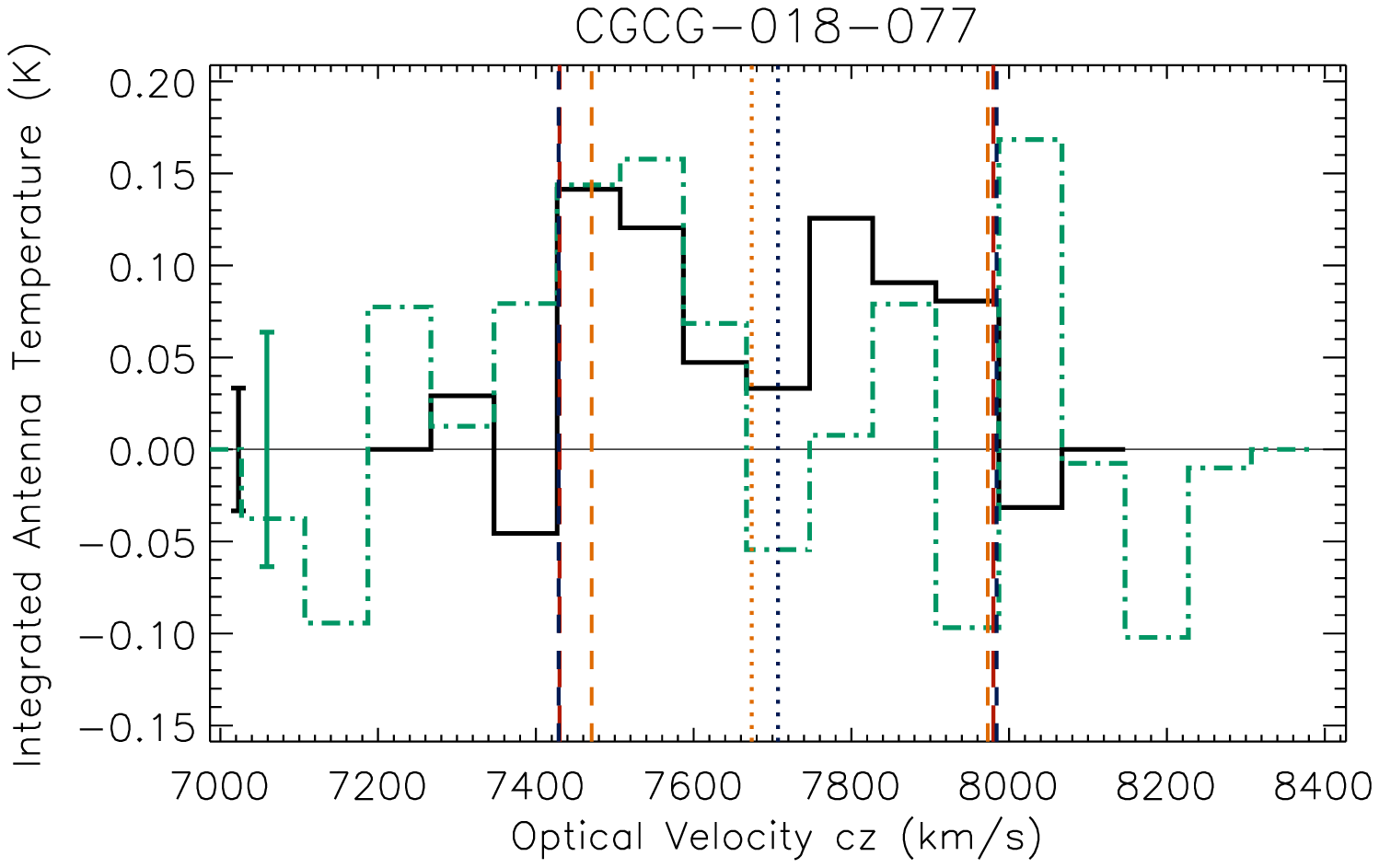}
\includegraphics[scale=0.55,clip=true,trim=1cm 0 0.5cm 0]{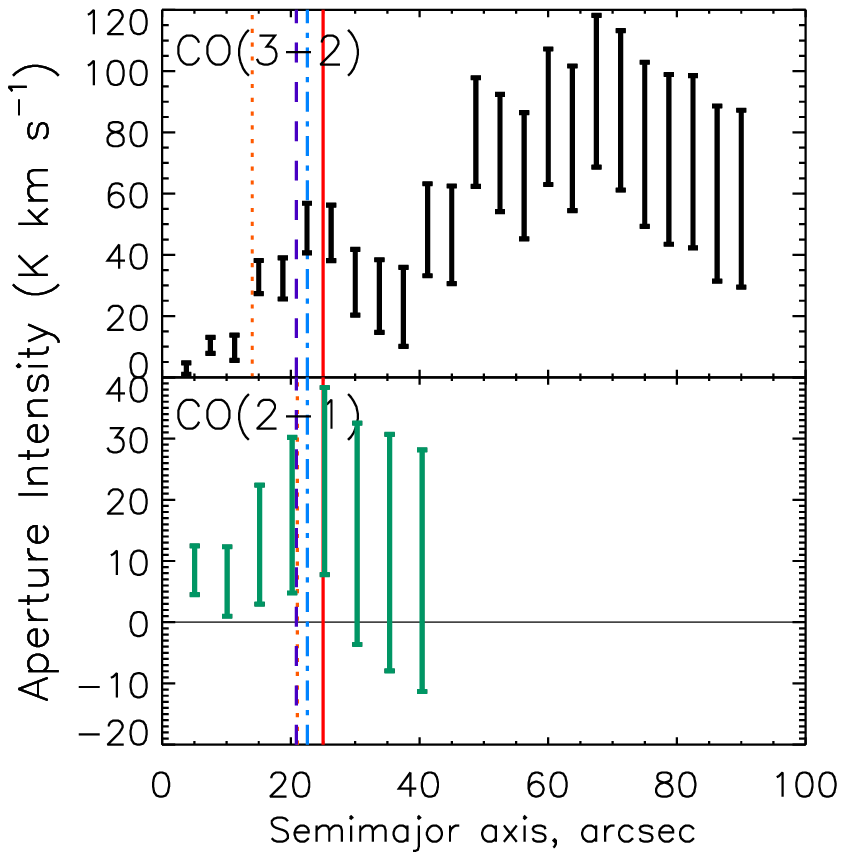}
\\
\raisebox{1cm}{\includegraphics[scale=0.253]{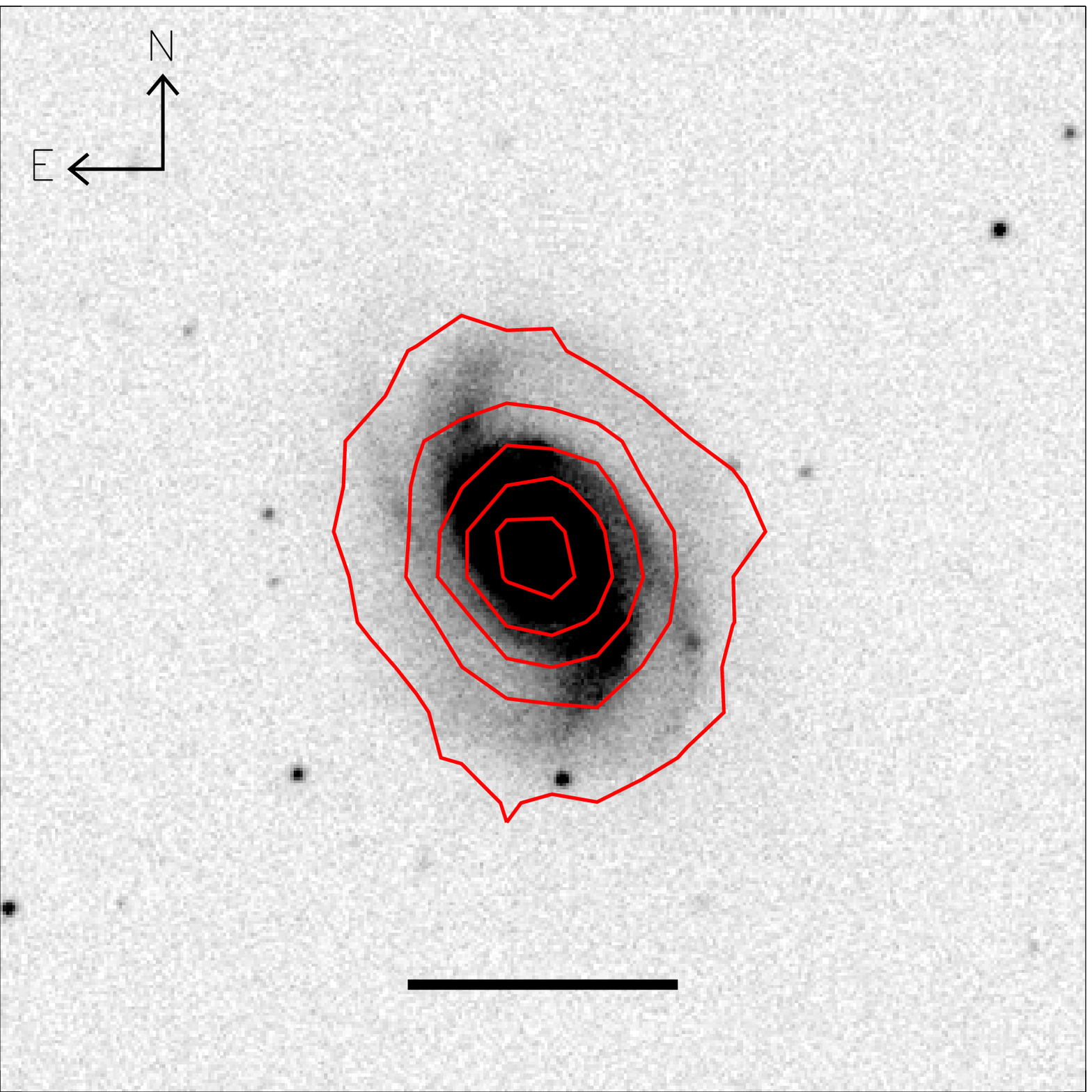}}
\includegraphics[scale=0.55,clip=true,trim=0.5cm 0 0.5cm 0]{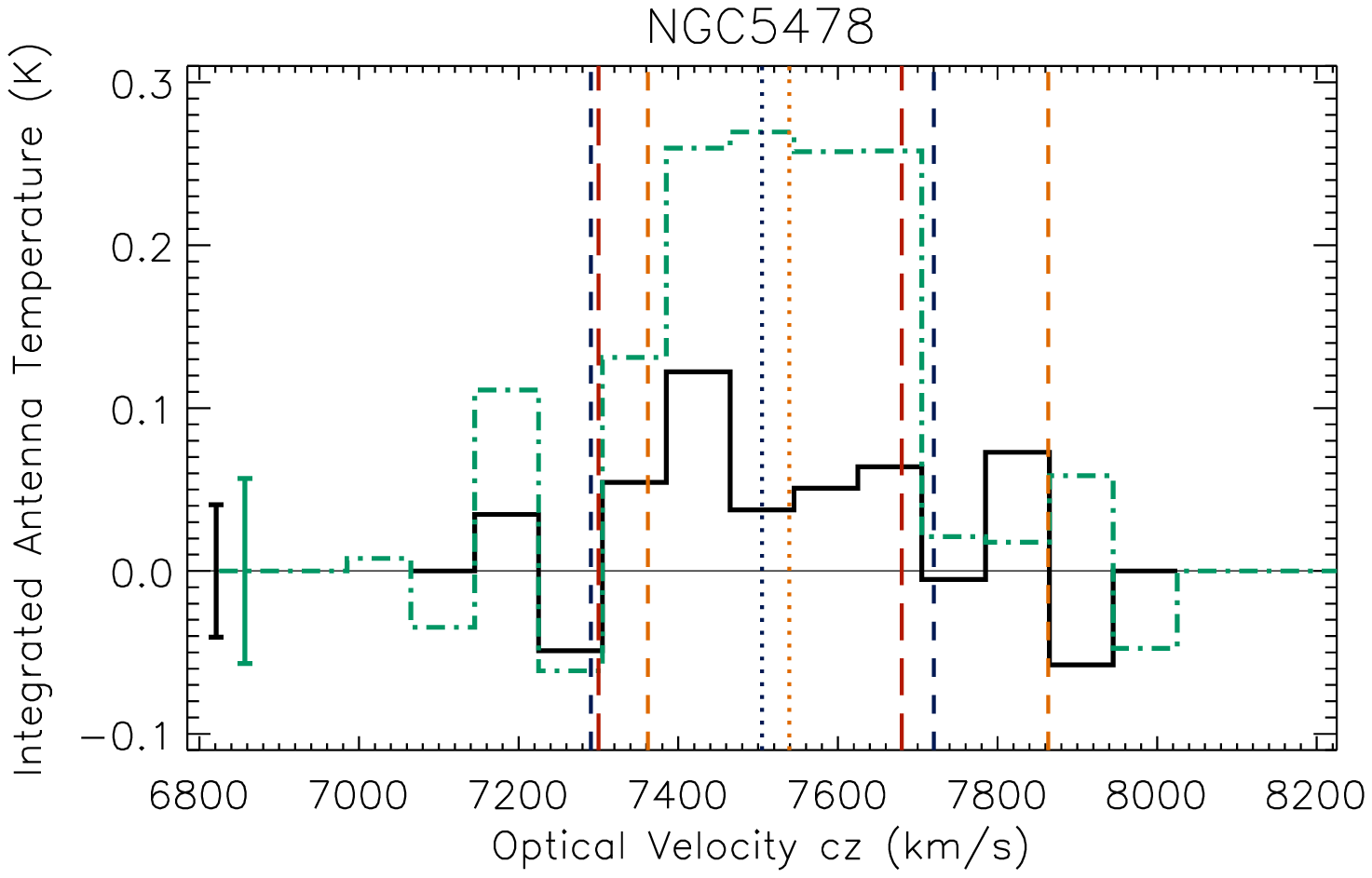}
\includegraphics[scale=0.55,clip=true,trim=1cm 0 0.5cm 0]{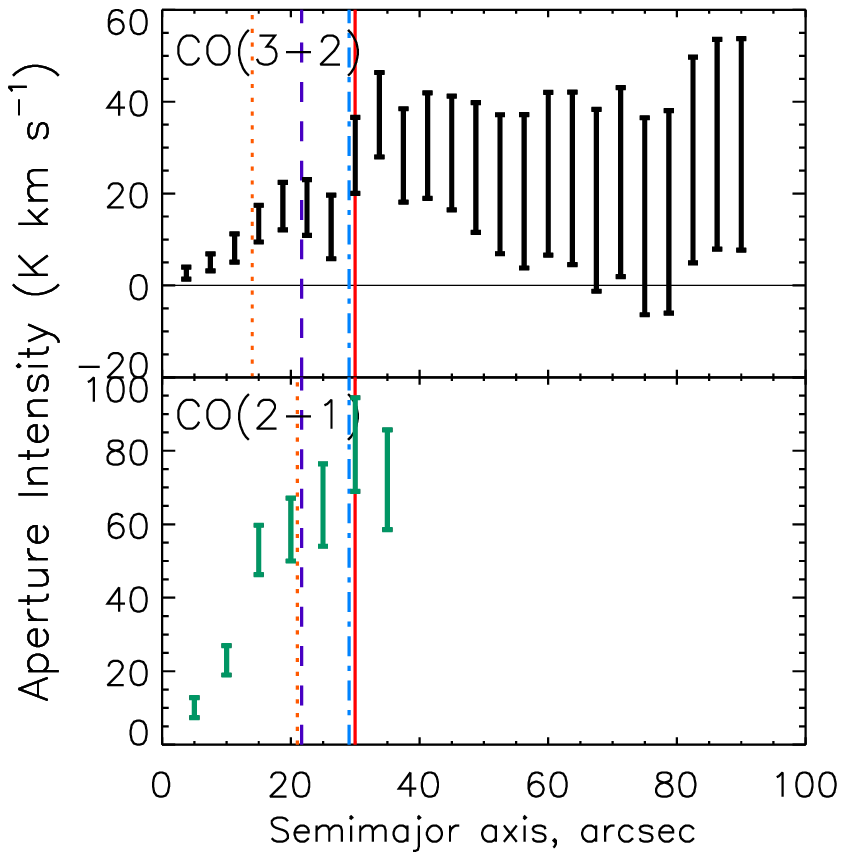}
\\
\centering {\bf Figure~1.} (continued)
\label{fig:spectra3}
\end{figure*}
\begin{figure*}
\raisebox{1cm}{\includegraphics[scale=0.253]{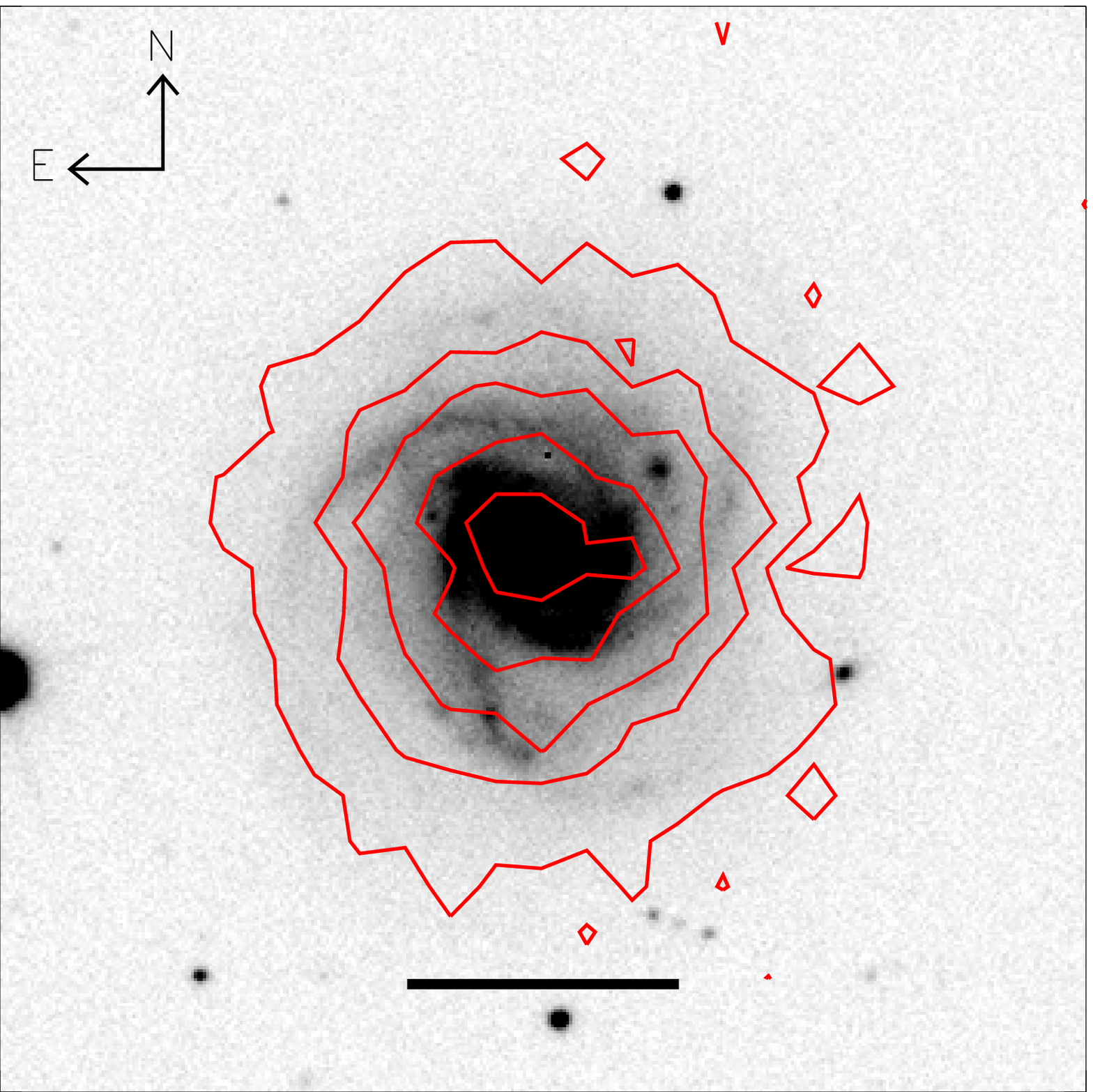}}
\includegraphics[scale=0.55,clip=true,trim=0.5cm 0 0.5cm 0]{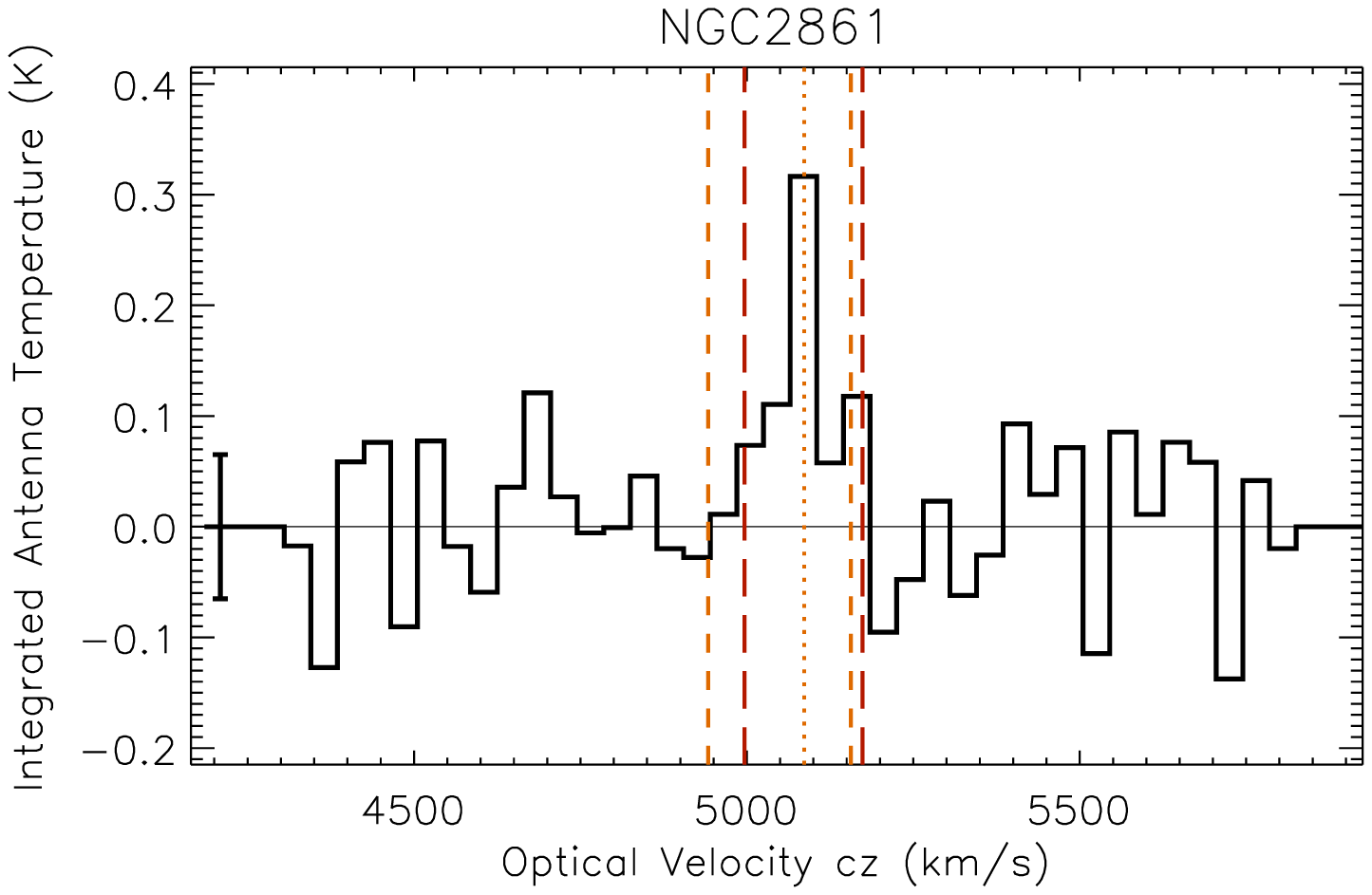}
\includegraphics[scale=0.55,clip=true,trim=1cm 0 0.5cm 0]{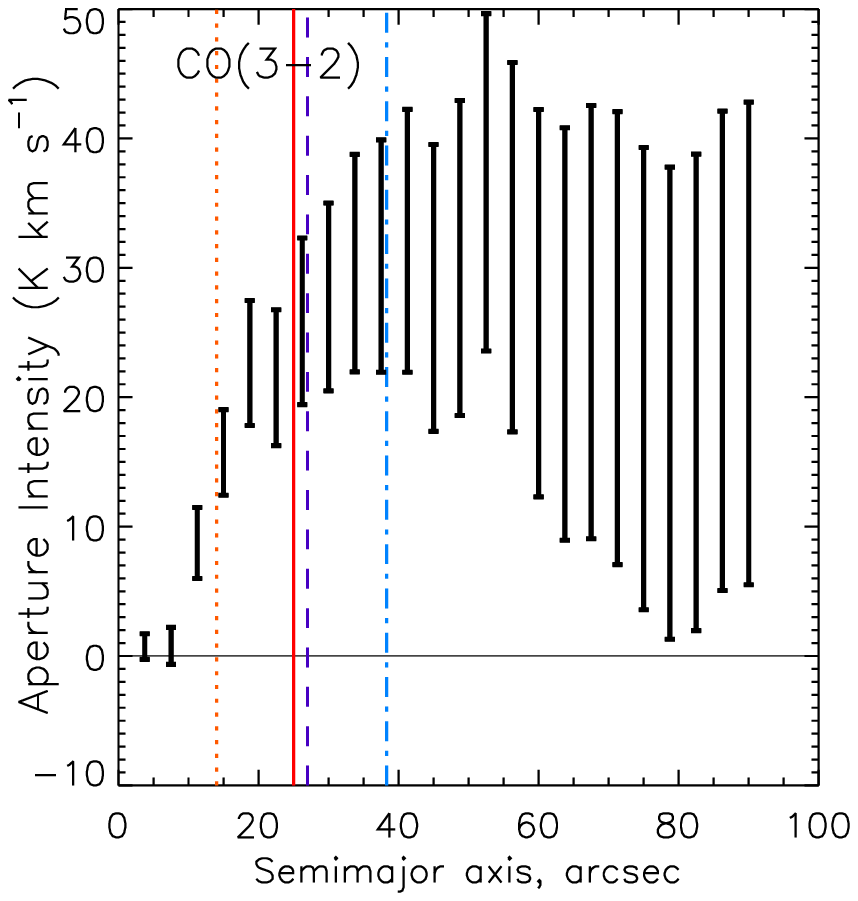}
\\
\raisebox{1cm}{\includegraphics[scale=0.253]{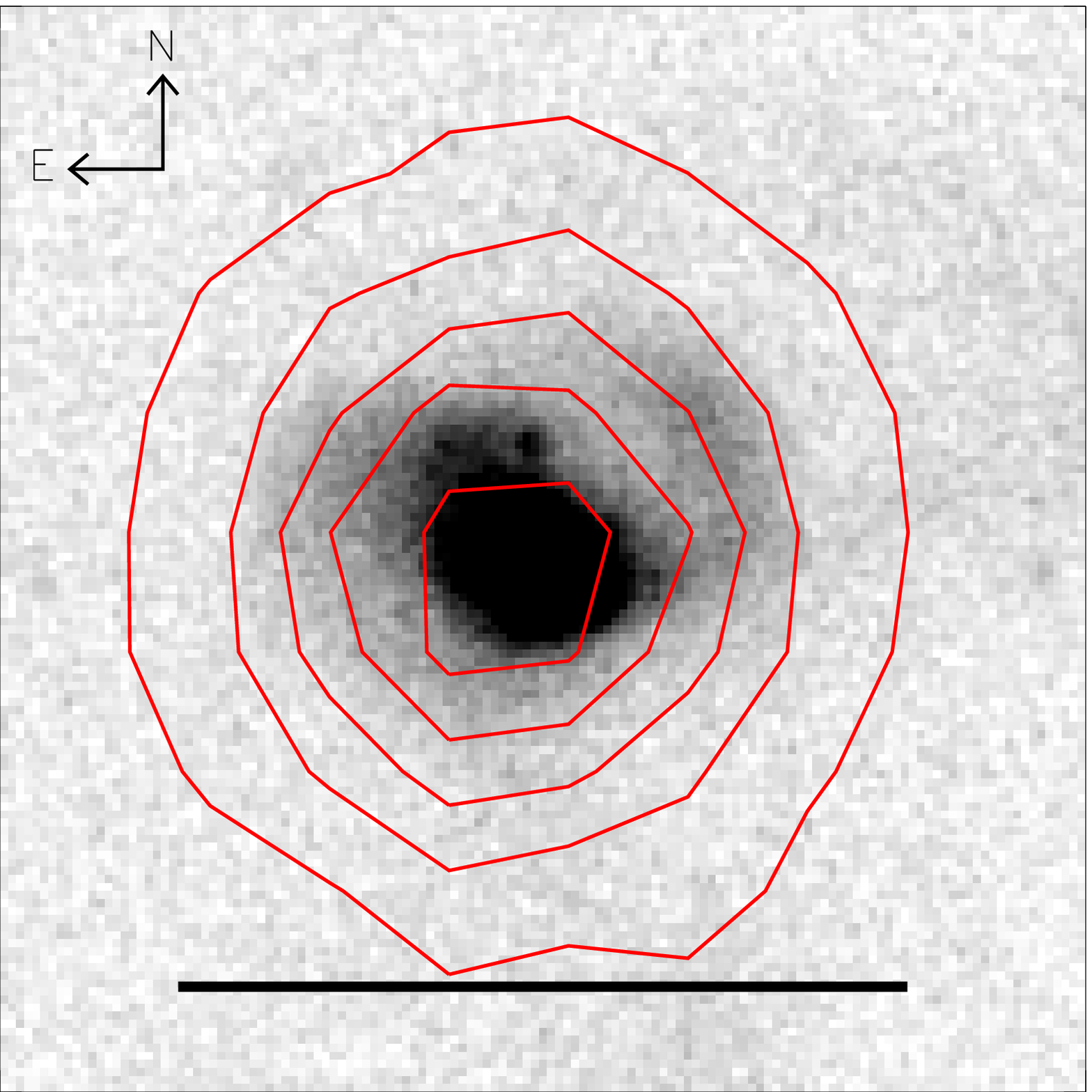}}
\includegraphics[scale=0.55,clip=true,trim=0.5cm 0 0.5cm 0]{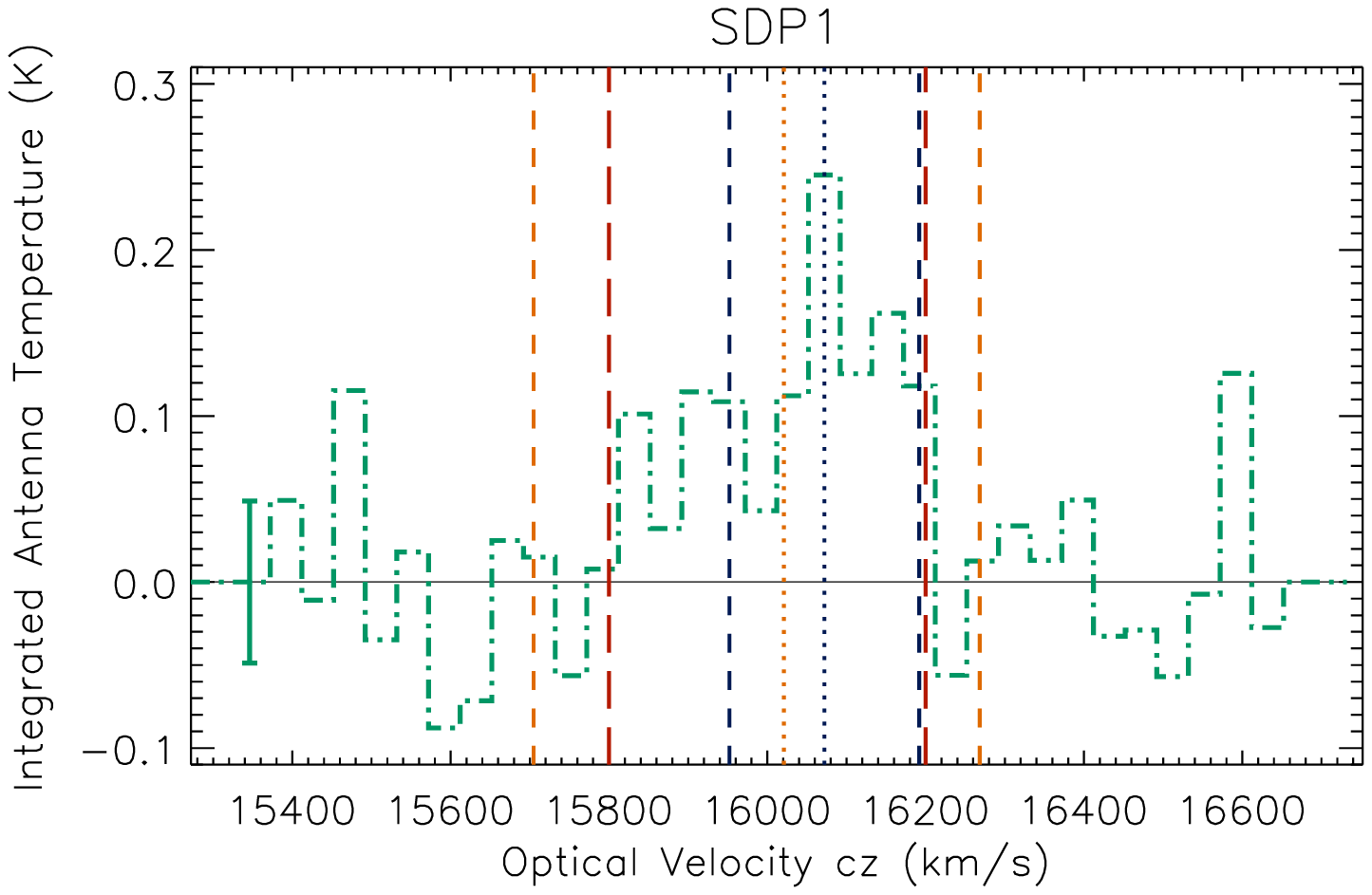}
\includegraphics[scale=0.55,clip=true,trim=1cm 0 0.5cm 0]{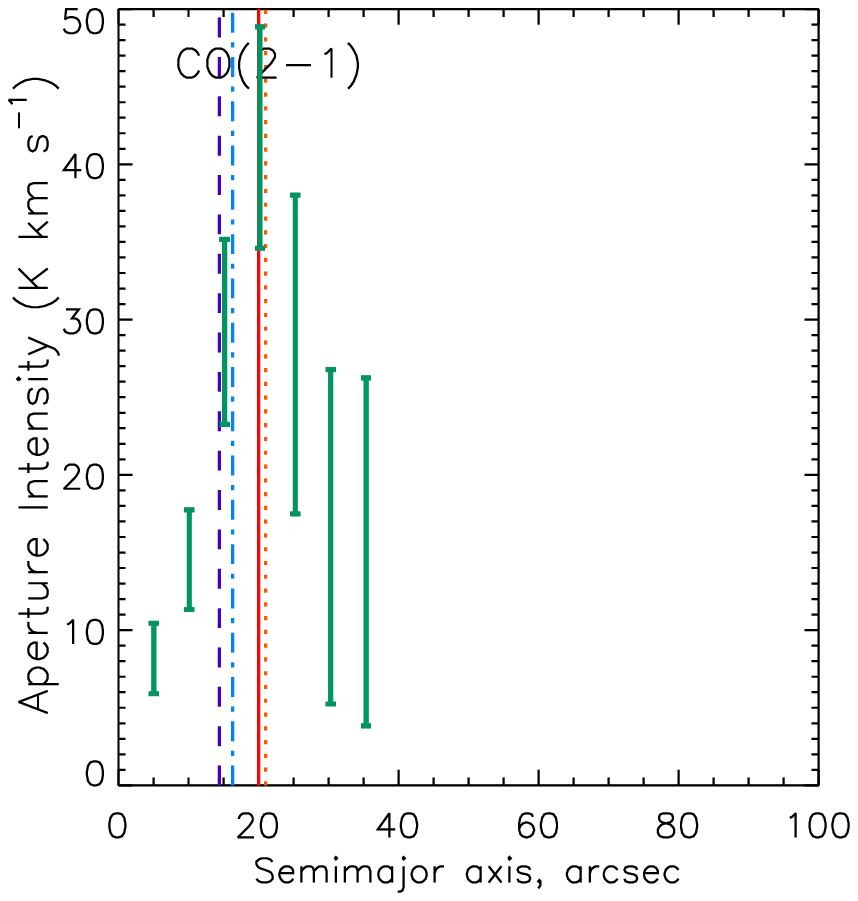}
\\
\raisebox{1cm}{\includegraphics[scale=0.253]{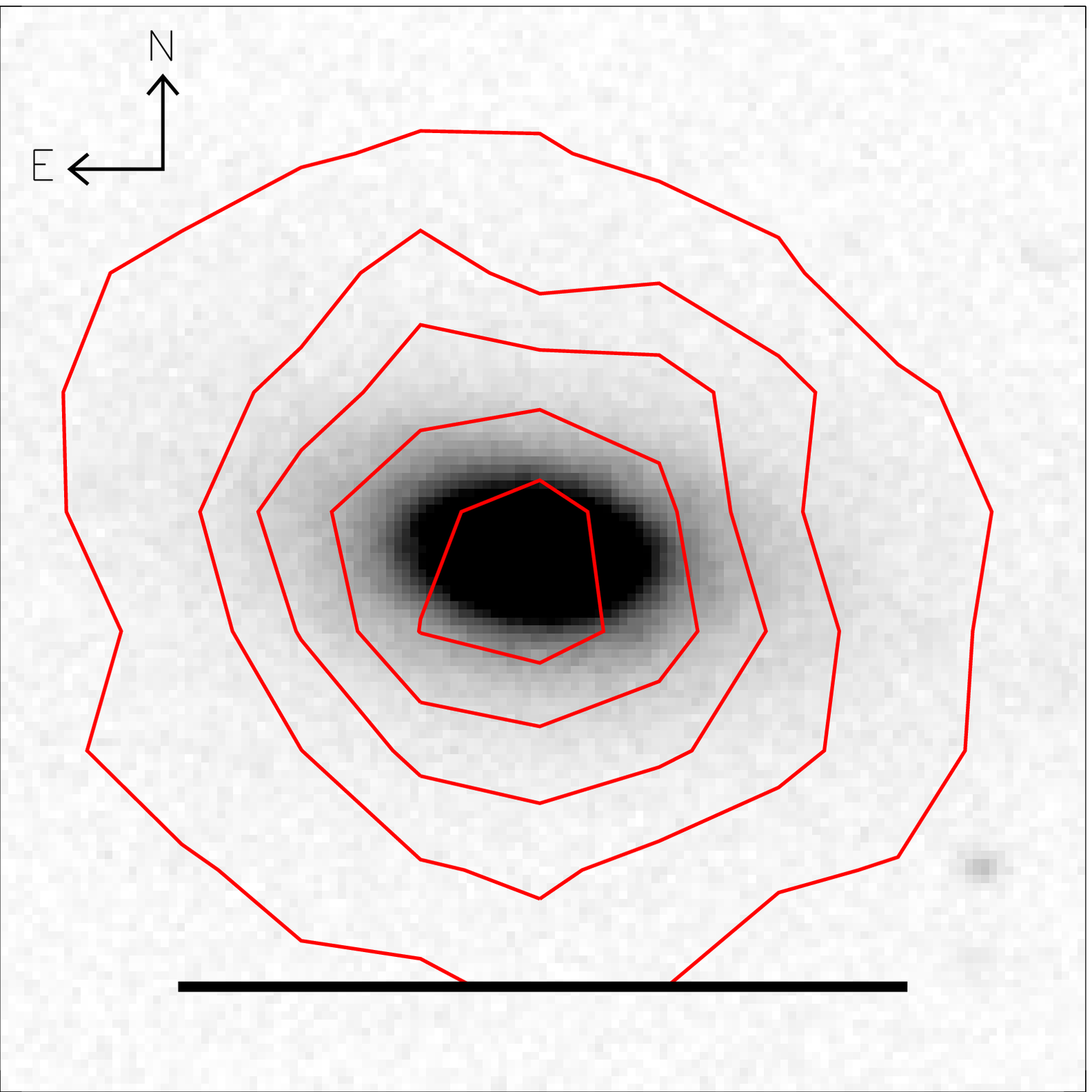}}
\includegraphics[scale=0.55,clip=true,trim=0cm 0 0.5cm 0]{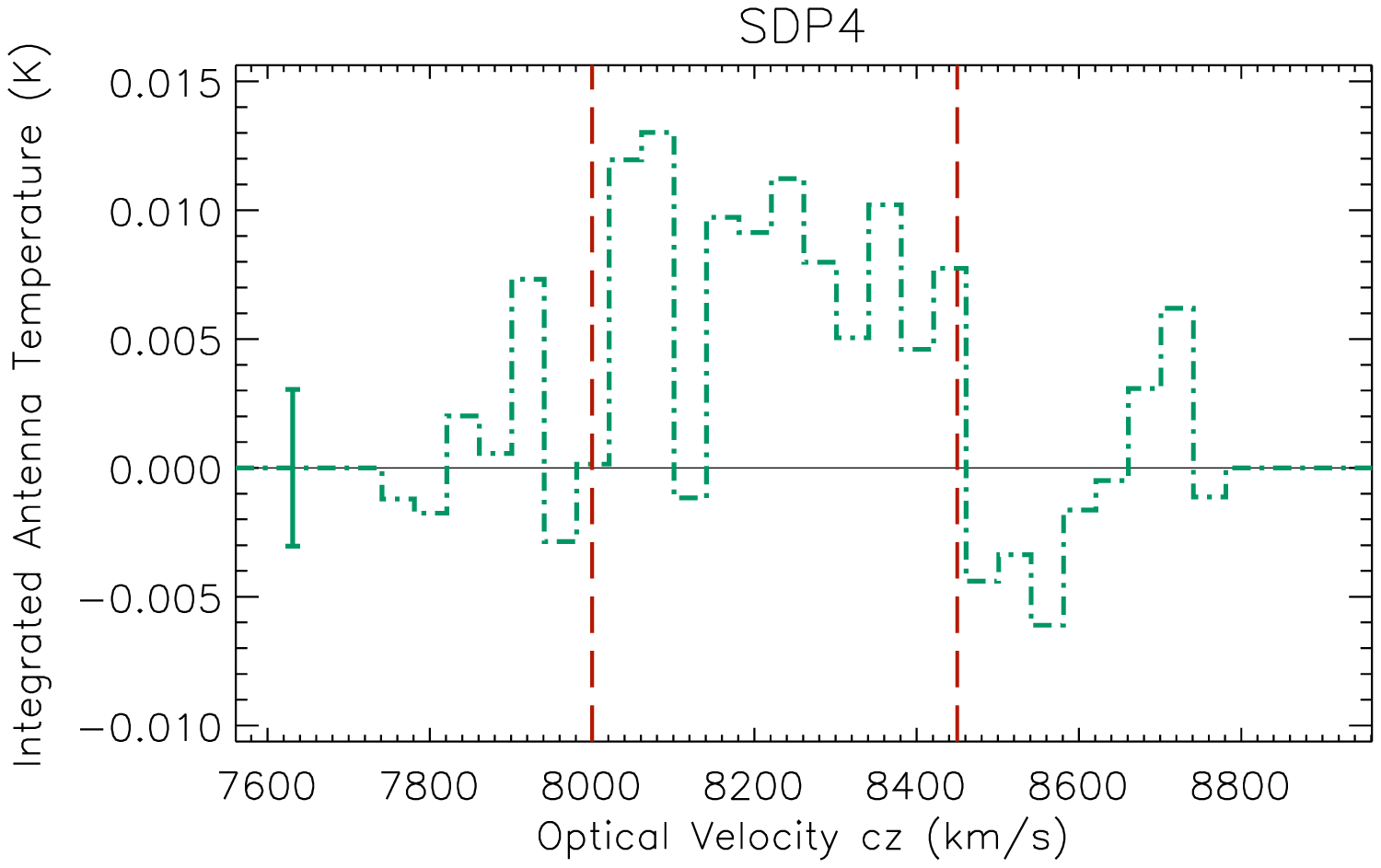}
\\
\raisebox{1cm}{\includegraphics[scale=0.253]{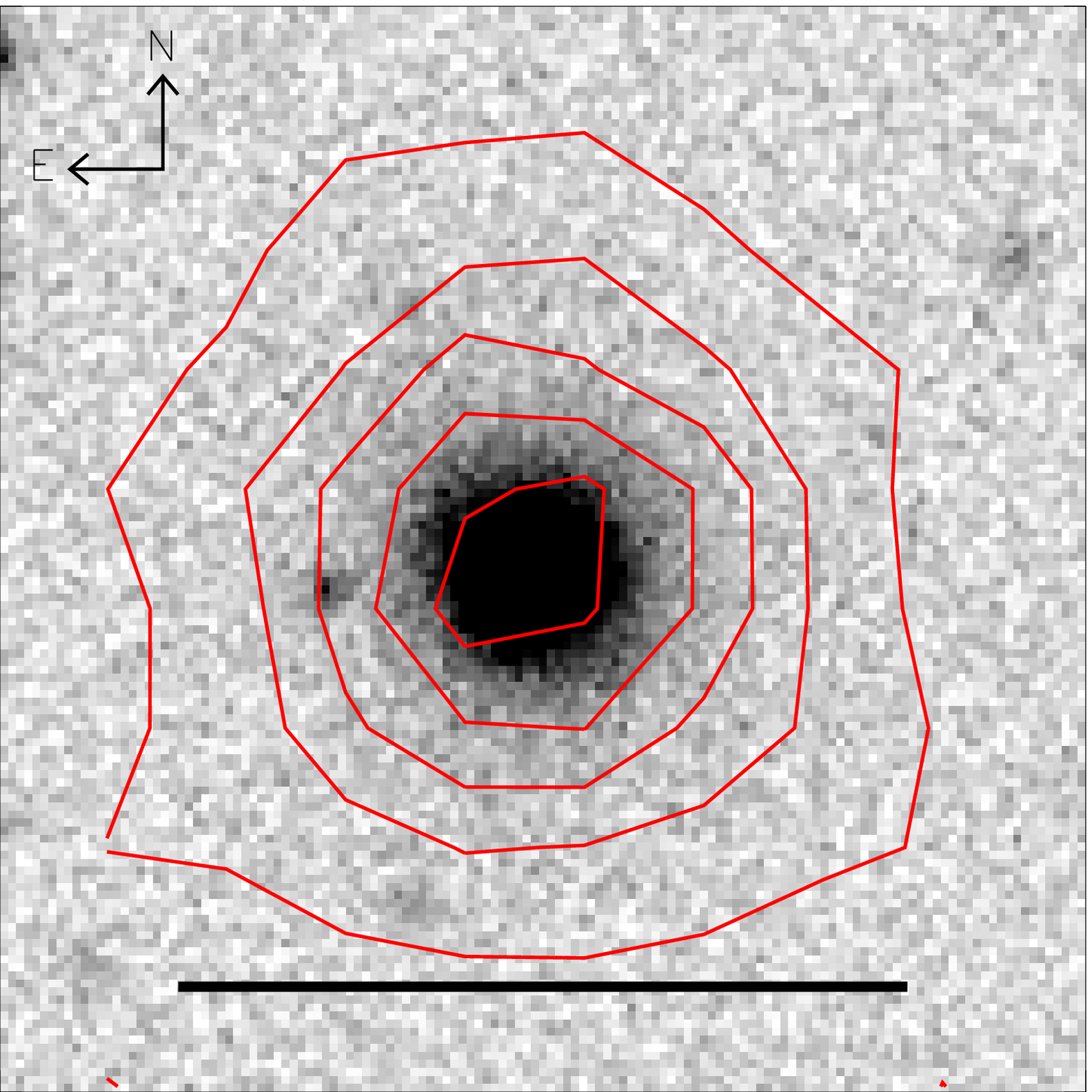}}
\includegraphics[scale=0.55,clip=true,trim=0cm 0 0.5cm 0]{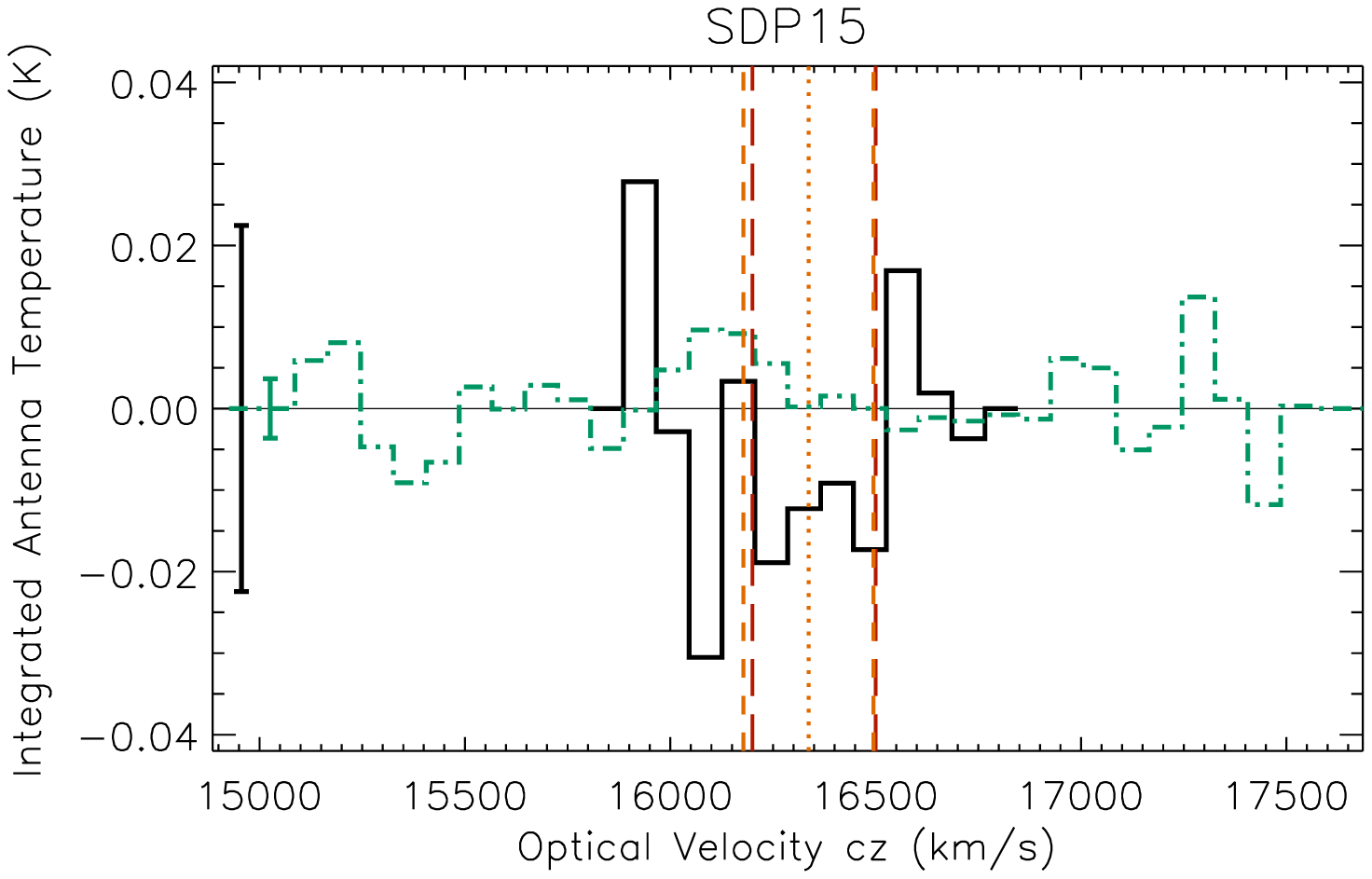}
\\
\centering {\bf Figure~1.} (continued)
\label{fig:spectra4}
\end{figure*}

\begin{sidewaystable*}
\caption[Details of the CO line measurements]{Details of the CO line measurements including FWZI line widths from literature data (see notes) and from our integrated data-cubes; elliptical aperture axis ratio ($a/b$) and position angle; semimajor axis, total intensity (or $3\sigma$ limit) and signal-to-noise ratio of each line determined from curves of growth.}
{\small
\begin{center}
\begin{tabular}{l r c c c c c c c c c c c c c}
\hline
        &       & \multicolumn{4}{c}{Line width (FWZI, \kms)}     & \multicolumn{2}{c}{Aperture} & \multicolumn{3}{c}{HARP cube \cott}      & \multicolumn{3}{c}{RxA cube \coto}      & Line ratio \\
 Galaxy & \multicolumn{1}{c}{$cz\,^\text{a}$} & \hi\,$^\text{b}$   & \cooz$^\text{c}$ & Lit.$^\text{d}$       & Cube    & $a/b$ & P.A.       & $a_{32}$ & $I_{32}$         & SNR        & $a_{21}$ & $I_{21}$         & SNR       & $I_{32}/I_{21}$ \\
                &     &                  &     &             &        &  & \degree\ E of N & \arcsec  & $\Kkms~(T_A^\star)$  &            & \arcsec  & $\Kkms~(T_A^\star)$  & & \\
\hline
  NGC 4030   &  1465 &  413\,~           & 474 &  360\,$^1$  &        & 1.2   &   33      & 53      & 490.3 $\pm$ 33.0$^\text{e}$ & 14.9  &            &       $   $      &         &                 \\
  NGC 5746   &  1723 &  633\,$^\text{L}$ & 227 &             &        & 8.5   &   170     &         &       $   $      &              &            &       $   $      &         &                 \\
  NGC 5713   &  1899 &  390\,$^\text{B}$ & 248 &  140\,$^2$  & 200    & 1.0   &   27     & 50       & 405.2 $\pm$ 13.1 &     30.8     &   40       & 496.2 $\pm$ 31.4 &    15.8 &  0.82$\pm$ 0.07  \\         
  NGC 5690   &  1732 &  374\,~           & 247 &  243\,$^3$  & 350    & 2.7   &   144    & 75       & 157.4 $\pm$ 20.3 &     7.8      &   45       & 197.3 $\pm$ 32.2 &    6.1  &  0.80$\pm$ 0.21 \\          
  NGC 5719   &  1756 &  660\,$^\text{B}$ & 516 &  400\,$^4$  & 460    & 2.3   &   112    & 55       & 149.3 $\pm$ 5.9  &     25.4     &            &       $   $      &    $ $  &                 \\                        
  NGC 5584   &  1652 &  267\,~           & 165 &  136\,$^5$  & 220    & 1.6   &   149    & 45       & 40.0  $\pm$ 6.4  &     6.3      &            &       $   $      &    $ $  &                 \\                        
  NGC 5740   &  1579 &  413\,~           & 350 &             & 320    & 1.7   &   163    & 30       & 28.2  $\pm$ 4.4  &     6.4      &            &       $   $      &    $ $  &                 \\                        
  NGC 5705   &  1732 &  320\,~           &     &             &        & 1.5   &   47      &         &       $   $      &              &            &       $   $      &         &                  \\                       
  UGC 09215  &  1313 &  466\,~           &     &             &        &  1.7  &   162     &         &       $   $      &              &            &       $   $      &         &                  \\                       
  NGC 5496   &  1559 &  373\,~           &     &             & 370    & 3.8   &   173    & 80       & $<$44.7          &              &   80       & 53.5  $\pm$ 20.9 &    2.6  &  $<$0.83         \\                  
  NGC 5750   &  1612 &  640\,~           & 309 &             & 600    & 1.6   &   71     & 42      &  44.0  $\pm$ 12.3 &     3.6      &   80       & $<$141.6         &         &  $>$0.31          \\                  
  NGC 5691   &  1879 &  214\,~           &     &             & 210    & 1.3   &   101    & 20      &  14.3  $\pm$ 3.4  &     4.2      &   35       & 59.7  $\pm$ 12.2 &    4.9  &  0.24$\pm$0.31   \\           
  C~013-010  & 11843 &                   &     &             & 220    & 2.0   &   144    & 30       & 47.9  $\pm$ 12.7 &     3.8      &   20       &  17.4 $\pm$ 7.0  &    2.5  &   2.76$\pm$ 0.48 \\         
  NGC 3907B  &  6228 &  408\,$^\text{L}$ & 497 &             & 400    & 2.2   &   72     & 25       & 25.8  $\pm$ 6.4  &     4.1      &   0        &       $   $      &    $ $  &                  \\                       
  C~018-077  &  7707 &  550\,~           & 502 &             & 550    & 3.1   &   87     & 25       & 47.2  $\pm$ 9.1  &     5.2      &   25       & $<$45.9          &         &  $>$1.03        \\                  
  NGC 5478   &  7505 &  430\,~           & 501 &             & 380    & 1.2   &   19     & 30       & 28.3  $\pm$ 8.3  &     3.4      &   30       & 81.7  $\pm$ 12.8 &    6.4  &  0.35$\pm$ 0.33 \\          
  NGC 2861   &  5085 &                   & 214 &             & 180    & 1.1   &   165    & 25       & 24.5  $\pm$ 6.2  &     3.9      &            &       $   $      &    $ $  &                  \\                       
  SDP~1       & 16072 &  240\,$^\text{L}$ & 563 &             & 400    & 1.3   &   45     &          &       $   $      &     $ $      &   20       & 45.4  $\pm$ 6.8  &    6.7  &                  \\                        
  SDP~4       &  8281 &                   &     &             & 450    & 1.1   &   79     &          &       $   $      &      $ $     &   0        &  3.6  $\pm$ 0.5  &     8.0 &                   \\        
  SDP~15      & 16326 &                   & 316 &             &        & 1.1   &   20     & 30       & $<$11.1          &              &   0        &  $<$1.8          &         &                  \\         
\hline
\end{tabular}
\end{center} 
}
{ \footnotesize Notes:                         
(a) Radial velocities ($cz$, \kms) are the centre of the \hi\ line and agree with the centre of the CO line in HARP and RxA. 
(b) \hi\ data labelled (L) are from HyperLEDA, and (B) from the HIPASS BGS \citep{Koribalski2004}; otherwise they are from the HIPASS cubes -- see Section~\ref{sec:hidata}.
(c) References for literature data:
(1) \citet{Young1995}; M.W.L.~Smith et al. in prep.
(2) \citet{Young1995, Yao2003, Albrecht2007}.
(3) \citet{Sauty2003}.
(4) \citet{Albrecht2007}.
(5) \citet{Sauty2003, Boker2003}.
(d) \cott\ measurement for NGC~4030 from the HRS data (Matthew~W.~L.~Smith, private communication).
Missing CO intensities from HARP and RxA indicate sources that were not observed for the reasons described in Section~\ref{sec:codata}.
}
\label{tab:aperdata}%
\end{sidewaystable*}

\subsection{HI data} 
\label{sec:hidata}

Most of the objects in the sample have \hi\ 21\,cm line detections in the literature. Data from many surveys are collected in the HyperLEDA database \citep{Paturel2003}, which lists \hi\ magnitudes for 15 galaxies in the sample.
In addition, many members of the sample have been covered by the \hi\ Parkes All-Sky Survey (HIPASS; \citealt{Barnes2001}).
The HIPASS tables were searched using the Vizier catalogue server\footnote{Vizier hosted by CDS, Strasbourg: \url{http://vizier.u-strasbg.fr/viz-bin/VizieR}}, revealing two matches in the HIPASS Bright Galaxy Catalogue \citep[BGC;][]{Koribalski2004}, seven in the main HIPASS catalogue \citep[HICAT;][]{Meyer2004}, and one in the Northern HIPASS catalogue \citep[NHIPASS;][]{Wong2006}.
Data are shown in Table~\ref{tab:hicomparison}.

Some discrepancies occur between data in the two catalogues. The homogenised photometry in HyperLEDA have been combined from a wide range of sources, and measurements on different telescopes can vary depending on the uncertainty of the absolute flux calibration, whether the source is fully sampled by the telescope beam, and whether nearby galaxies at the same redshift are blended within a single beam. 
The average HPBW of the Parkes telescope in multi-beam mode (used for HIPASS) is 14.3~arcmin \citep{Barnes2001}, so all sources in the sample are point sources, and some are potentially blended.
NGC~5713 is blended with NGC~5719, and NGC~5746 with NGC~5740. 
NGC~4030 is 16.7~arcmin away from UGC~07000 at the same redshift, but this galaxy is 3 magnitudes fainter in the optical and has an \hi\ flux of 5.9\Jykms\ in HyperLEDA (6 times fainter than the HyperLEDA flux of NGC~4030), so it will not significantly bias the NGC~4030 measurement.

The best solution to obtain directly comparable results for a sample is to draw data from a single survey, measured on the same telescope at similar times, and reduced and calibrated in a self-similar way. We therefore used \hi\ fluxes from HIPASS where they were available and not blended. For the brightest sources these were available from the catalogues, as shown in Table~\ref{tab:hicomparison}. Additional data for all sources were obtained from the public HIPASS archive via the ATNF \hi\ Gateway\footnote{ATNF \hi\ Survey Gateway hosted by CSIRO, Australia: \url{http://www.atnf.csiro.au/research/HI/hipass/}}. 
These extend to a maximum radial velocity of $cz=12,000\kms$, so the two sources at $z>0.04$ (SDP~1 and SDP~15) are not covered. 
We used the {\sc miriad} software\footnote{Available from \url{http://carma.astro.umd.edu/miriad/}} to measure point-source spectra at the required positions in data cubes downloaded from the server. 
Baselines were subtracted from the spectra, and lines were measured between the zero-crossing points. 
In most cases a first-order polynomial baseline was adequate, except for NGC~2861, NGC~5478, and SDP~4 which required fourth-order fits to remove baseline ripples \citep[c.f.][]{Koribalski2004}. Note that this was possible in these spectra, but not in the CO spectra, because the baseline range was wide enough to fit a higher order polynomial.
Results are included in Table~\ref{tab:hicomparison}. 
In general the results obtained from the cubes are similar to those in the HIPASS catalogues, with three notable exceptions. NGC~5713 and NGC~5719 are blended and at the same velocity, so the measured fluxes will depend on the pixels included in the weighted spectrum and on the line range chosen. 
The flux we have measured between the zero-crossing points is therefore an overestimate in these cases, but the BGC measurements used a smaller velocity range so are likely to be more reliable.
The third exception is NGC~4030, which is not significantly blended or extended, so it is unclear why our measurement misses 20 per cent of the flux reported in HICAT.

To summarise, we used HIPASS catalogue measurements where they were available and not blended, and measured fluxes in the HIPASS cubes for sources not in the catalogues. We used HyperLEDA fluxes for galaxies that were not detected in the cubes or were blended in the Parkes beam. This left two sources (CGCG~013-010 and SDP~4) with only upper limits from the cubes, and one (SDP~15) with no data at all.

\subsubsection{\hi\ self-absorption corrections}
\label{sec:hiselfabs}
\hi\ fluxes and masses will be underestimated if there is sufficient optical depth in the line of sight for \hi\ clouds in the source galaxy to absorb emission from those behind them. This optical depth can depend on the inclination of the galaxy, its morphological type, velocity dispersion and the thickness of the \hi\ disk \citep[\eg][]{Heidmann1972}, and also on the column density of individual \hi\ clouds \citep{Braun2012}.
Some of the galaxies in this sample are highly inclined and all are dust-rich, so the density of gas is likely to be high.
It is therefore prudent to consider the possibility of \hi\ self-absorption (and especially the uncertainty that this introduces) which could affect measured correlations between the \hi\ flux and optically thin emission such as the submm.

Various approaches have been used to correct measured fluxes for self-absorption \citep[\eg][]{Heidmann1972,Haynes1984,Zwaan1997,Lang2003}, and their predictions can vary significantly.
We follow the method of \citet{Haynes1984}, which is an empirical correction based on observations of 1500 galaxies. The correction factor is a function of axis ratio and morphological type, and is given by
\begin{equation}
 f_{\hi} = r_i^{-c}
\label{eqn:hiselfabs}
\end{equation}
where $r_i\equiv \cos{i}$ is the observed axis ratio, and $c$ varies from -0.02 to 0.16 depending on morphological type $t$ \citep{deVaucouleurs1991}.
Using this formula, and type information from HyperLEDA, we obtained the corrections shown in Table~\ref{tab:hicomparison}.
To account for the uncertainty of the correction, we conservatively allow a $100\%$ error on the estimated self-absorbed fraction itself (\ie\ a correction of 1.3 has the error $\pm0.3$, such that the error bar encompasses a null correction).

\begin{sidewaystable*}
\begin{center}
{\footnotesize%
\caption[Compilation of \hi\ line measurements and self-absorption corrections]{Compilation of \hi\ line measurements from HyperLEDA, HIPASS catalogues and HIPASS cubes, with self-absorption corrections (see notes below).}%
\label{tab:hicomparison}%
\begin{tabular}{l r r c r r c r r c c c c c c c}
\hline
Galaxy     & \multicolumn{3}{c}{HyperLEDA} &    \multicolumn{3}{c}{HIPASS catalogues}        & \multicolumn{3}{c}{HIPASS cubes}       &  \multicolumn{4}{c}{Self-absorption correction} & Final\\
           & $v_\text{rad}$ & $\Delta v$ & $S_\hi$ & $v_\text{rad}$ & $\Delta v$ & $S_\hi$ & $v_\text{rad}$ & $\Delta v$ & $S_\hi$ & $i (\degree)$ & $t$ & $c$ & $f_{\hi}$ & $S_{\hi}^\text{corr}$ \\
\hline
NGC~4030   &   1463   &   295 &  {   37.9  $\pm$ 1.6  } &  1465  & 441     & {\bf72.0 $\pm$ 0.9    }  &  1465 &   413 & {     56.9 $\pm$  1.5   } &  40 & $4.0\pm0.2$ & 0.04                   & $1.01\pm0.01$ &          $72.8\pm7.4$   \\     
NGC~5746   &    1723  &   633 &  {\bf26.3  $\pm$ 1.4  } &        &          &{                     }  &  1724 &   721 & {     29.7 $\pm $ 1.5   } &  84 & $3.0\pm0.3$ & 0.04                   & $1.09\pm0.09$ &          $28.7\pm3.3$   \\
NGC~5713   &    1899  &   160 &  {\bf35.8  $\pm$ 1.8  } &   1899 &  390     &{   48.8 $\pm$ 5.1$^B$}  &  1899 &   587 & {     73.6 $\pm $ 1.4   } &  48 & $4.0\pm0.3$ & 0.04                   & $1.02\pm0.02$ &          $36.5\pm4.1$   \\   
NGC~5690   &    1753  &   266 &  {   19.88 $\pm$ 0.68 } &  1754  & 398     & {\bf25.6 $\pm$ 0.5$^N$}  &  1753 &   374 & {     21.89$ \pm$  0.89 } &  76 & $5.4\pm0.5$ & 0.16                   & $1.25\pm0.25$ &          $32.1\pm7.3$   \\    
NGC~5719   &    1732  &   355 &  {\bf31.7  $\pm$ 1.5  } &   1756 &  660     &{   68.3 $\pm$ 5.1$^B$}  &  1733 &   587 & {     74.0 $\pm $ 1.2   } &  71 & $2.4\pm0.6$ & 0.00                   & $1.00\pm0.00$ &          $31.7\pm3.5$   \\  
NGC~5584   &    1638  &   166 &  {   21.76 $\pm$ 0.92 } &   1638 &  280     &{\bf27.1 $\pm$ 0.4    }  &  1638 &   267 & {     25.54$ \pm$  0.62 } &  42 & $6.0\pm0.3$ & 0.14                   & $1.04\pm0.04$ &          $28.3\pm3.1$     \\  
NGC~5740   &    1571  &   308 &  {\bf21.90 $\pm$ 0.88 } &  1572  & 409     & {   33.3 $\pm$ 0.5    }  &  1572 &   413 & {     29.63$ \pm$  0.80 } &  55 & $3.0\pm0.4$ & 0.04                   & $1.02\pm0.02$ &          $22.3\pm2.4$    \\ 
NGC~5705   &    1760  &   168 &  {   18.7  $\pm$ 1.2  } &  1758  & 279     & {\bf27.9 $\pm$ 0.4    }  &  1758 &   320 & {     26.25$ \pm$  0.72 } &  65 & $6.5\pm1.1$ & 0.14                   & $1.13\pm0.13$ &          $31.5\pm4.8$     \\  
UGC~09215  &    1389  &   162 &  {   16.3  $\pm$ 1.0  } &  1387  & 296     & {\bf23.7 $\pm$ 0.5    }  &  1397 &   280 & {     21.12$ \pm$  0.57 } &  64 & $6.4\pm0.9$ & 0.14                   & $1.12\pm0.12$ &          $26.6\pm4.0$     \\   
NGC~5496   &    1541  &   212 &  {   39.4  $\pm$ 1.4  } &  1541  & 335     & {\bf60.9 $\pm$ 0.5    }  &  1541 &   373 & {     56.72$ \pm$  0.73 } &  82 & $6.5\pm0.8$ & 0.14                   & $1.32\pm0.32$ &          $80.3\pm21.0$     \\    
NGC~5750   &    1687  &   374 &  {   5.17  $\pm$ 0.30 } &        &         & {                     }  &  1687 &   640 & {\bf  13.8 $\pm $ 1.1   } &  66 & $0.4\pm0.9$ & 0.00                   & $1.00\pm0.00$ &          $13.8\pm1.7$     \\    
NGC~5691   &    1871  &   130 &  {   7.66  $\pm$ 0.28 } &  1881  & 219     & {\bf5.5  $\pm$ 0.3    }  &  1870 &   213 & {     5.63 $\pm $ 0.71  } &  44 & $1.2\pm0.6$ & 0.00                   & $1.00\pm0.00$ &          $5.5 \pm0.6$     \\                 
C~013-010  &   11843  &       &  {                    } &        &         & {                     }  & 11843 &   500 & {\bf  $<$5.07           } &  65 & $1.5\pm2.9$ & $0.00^{+0.04}_{-0.00}$ & $1.00^{+0.04}_{-0.00}$ & $<$5.07        \\
NGC~3907B  &   6252   &   408 &  {\bf3.28  $\pm$ 0.12 } &        &         & {                     }  &  6228 &   500 & {     $<$5.81           } &  84 & $3.1\pm0.4$ & 0.04                   & $1.09\pm0.09$ &          $3.6 \pm1.0$   \\  
C~018-077  &    7676  &       &  {                    } &        &         & {                     }  &  7674 &   500 & {\bf  5.44 $\pm $ 0.93  } &  74 & $6.0\pm2.0$ & $0.14^{+0.02}_{-0.10}$ & $1.20\pm0.20$ &          $6.5 \pm1.8$   \\
NGC~5478   &    7530  &       &  {                    } &        &         & {                     }  &  7539 &   430 & {\bf  6.03 $\pm$  0.78  } &  49 & $3.6\pm0.6$ & 0.04                   & $1.02\pm0.02$ &          $6.1 \pm1.0$   \\ 
NGC~2861   &   5134   &   120 &  {\bf2.49  $\pm$ 0.09 } &        &         & {                     }  &  5086 &   500 & {     $<$6.22           } &  25 & $3.7\pm0.6$ & 0.04                   & $1.00\pm0.00$ &          $2.5 \pm0.3$   \\  
SDP~1      &   9174   &   240 &  {\bf0.53  $\pm$ 0.09 } &        &         & {                     }  &       &       & {                       } &  42 & $0.4\pm5.0$ & $0.00^{+0.16}_{-0.00}$ & $1.00^{+0.05}_{-0.00}$ & $0.5 \pm0.1$   \\
SDP~4      &   8281   &       &  {                    } &        &         & {                     }  &  8241 &   500 & {\bf  $<$4.82           } &  75 & $0.1\pm2.0$ & 0.00                   & $1.00\pm0.00$ &          $<$4.82         \\  
SDP~15     &   16326  &       &  {                    } &        &         & {                     }  &       &       & {                       } &  28 & $4.5\pm5.0$ & $0.04^{+0.12}_{-0.04}$ & $1.00^{+0.02}_{-0.00}$ &                \\
\hline
\end{tabular}
}
\end{center}
{\footnotesize Notes:
All fluxes are \Jykms; the final flux used for analysis is highlighted in boldface.
HyperLEDA fluxes are calculated from magnitudes in the database using \mbox{$\log S_{\hi} = (17.4-m)/2.5$}.
HIPASS catalogue fluxes are from HICAT except where marked B (BGC) or N (NHIPASS).
Radial velocities ($cz$, \kms) from HyperLEDA are taken from the radio measurements (the {\tt Vrad} parameter), while the line width ($\Delta v$) is estimated from the maximum velocity of the gas ({\tt Vmaxg}$\times2$). Velocities from the HIPASS catalogues are given by the {\tt Vsys} or {\tt RV50max} keywords, and widths are \mbox{{\tt RV2} -- {\tt RV1}} (which generally exceeds {\tt Vmaxg}$\times2$).
Fluxes were measured in the HIPASS cubes between the zero-crossing points, and where no line was detected a $5\sigma$ upper limit was measured assuming a width of 500\kms.
Corrections for \hi\ self-absorption ($f_{\hi}$) are given by equation~(\ref{eqn:hiselfabs}), as a function of inclination angle $i$ and de~Vaucouleurs type $t$, following \citet{Haynes1984}. Errors on $f_{\hi}$ include both the uncertainty on $t$ (from HyperLEDA) and a $100\%$ uncertainty on the self-absorbed fraction (see text). Errors on the final corrected fluxes ($S_\hi^\text{corr}$) include the error on $f_{\hi}$ as well as measurement errors from on the flux and a $10\%$ calibration error added in quadrature.
}
\end{sidewaystable*}

\section{Results and Analysis}
\label{sec:analysis}

\subsection{Properties of the sample and line ratios}
The 500\mum-selected sample consists of a diverse range of galaxy types, shown by the images in Figure~\ref{fig:spectra}. The sample is dominated by red, dusty spirals such as NGC~2861, 3907B, 4030, 5478, 5690, 5740, CGCG~013-010, 018-077; but also contains blue low-surface-brightness disks (NGC~5496, 5584, 5691, 5705, UGC~09215), dust-lane early-types (NGC~5719, 5750), and the starburst/LIRG NGC~5713.  
The additional sources that are not part of the flux-limited sample comprise two further early-types (SDP~4, 15) and another LIRG (SDP~1).

Our data, while not reaching sufficient rms to map the emission in most cases, nevertheless demonstrate that emission in \cott\ and \coto\ is often extended throughout the disk, as can be seen in the rising curves of growth and broad spectra in Figure~\ref{fig:spectra}.
We detect line emission in either \cott\ or \coto\ in most of the sample, including three of the four early-type galaxies (NGC~5719, 5750, SDP~4), as shown in Table~\ref{tab:aperdata}.
NGC~5719 is noteworthy since its warped dust lane and counter-rotating gas and stellar disks indicate that dust and gas were accreted from an interaction with NGC~5713 \citep{Vergani2007,Coccato2011}. Our \cott\ moment map shows that rotation of the molecular gas follows that of the counter-rotating disk and is therefore also likely to have been accreted. 

Although we only detect both CO lines in five galaxies, a wide range of line ratios ($R_{32}=I_{32}/I_{21}$) is apparent in Table~\ref{tab:aperdata}, consistent with the diversity of the sample. 
Similarly broad ranges of $R_{31}=I_{32}/I_{10}$ were reported in the nuclei of SLUGS galaxies by \citet{Yao2003} and in the large sample of nearby galaxies by \citet{Mao2010}. In contrast, \citet{Wilson2012} found a relatively small range of $R_{32}$ and $R_{31}$ in global line intensities of the \hi-selected NGLS sample, which may suggest that submm selection probes a wider range of star-forming properties than \hi\ selection.
The highest detected $R_{32}$ in our sample belongs to CGCG~013-010 ($R_{32}=2.76\pm0.48$), indicating high excitation perhaps by an AGN or nuclear starburst.
The \coto\ upper limit on CGCG~018-077 also implies a high ratio greater than 1.0.
Perhaps surprisingly, the starburst NGC~5713 has a global $R_{32}$ of just $0.82\pm0.07$, although this is likely to be much higher in the nucleus \citep[measured a high excitation of $R_{31}=1.65^{+0.22}_{-0.29}$ in the central 15\arcsec]{Yao2003}. 
We measure a similar value of $R_{32}$ in the star-forming spiral NGC~5690, while lower values are measured in NGC~5691 and 5478.

\subsection{Correlations between global fluxes}
\label{sec:correlations}
\label{sec:total_ico_results}

The total CO line intensity in a galaxy is correlated with FIR flux as encapsulated by the SK law, and correlations have also been observed in the submm in low-redshift galaxies \citep{Dunne2000,Yao2003,Corbelli2012,Wilson2012} and high-redshift SMGs \citep{Greve2005,Iono2009,Engel2010,Ivison2011,Magdis2011b,Bothwell2012}. 
The current sample is novel because it was selected from a blind survey of the local Universe at 500\mum, and is selected primarily on cold dust mass rather than FIR luminosity or SFR. Using our coverage of the IR SED, we can explore the correlation between the integrated line emission and IR fluxes at various wavelengths, comparing the FIR ($\lambda\lesssim200\mum$) with the submm ($\lambda\gtrsim200\mum$).

\begin{figure*}
\begin{center}
\includegraphics[width=0.9\textwidth]{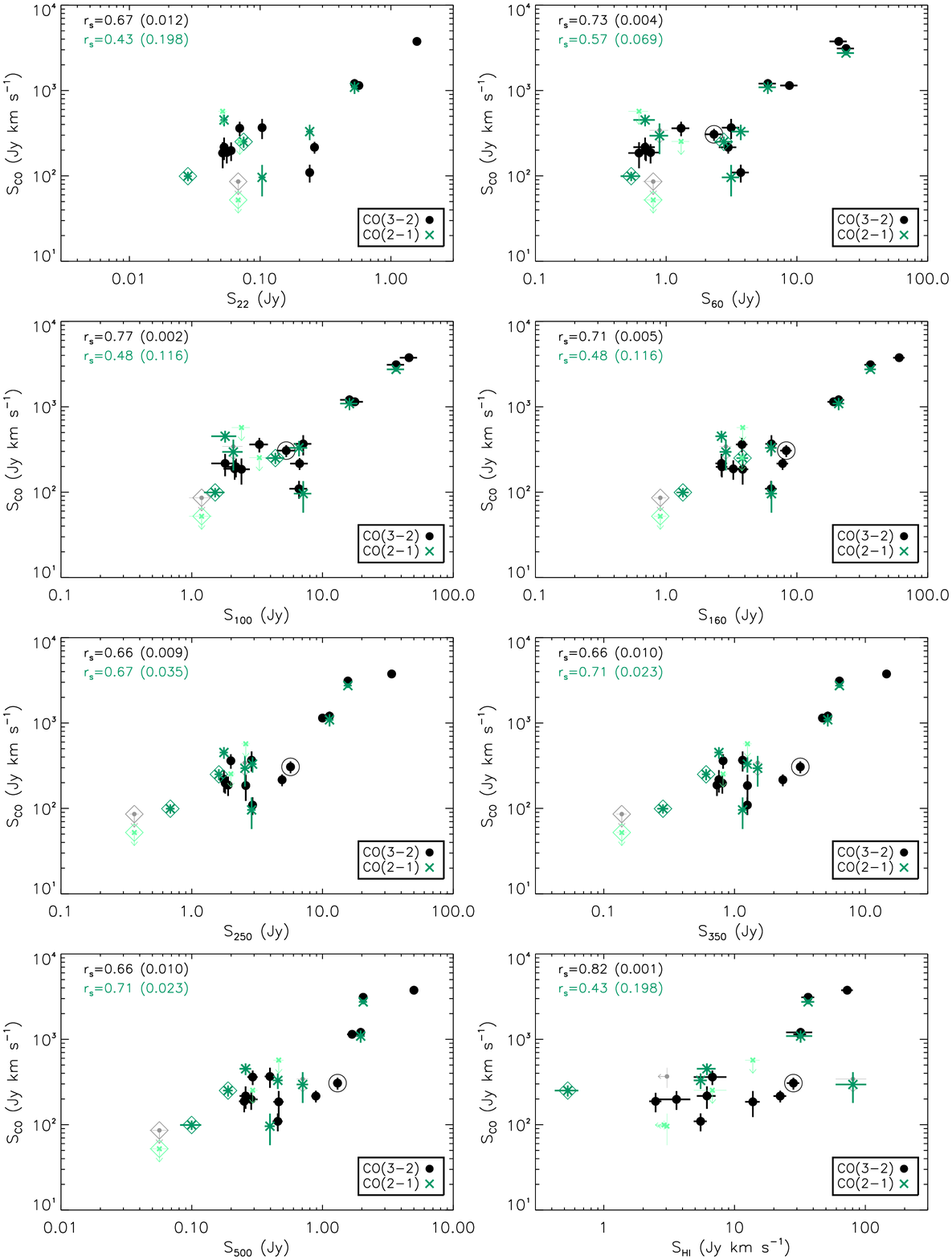}
\caption[Scatter plots of global fluxes in each of the CO lines as a function of IR fluxes from \WISE, \IRAS, PACS and SPIRE, and \hi\ flux]{Scatter plots of global fluxes in each of the CO lines as a function of IR fluxes from \WISE\ 22\mum, \IRAS\ 60, 100\mum, PACS 160\mum, SPIRE 250, 350, 500\mum, and finally \hi\ flux. 
Data from HARP and RxA are plotted as filled black circles and green crosses respectively. 
$3\sigma$ upper limits on CO and $5\sigma$ upper limits on \hi\ fluxes are also shown by arrows in a ligher shade.
Large diamonds mark galaxies that are not part of the flux-limited sample (see Table~\ref{tab:phottable}).
The circled point is the HARP measurement of NGC~5584, which may be inaccurate as described in Section~\ref{sec:codata}, and is therefore excluded from the correlation analysis.
Spearman's rank correlation coefficients are printed in the top-left corner of each panel, with $p$-values in parentheses.}
\label{fig:corr_fir_sco}
\end{center}
\end{figure*}

Figure~\ref{fig:corr_fir_sco} compares the global fluxes in the CO lines to the global fluxes in the IR from 22 to 500\mum, and to the \hi\ fluxes.
The total integrated intensities (\Kkms\ $T_A^\star$) in Table~\ref{tab:aperdata} have been converted to flux in \Jykms\ using 
\begin{equation}
 S_\text{CO} = \dfrac{2k_B \Omega I_\text{CO}}{\eta_\text{mb}\lambda^2}
\label{eqn:Snu_Tmb_general}
\end{equation}
\citep[\eg][]{Rohlfs2006}, where $\eta_\text{mb}$ is the main-beam efficiency, which is taken to be 0.63 for HARP and 0.69 for RxA.\footnote{See the JCMT Guide to Spectral Line Observing: \url{http://www.jach.hawaii.edu/JCMT/spectral_line/}}
In this equation, $\Omega$ is the solid angle (in steradians) over which the intensity is averaged. For point sources measured in a single beam, this is the beam area, but for extended sources measured in apertures, the relevant solid angle is the pixel area, which is (7.5~arcsec)$^2$ for HARP and (10.0~arcsec)$^2$ for RxA maps.

\begin{figure*}
\begin{center}
\includegraphics[width=0.9\textwidth]{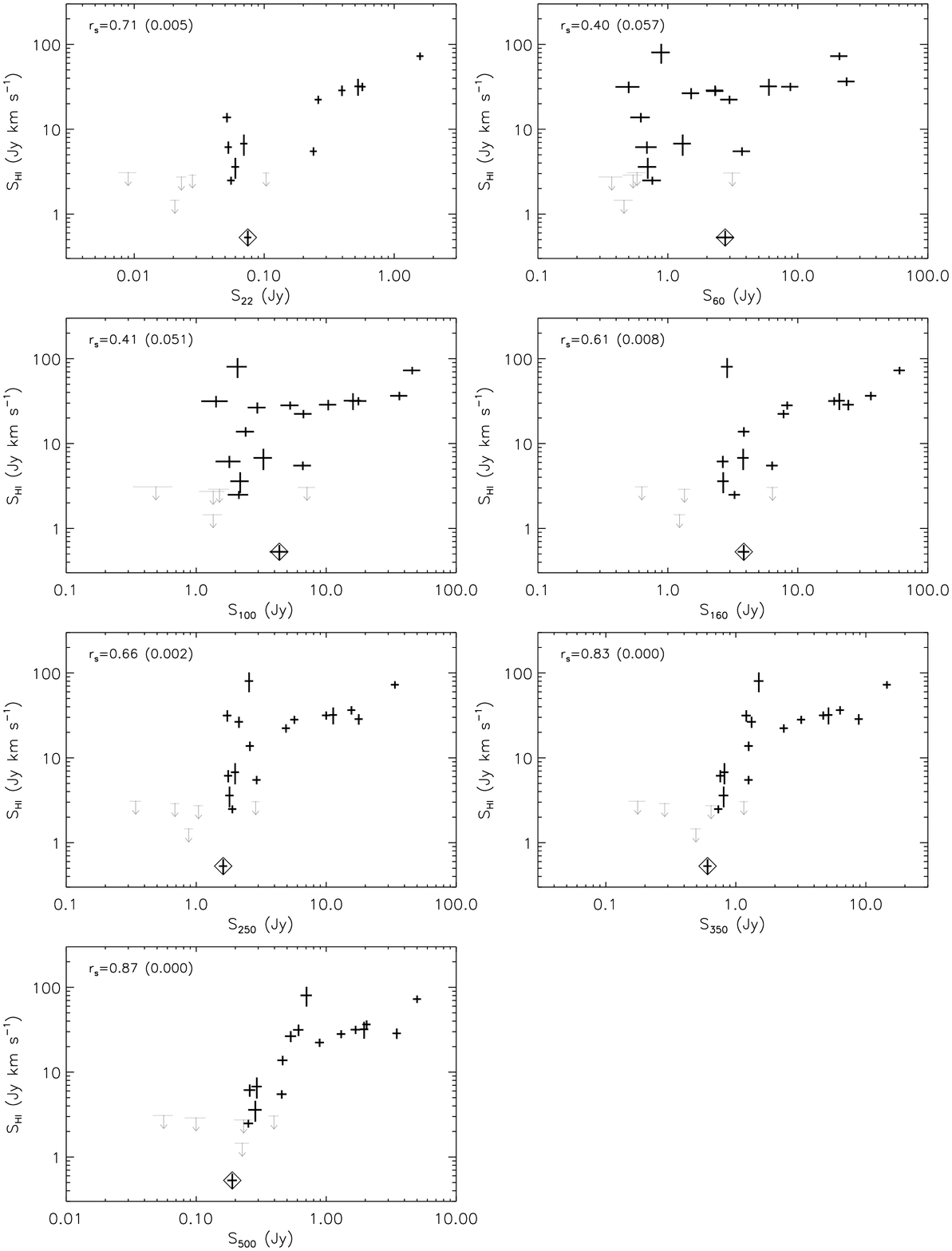}
\caption[Scatter plots of global fluxes in \hi\ as a function of IR fluxes from \IRAS, PACS and SPIRE]{Scatter plots of global fluxes in \hi\ and the IR bands (as in Figure~\ref{fig:corr_fir_sco}). Spearman's rank correlation coefficients are printed in the top-left corner of each panel, with $p$-values in parentheses. For galaxies with no flux in the literature and no detection in the HIPASS cubes, $5\sigma$ upper limits estimated from the cubes are plotted as thin arrows.
The large diamond marks SDP~1, the only galaxy that is not part of the flux-limited sample to have an \hi\ detection.
}
\label{fig:corr_fir_shi}
\end{center}
\end{figure*}

Correlations clearly exist between the fluxes in both CO lines and all IR bands in Figure~\ref{fig:corr_fir_sco}, but there is considerable scatter outside of the error bars, indicating intrinsic variation within the sample. 
A correlation also exists between \cott\ and \hi\ ($p=0.001$), while \coto\ and \hi\ are not significantly correlated, although only six sources have both lines detected. 
Figure~\ref{fig:corr_fir_shi} shows the correlations between the \hi\ fluxes and the same IR bands. Scatter is particularly high in the correlation with the \WISE\ and \IRAS\ bands, but shows a marked reduction as the IR wavelength is increased. Two notable outliers persist through each of the plots in Figure~\ref{fig:corr_fir_shi}, which are SDP~1 (with the lowest \hi\ flux) and NGC~5496 (with the highest). These two are also outliers in the CO--\hi\ correlation (Figure~\ref{fig:corr_fir_sco}, last panel), with SDP~1 having a very high \coto/\hi\ ratio, and NGC~5496 having low \coto/\hi\ and \cott/\hi\ ratios (the latter being an upper limit).
These outliers will be revisited in Section~\ref{sec:outliers}.

\subsection{Strength of the correlations}
\label{sec:r_s}
We use the non-parametric Spearman's rank coefficient $r_s$ to compare the relative strengths of the correlations in Figures~\ref{fig:corr_fir_sco} and \ref{fig:corr_fir_shi}. These coefficients, which are printed in the upper left of each panel, are calculated using all detected fluxes (not including upper limits or the uncertain detection of NGC~5584). 
Alongside the $r_s$ coefficients of each correlation, we also show the associated $p$-values (one-sided probability of measuring the given correlation strength under the null hypothesis). These indicate that the correlations between all IR bands and \cott\ or \hi\ fluxes are significant ($p<0.05$), but that correlations between FIR bands at $\lambda\leq160\mum$ and \coto\ are not.
If we exclude the galaxies which are not part of the flux-limited sample (see Table~\ref{tab:phottable}), results are consistent within errors, and the $p$-values for \cott\ and \hi\ correlations remain significant ($p<0.05$).

The $r_s$ values show an increase in the level of correlation between the IR and \cott\ flux as the IR wavelength decreases from 500 to 100\mum. This indicates that \cott\ is more strongly associated with warmer dust, which increasingly dominates the SED at shorter wavelengths.
The decreasing trend at $\lambda\geq100\mum$ is inverted in the correlations with \hi\ fluxes, which show a strong increase in $r_s$ with wavelength. The correlation of IR flux with the \coto\ line also shows a general increase in strength with wavelength, indicating that the emission in this line may be more closely related to cold dust, and possibly \hi, than the \cott. However, most of the correlations with \coto\ are not significant.
Weaker correlations between \cott\ and the 22 and 60\mum\ bands break the trend that is seen at longer wavelengths, as does the relatively strong correlation between \hi\ and 22\mum. 
This may result from a contribution to these wavelengths from stochastically-heated very small grains (VSGs), which may be more strongly associated with cold cirrus dust than warm dust in star-forming regions  \citep{Draine1985,Walterbos1987,Xu1994,Compiegne2010a}.

The trends are visualised in Figure~\ref{fig:chisq_sco} (left-hand panel), in which the uncertainties on $r_s$ are estimated from 500 jackknife resamples of the data, accounting for the potential bias of extreme data points on the measured correlation.
We also estimated the uncertainties on the $r_s$ parameters resulting from the flux error bars, by measuring the standard deviation in the measured parameter in 500 Monte Carlo realisations. In each realisation the fluxes were randomly perturbed with a Gaussian probability distribution of width $\sigma$ equal to the flux error bar. The uncertainty on $r_s$ resulting from the flux errors is on average a factor 2.2 smaller than the uncertainty given by the jackknife method, so we adopted the jackknife errors on $r_s$ as the error bars in Figure~\ref{fig:chisq_sco}.

With these uncertainties on $r_s$ it is apparent that the evolution in the \hi--IR correlation is significant, but that in the \cott--IR and \coto--IR correlations is less so. It is also significant that \cott\ is better correlated with 100\mum\ than either \coto\ or \hi, and \hi\ is better correlated with the submm (350 and 500\mum) than either CO line.
The same trends are observed using Pearson's coefficient $r_p$ and Kendall's $\tau$ parameter as alternative measures of correlation.

\subsection{Parameterizing the scatter}
\begin{figure*}
\begin{center}
\includegraphics[width=\textwidth]{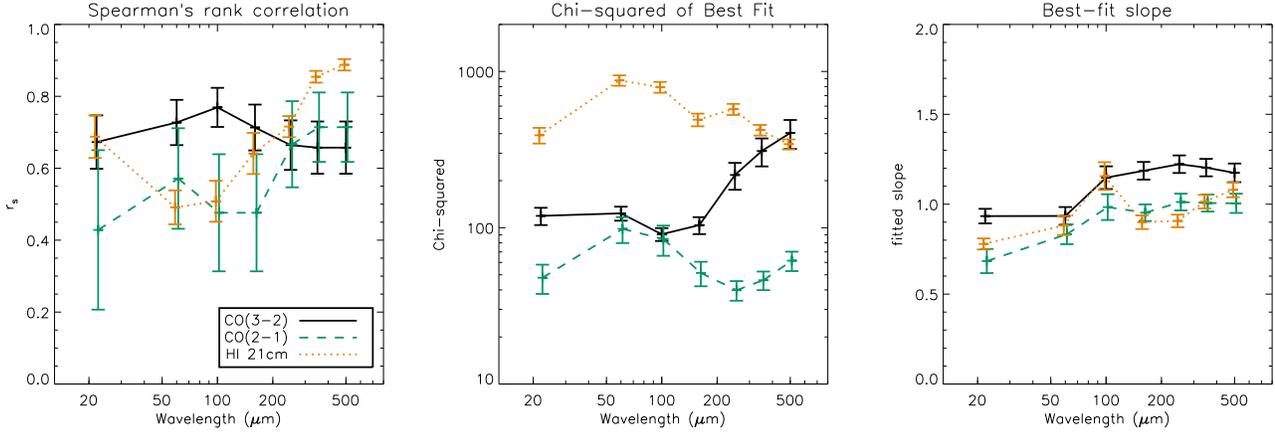}
\caption{Spearman's rank correlation, \chisq\ of the best fit, and slope of the best fit, for the relationship between line fluxes and IR fluxes as a function of IR wavelength.
Here, the slope $m$ is defined such that $S_\text{line} \propto S_\text{IR}^m$.
}
\label{fig:chisq_sco}
\end{center}
\end{figure*}

In order to understand the intrinsic scatter in the flux correlations we need an estimate of the additional scatter above that expected from the size of the error bars. 
For this we use \chisq, which explicitly describes the measured deviation of the data from a given model, in relation to the expected variance due to errors. 
We measured \chisq\ using the {\sc fitexy} routine from the NASA IDL library\footnote{\url{http://idlastro.gsfc.nasa.gov/}}, which takes into account errors on both axes following \citet{Press2007}.
Unlike non-parametric statistics such as Spearman's rank coefficient, \chisq\ depends on the model that the data are compared to, but if the model is plausible then \chisq\ is a good measure of intrinsic scatter around that trend.
We might expect the various IR bands to be linearly correlated with the gas mass, either as a result of the SK law (assuming the FIR bands trace SFR) or as a result of a constant gas-to-dust ratio (assuming the submm bands trace dust mass). The relationship can vary, however, if these assumptions are not valid, and depending on the linearity of the conversion from the CO tracer to the mass of star-forming gas.
We fitted power-laws to the flux scatter plots, weighting by errors on both line fluxes and IR fluxes.
Slopes close to unity were obtained in all cases (see Figure~\ref{fig:chisq_sco}, right-hand panel), although there is some variation between different lines/bands, which may be genuine or may be biased by outliers in the individual plots.

The \chisq\ values of the fits are shown in Figure~\ref{fig:chisq_sco} (middle), with errors estimated from a jackknife resampling of the data. These \chisq\ values pertain to the best-fit slopes which vary between different correlations, but indistinguishable results were obtained using a model with the slope fixed to unity. The \chisq\ values increase with IR wavelength from 100 to 500\mum\ for the \cott\ correlations, but decrease for the \hi\ correlations over the same wavelength range. This is consistent with the inverse trends measured in the correlation coefficients. These trends indicate that the \cott\ emission, tracing dense molecular gas in star-forming regions, is better correlated with 100\mum\ fluxes (at the peak of the SED) than fluxes at longer wavelengths (which are increasingly dominated by emission from cold dust).  
The opposite trend is implied by the \hi\ correlations, and by \coto\ between 60 and 250\mum\ only, suggesting that these two transitions trace different components of the ISM. We will discuss the interpretation of this in Section~\ref{sec:interpretation}.

If we exclude from the analysis the galaxies which are not part of the flux-limited sample (see Table~\ref{tab:phottable}), the \chisq\ of all correlations with \hi\ are reduced by around a factor 0.6 due to the removal of the outlier SDP~1, but the trend remains the same. Correlations between \coto\ and FIR bands with $\lambda\leq100\mum$ also have a lower \chisq\ due to the removal of SDP~1 and SDP~4, but correlations with longer-wavelength bands are unaffected, and the negative trend with wavelength up to 250\mum\ remains. 
Correlations involving \cott\ are unaffected because none of these galaxies is detected in this line.
Regardless of whether we include all galaxies in the sample, fluxes at 22\mum\ exhibit less scatter with \coto\ and \hi\ than fluxes at $60-160\mum$ (closer to the peak), which suggests that dust emission at 22\mum\ is more correlated with the diffuse ISM, consistent with the $r_s$ results in Section~\ref{sec:r_s}. 

A more ambiguous observation is that the \chisq\ values for the \hi--submm correlations imply a similar level of scatter as in the \cott--submm correlations, while the $r_s$ results imply less scatter in the \hi--submm (see black and yellow lines in Figure~\ref{fig:chisq_sco}). 
The \chisq\ scales with the number of degrees of freedom: $dof=N_\text{data}-N_\text{model}$. For the \hi--SPIRE correlations, $dof=17-2=15$, while for the \cott--SPIRE correlations, $dof=12-2=10$. This relative difference of $1.5$ partly explains the discrepancy, but it is also possible that the systematic errors on \hi\ fluxes have been under-estimated since the data are from multiple sources, unlike the CO. Under-estimated errors would lead to higher \chisq\ overall (since \chisq\ measures scatter beyond the errors).

A simple way to quantify the confidence of the trends with wavelength is a linear least-squares fit to the data in the left and middle panels of Figure~\ref{fig:chisq_sco}, resulting in an estimate of the slope and its uncertainty (from the jackknife error bars), assuming a linear trend. In fact the trend is unlikely to be linear, since there are multiple dust components whose relative luminosities vary as a function of wavelength according to grain size and temperature \citep[\eg][]{Draine1985, Compiegne2010a}, 
but in the absence of a more realistic model this is the simplest assumption. 
We chose to exclude the 22\mum\ data from these fits since it clearly violates the monotonic trend with wavelength, which we interpret to be a result of the VSG contribution. Assuming that the other bands are dominated by emission from large grains in thermal equilibrium, we can assume that shorter wavelengths trace warm dust, and longer wavelengths trace colder dust. The results of the linear fits are shown in Table~\ref{tab:slopesignificance}.
Thus, according to the \chisq\ results, the correlation of \cott\ with the IR becomes tighter with decreasing wavelength, while both \chisq\ and $r_s$ indicate that the correlation of \hi\ with IR becomes tighter with increasing wavelength. The correlation of \coto\ with IR may also become tighter with increasing wavelength, but the upturn in \chisq\ at 350 and 500\mum\ reduces our confidence in this observation, while the low significance of the $r_s$ coefficients shows that more data are required to understand this correlation fully.

\begin{table}
\caption{Confidence levels in slopes of the correlation coefficients and chi-squared with wavelength  ($\lambda\geq60\mum$), from Figure~\ref{fig:chisq_sco}.}
\begin{center}
 \begin{tabular}{l c c}
\hline
  Line & Parameter & Deviation of slope from zero\\
\hline
 \cott & $r_s(\lambda)$ & $-1.4\sigma$\\
 \cott & $\chisq(\lambda)$ & $+3.8\sigma$\\
 \coto & $r_s(\lambda)$ & $+1.5\sigma$\\
 \coto & $\chisq(\lambda)$ & $-0.8\sigma$\\
 \hi   & $r_s(\lambda)$ & $+10.2\sigma$\\
 \hi   & $\chisq(\lambda)$ & $-8.2\sigma$\\
\hline
 \end{tabular}
\end{center}
\label{tab:slopesignificance}
\end{table}

\section{Discussion} 
\label{sec:disc}
\subsection{Relationships between ISM tracers}
\label{sec:interpretation}
In Section~\ref{sec:analysis} we observed an increase in the scatter between global \cott\ and IR flux densities with increasing IR wavelengths ($\lambda\geq100\mum$), which may indicate that the amount of dense gas (which emits the higher excitation CO lines) is more strongly associated with warm dust (which dominates at $\sim100\mum$) than cold dust (which dominates at longer wavelengths). This is consistent with the notion of dense gas being the fuel for ongoing star formation, {and that star formation is directly traced by emission from warm dust (see Section~\ref{sec:seds}).}
Applying similar arguments to the scatter in the \hi/IR [and \coto/IR] indicates that atomic gas (and more diffuse molecular gas) is more strongly associated with cold dust, although the situation is not clear-cut with \coto\ due to insufficient data. 
In comparison, \citet{Corbelli2012} used the \cooz\ line to trace the molecular gas content in Virgo spirals with FIR-submm data from the \Herschel\ Virgo Cluster Survey \citep[HeViCS][]{Davies2010}. This line traces the total molecular gas mass (modulo uncertainties in the CO/\htwo\ conversion), and those authors found that it was better correlated with cold dust than warm dust. 

Since \coto\ tends to be more closely associated with \cooz\ than \cott\ \citep{Israel1984,Braine1993,Sakamoto1995}, one would expect correlations involving \coto\ to follow the same trends, as is suggested (but not proven) in the current sample.
The excitation temperature of \coto\ is intermediate between that of the (1--0) and (3--2) lines, so it is not surprising that the results of correlations involving this line are somewhat ambiguous. Nevertheless, the kinetic temperature needed to excite \cott, $T_k\sim E/k_B=33$\,K, is about twice that for \coto.
Moreover, the critical density required to thermalise the lines is much higher for the (3--2) transition ($n_\text{crit}\sim10^5$\,cm$^{-3}$) compared to the (2--1) transition ($n_\text{crit}\sim10^4$\,cm$^{-3}$). 
So it is reasonable to assume that the \cott\ transition traces warm, dense gas in star-forming regions while \coto\ traces more diffuse \htwo.

There are also parallels in the results of \citet{Wilson2009}, who mapped three spirals in the Virgo cluster in \cott\ and \cooz, and concluded that the \cott\ transition was better correlated with two different SFR indicators -- 24\mum, tracing warm dust, and H$\alpha$, tracing ionising radiation from OB stars. Likewise, \citet{Iono2009} and \citet{Gao2004a} found strong correlations between dense gas tracers [\cott\ and HCN respectively] and the FIR, while \citet{Wilson2012} observed a strong \cott--FIR correlation in low redshift \hi-selected galaxies from NGLS.
By analogy, \citet{Calzetti2010} showed that the correlation between the FIR and SFR (derived from H$\alpha$ and 24\mum) was shown to have increasing scatter with increasing FIR wavelength (70 to 160\mum) as a result of both variable metallicity and dust heated by old stellar populations.

The scatter in the correlation between \cott\ and IR fluxes at $\lambda<100\mum$ does not follow the trend of the other wavelengths, suggesting that the 22 and 60\mum\ bands contain some contribution that is not correlated with star formation and dense gas. A likely candidate is emission from VSGs, whose emission spectrum depends on transient heating by individual photons \citep{Draine1985} and can be important at both 22 and 60\mum\ in low-intensity radiation fields \citep{Compiegne2010a}. The incidence of VSG emission varies between different spiral galaxies and may well be better correlated with cold cirrus dust rather than warm dust and SFR \citep{Draine1985,Walterbos1987,Xu1994}. The relatively low \chisq\ and high $r_s$ of the correlation between \hi\ and 22\mum\ flux indicates that the FIR emission at that wavelength is dominated by dust associated with the diffuse neutral medium. Moving up to 60 and 100\mum, the results in Figure~\ref{fig:chisq_sco} indicate that this contribution weakens relative to 
dust associated with denser gas.

Considering large grains in thermal equilibrium (which dominate the emission at $\lambda\gtrsim60\mum$),
our results, together with the parallels in the literature described above, constitute strong evidence that dust at different temperatures inhabits different phases of the ISM.
Correlations with dense molecular gas tracers such as \cott\ and SFR tracers such as H$\alpha$ indicate that warm dust is associated with dense molecular clouds, while correlations with \hi\ [and \coto] indicate that cold dust is more strongly associated with the diffuse phase of the ISM. 
It also directly implies that the dominant component of cold dust is less strongly associated with star formation than warm dust, since star formation is fueled by the dense gas. There is a component of cold dust in the cores of dense molecular clouds, where it is shielded from the UV radiation by high optical depths, but the warm dust in those regions dominates the emission. The diffuse ISM is likely to contain a much higher mass of cold dust which dominates the emission at submm wavelengths.

This has implications for the dust heating mechanism. Cold cirrus dust may be heated by the general radiation field, or by UV photons escaping from star-forming molecular clouds \citep{Boulanger1988}; however, the increased scatter between the submm and {both SFR and dense gas} suggests that it may be heated by sources other than young OB stars. 
AGN are a potential source of dust heating, {although evidence suggests that the FIR/submm SEDs of AGN hosts are dominated by dust heated by star formation} \citep{Schweitzer2006,Gallimore2010,Hatziminaoglou2010}. \citet{Bendo2010} studied the 70--500\mum\ SED of the low-luminosity AGN host M81, and found evidence for synchrotron emission from the AGN contributing to the 500\mum\ flux, but only within the central $\sim2$~kpc. The majority of our sample (apart from SDP~1 and SDP~15) have no recorded evidence of an AGN (e.g. in SDSS spectra) so synchrotron emission is unlikely to affect the measured correlations. 
{We tested this by extracting $1.4\,$GHz fluxes from the NVSS \citep{Condon1998}, using a 30-arcsec matching radius. The radio fluxes appear to be dominated by star formation rather than AGN (see the FIR--radio correlation in Section~\ref{sec:FRC}). To test whether synchrotron or free-free emission related to star formation could bias the 500\mum\ fluxes,
we assumed a synchrotron spectrum $S_\nu\propto\nu^{-0.8}$, and a free-free spectrum $S_\nu\propto\nu^{-0.1}$ \citep{Condon1992}. Adopting the free-free fraction of 0.76 at $33\,$GHz \citep{Murphy2012}, we extrapolated from $1.4\,$GHz to 500\mum\ and found that the contribution of free-free plus synchrotron emission is less than 2 per cent of the observed 500\mum\ fluxes, so will have a negligible effect on the correlations.

Having ruled out AGN dust-heating and radio emission affecting the submm--CO correlations, it is likely that the scatter is increased by cold dust being heated at least partially by old stars \citep[\eg][]{Helou1986,Walterbos1996}.}
There is much observational evidence to support this theory from the correlations between FIR colours and photometric tracers of old and young stars, both in resolved regions within galaxies \citep{Boquien2011,Komugi2011,Bendo2010,Bendo2011,Groves2012,Smith2012a} as well as in the global properties of galaxies \citep{Rowan-Robinson2010a,Totani2011,Boselli2012}. 
The high level of scatter in the correlations between dense CO and the long wavelength bands in particular shows that these submm wavelengths do not directly trace SFR, but rather trace diffuse dust (and gas) which may not be associated with star-formation. 
Even the 100\mum\ band shows considerable scatter ($r_s=0.77\pm0.06$; $\chisq=91\pm 9$) suggesting a significant contribution from cold dust not correlated with the SFR. Reliable IR SFRs can only be obtained from full SED fits to data covering the peak, and allowing for emission from at least two components of dust.

\subsection{Infrared SED fitting}
\label{sec:seds}

We used the six photometric data points to fit a two-component SED model for each galaxy:
\begin{equation}
\begin{array}{r}   
L_\nu(\nu) = 4\pi\,\kappa_{850} \left[ \left(\dfrac{\nu}{\nu_{850}}\right)^{\beta_w}M_w B_\nu(\nu,T_w)\;  \hspace{1cm} \right. \\
\left. + \left(\dfrac{\nu}{\nu_{850}}\right)^{\beta_c} M_c B_\nu(\nu,T_c) \right].
\end{array}
\label{eqn:twocomptsed}
\end{equation}
The 22\mum\ fluxes were treated as upper limits to allow for the possibility of a VSG emission component, over that from the thermal hot dust, which cannot be modelled using modified blackbodies.
The dust emissivity term \mbox{$\kappa_\nu\propto\nu^{\beta}$} was normalised to \mbox{$\kappa_{850} = 0.077\,\text{m}^2\,\text{kg}^{-1}$} at~$\nu_{850}=c/(850\mum)$ \citep{Dunne2000}. Uncertainty in this value leads to a systematic uncertainty in the normalisation of dust masses, but assuming that large dust grains (which dominate at $\lambda\gtrsim60\mum$) have similar emission properties in different galaxies, the trends observed in the results will be robust. 
The spectral index of emissivity was fixed at $\beta_c=2$ for the cold dust component and $\beta_w=1.5$ for the warm \citep[e.g.][]{Li2001,Dale2002,daCunha2008}. This parameter depends on distributions of grain size and temperature, which motivates the choice of a higher value for the cold, cirrus dust component; values between 1.5 and 2 are generally measured in external galaxies \citep[see also][]{Dunne2001}. Using $\beta_c=2$ for cold dust provides a good fit to nearby galaxies observed with \Herschel\ \citep{Davies2011,Smith2012d}, at least for metal-rich, high surface brightness galaxies \citep{Boselli2012}.

\begin{figure}
\includegraphics[height=0.95\textheight, width=0.45\textwidth]{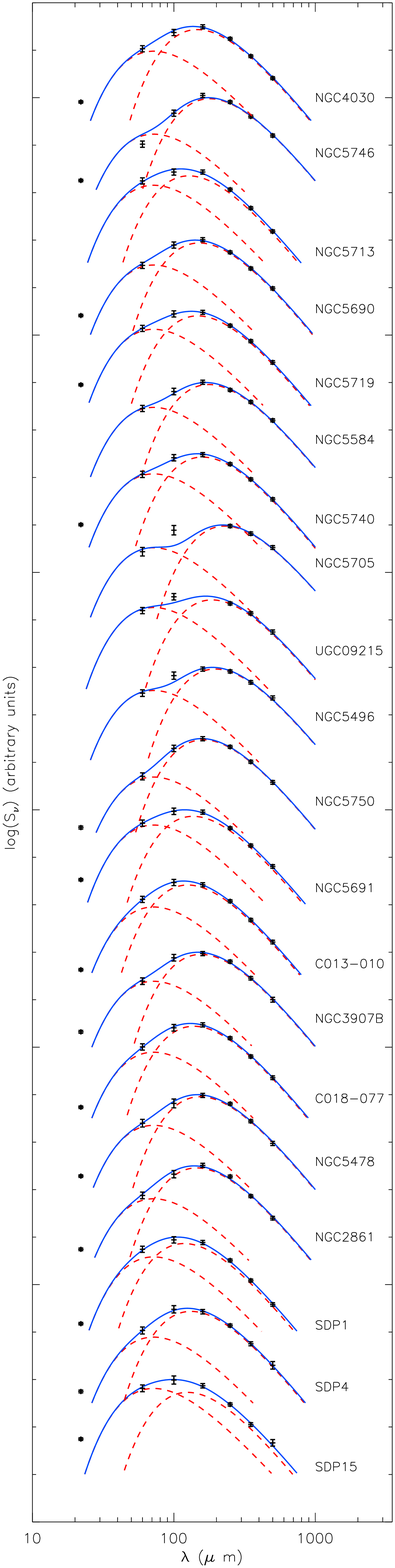}
\caption{Photometry and best-fit two-component SED models for galaxies in the sample. Note that the 22\mum\ data were used as upper limits to allow for a contribution from very small grains that cannot be modelled as a modified blackbody.}
\label{fig:sedfits}
\end{figure}

The remaining free parameters are the masses of warm and cold dust, $M_w$ and $M_c$, and their temperatures, $T_w$ and $T_c$. 
{We were unable to adequately constrain $T_w$ with the 22\mum\ data as an upper limit only,
but $M_c$ and $T_c$ are well constrained after fixing $T_w$. Repeating the fitting with $T_w$ fixed at a range of values between 30 and 60\,K, we found that 45\,K gave the minimum \chisq, and so this value was assumed. }
The potential for a varying $\beta_c$ leads to a small uncertainty in $T_c$ and $M_c$; for example using $\beta_c=1.5$ leads to a temperature increase by 15 -- 30 per cent (2 -- 7\,K) and a change in mass of up to $\pm$10 per cent.\footnote{We acknowledge the possibility that dust masses measured by fitting global photometry in this way can be biased compared with spatially resolved measurements which separate the dense and diffuse phases of the ISM \citep[\eg][]{Bendo2011}. For example, global photometry can under-estimate the contribution of the warm component at long wavelengths, leading to an over-estimate of its temperature and an under-estimate of its mass. Although we cannot be sure of the size of this bias using the current data, we note that our excellent photometric coverage of the SED should give reliable constraints on the dust masses under a two-component assumption. As always, absolute values of dust masses should be treated with caution considering the assumptions in the model.}
Derived values are collected in Table~\ref{tab:derived} and the models are shown in Figure~\ref{fig:sedfits}. Dust temperatures are typical of the full \hatlas\ sample, while luminosities and masses are consistent with the low end of the \hatlas\ sample \citep{Smith2012c}, which is explained by the selection of very nearby galaxies in the current sample. The SEDs do not show any evidence for an additional component of cold dust at $T\sim10$\,K  \citep[\eg][]{Galametz2011}.

\begin{figure}
\includegraphics[width=0.45\textwidth]{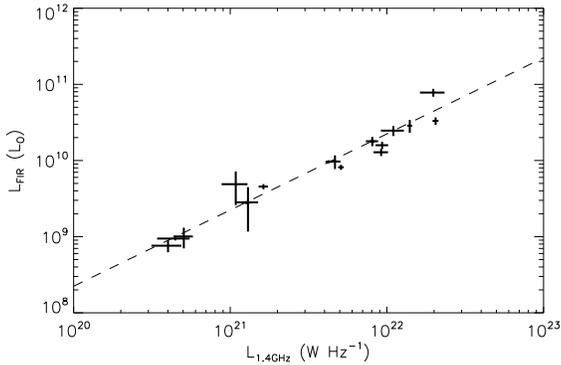}
\caption{The correlation between integrated $40-120\mum$ FIR and $1.4$\,GHz radio continuum luminosities. The dashed line is the correlation from \citet{Bell2003}.}
\label{fig:FRC}
\end{figure}

{We integrated the model SED to obtain $\Lfir=L_{40-120\mum}$, a commonly-used estimate of SFR. This integral is not dependent on the assumed $\beta_w$ or $T_w$ since the normalisation is constrained by the data at 60 and 100\mum. Similarly, it should not be strongly affected by VSG emission (which can be important at $\lambda \lesssim60\mum$).}
\label{sec:FRC}
{We can test the assumption that \Lfir\ traces SFR by comparing to radio continuum luminosities from the NVSS \citep{Condon1998}. We matched to the NVSS catalogue on Vizier using a 30 arcsec matching radius, and found 1.4\,GHz fluxes for all but NGC~5705. In Figure~\ref{fig:FRC} we plot the FIR--radio correlation, showing that the FIR luminosities are consistent with the expected correlation for star-forming galaxies from \citet{Bell2003}. The tightness of this correlation indicates that the FIR luminosities do not significantly overestimate the SFR assuming that $1.4\,$GHz is a linear tracer of SFR (in fact it is not quite linear for $L_{1.4}<6.4\times10^{21}\,\text{W\,Hz}^{-1}$ as discussed by \citet{Bell2003}, but the deviation is small compared to the errors in this sample).}

\begin{figure*}
\begin{minipage}[t]{\linewidth}
\begin{center}
\includegraphics[width=0.9\textwidth]{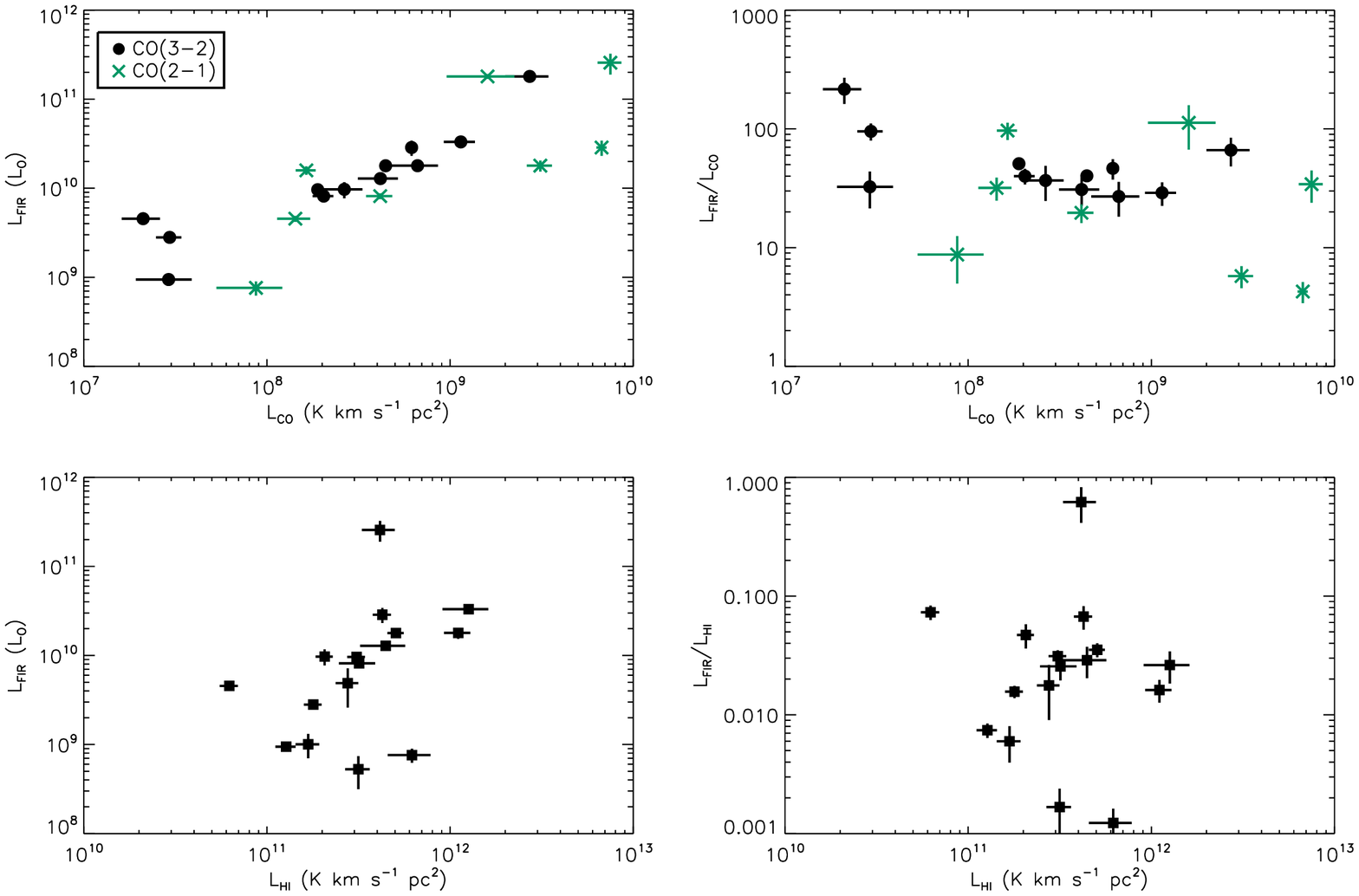}
\caption[Integrated $\Lfir$ and $\Lfir/L_\text{CO}$ versus CO and \hi\ luminosities]{Integrated FIR luminosity \Lfir\ from two-component SED fits versus CO and \hi\ luminosities (left), and the ratios $\Lfir/L_\text{CO}$ and $\Lfir/L_\text{HI}$ (right).}
\label{fig:lfir_lco}
\end{center}
\end{minipage}
\begin{minipage}[t]{\linewidth}
\begin{center}
\includegraphics[width=0.9\textwidth]{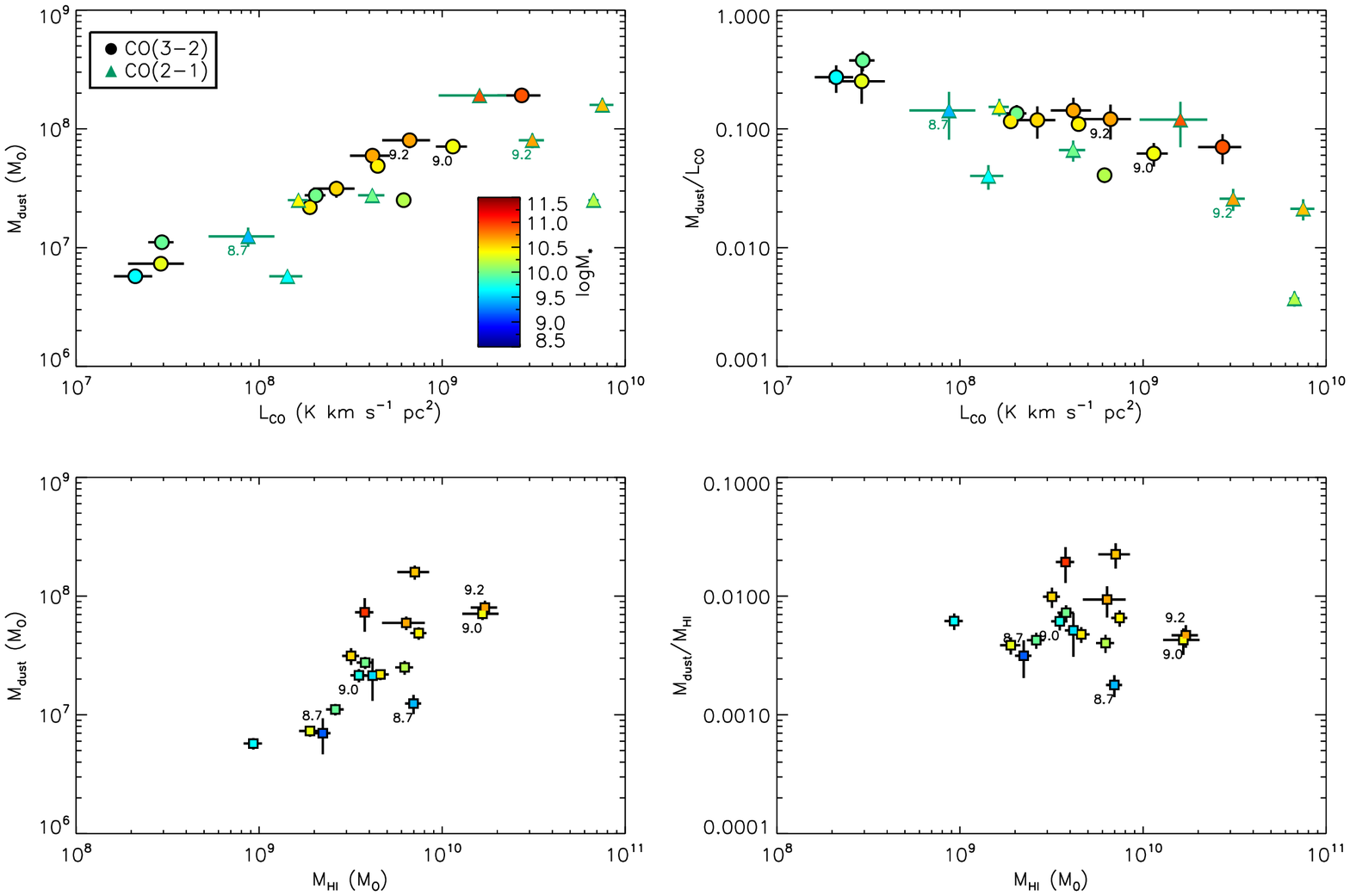}
\caption[Total dust mass and $\Mdust/L_\text{CO}$ versus CO luminosity, and dust mass and $\Mdust/M_\text{HI}$ versus \hi\ mass]{Total dust mass from two-component SED fits versus CO luminosity (top left), and the ratio $\Mdust/L_\text{CO}$ versus $L_\text{CO}$ (top right); dust mass versus \hi\ mass (bottom left) and $\Mdust/M_\text{HI}$ versus $M_\text{HI}$ (bottom right). Points are coloured according to the stellar mass derived from multiwavelength SED fitting as described in the text. Labels show the metallicities [12+log(O/H)] for galaxies with spectroscopic measurements in the MPA-JHU catalogue.}
\label{fig:mdust_lco_lhi}
\end{center}
\end{minipage}
\end{figure*}

In Figure~\ref{fig:lfir_lco} we plot our estimate of \Lfir\ against CO luminosity in each of the two emission lines, defined as
\begin{equation}
 L_\text{CO} = \dfrac{I_\text{CO}\, \Omega\, d_L^2}{(1+z)^3}   \;  \Kkmspc
\label{eqn:Lco_Ico}
\end{equation}
where $d_L$ is the luminosity distance in pc.
In order to remove the confounding effect of the $d^2$ factor, which can accentuate correlations between luminosities, we plot the ratio of luminosities against $L_\text{CO}$ in the right-hand panel of Figure~\ref{fig:lfir_lco}.
The scatter in these plots confirms that \cott\ is better correlated with the integrated FIR than \coto, as was also seen in the flux correlations.
The \Lfir/$L\subcott$ ratio appears to be roughly constant across the sample, although one galaxy is a marginal outlier (NGC~5691, which has the lowest $L\subcott$ and the highest ratio). This is a very blue spiral galaxy which appears to be strongly star-forming; it also has a relatively high $M_w/M_c$ ratio from the SED fit, indicative of a hot SED, so it is possible that the gas excitation conditions are different or that there is another contribution to the integrated \Lfir\ in this source.
Figure~\ref{fig:lfir_lco} also shows \Lfir\ against the \hi\ luminosity, in which the correlation is very weak.
{If \Lfir\ traces the SFR, this indicates a surprisingly weak relationship between the atomic gas reservoir and the SFR, albeit over a relatively small luminosity range. This could be due to a large fraction of the $40-120\mum$ emission arising from dust heated by old stars in this submm-selected sample, or it may be that the galaxies with high FIR/\hi\ ratios have a greater fraction of gas in the molecular phase. }

\begin{sidewaystable*}
\caption{Physical parameters of the sample: stellar masses from \textsc{magphys} SED models; metallicities from MPA-JHU; CO luminosities and \hi\ masses; and dust properties from the two-component SED models.}
{\small
\setlength{\tabcolsep}{4pt}
\begin{center}
\begin{tabular}{l c c c c c c c c c c}
\hline
  Galaxy          & $M_{\star}$       &  $Z$             &    $L_{32}$        &    $L_{21}$       & $M_\text{HI}$        & $M_\text{dust}$&    $T_{c}$     &     $M_w/M_c$           &  $L_\text{8-1000\mum}$                               \\
                  & ($10^{10}\Msun$)   &  12+log(O/H)    & ($10^8\Kkms pc^2$) & ($10^8\Kkms pc^2$)& ($10^{9}\Msun$)      & ($10^7\Msun$)  &      (K)       &                         &  ($10^{10}\Lsun$) \\
\hline
  NGC 4030      & $2.78  \pm 0.26  $ &  $             $ &  $4.45 \pm 0.30  $ & $               $ & $7.45 \pm 0.75$ & $4.88\pm0.32 $   & $19.8   \pm    0.4 $ & $(2.05\pm 0.16)  \times10^{-2}$ &  $3.02  \pm 0.27  $  \\
  NGC 5746      & $10.4  \pm 0.96  $ &  $             $ &  $               $ & $               $ & $3.77 \pm 0.43$ & $7.3 \pm2.2  $   & $16.2   \pm    1.3 $ & $(3.79\pm 1.35)  \times10^{-3}$ &  $1.13  \pm 0.55  $  \\
  NGC 5713      & $1.53  \pm 0.14  $ &  $             $ &  $6.16 \pm 0.20  $ & $ 67.2 \pm 4.2  $ & $6.24 \pm 0.70$ & $2.51\pm0.22 $   & $22.7   \pm    1.0 $ & $(7.34\pm 1.73)  \times10^{-2}$ &  $4.18  \pm 0.78  $  \\
  NGC 5690      & $0.989 \pm 0.091 $ &  $             $ &  $2.04 \pm 0.26  $ & $ 4.15 \pm 0.67 $ & $3.80 \pm 0.39$ & $2.75\pm0.16 $   & $19.2   \pm    0.4 $ & $(1.71\pm 0.12)  \times10^{-2}$ &  $1.42  \pm 0.11  $  \\
  NGC 5719      & $2.94  \pm 0.27  $ &  $             $ &  $1.89 \pm 0.075 $ & $               $ & $4.60 \pm 0.51$ & $2.188\pm0.064$  & $19.6   \pm    0.2 $ & $(2.93\pm 0.16)  \times10^{-2}$ &  $1.55  \pm 0.085 $  \\
  NGC 5584      & $0.532 \pm 0.049 $ &  $9.01 \pm 0.01$ &  $0.453\pm 0.072 $ & $               $ & $3.51 \pm 0.35$ & $2.15\pm0.17 $   & $16.4   \pm    0.4 $ & $(7.60\pm 0.99)  \times10^{-3}$ &  $0.466 \pm 0.054 $  \\
  NGC 5740      & $0.910 \pm 0.084 $ &  $             $ &  $0.294\pm 0.046 $ & $               $ & $2.61 \pm 0.29$ & $1.109\pm0.047$  & $18.2   \pm    0.2 $ & $(1.71\pm 0.08)  \times10^{-2}$ &  $0.485 \pm 0.024 $  \\
  NGC 5705      & $0.348 \pm 0.032 $ &  $             $ &  $               $ & $               $ & $4.16 \pm 0.42$ & $2.14\pm0.80 $   & $12.6   \pm    1.3 $ & $(2.17\pm 1.06)  \times10^{-3}$ &  $0.113 \pm 0.061 $  \\
  UGC 09215     & $0.144 \pm 0.013 $ &  $8.68 \pm 0.01$ &  $               $ & $               $ & $2.23 \pm 0.28$ & $0.700\pm0.22 $  & $15.6   \pm    1.6 $ & $(1.22\pm 0.47)  \times10^{-2}$ &  $0.171 \pm 0.064 $  \\
  NGC 5496      & $0.269 \pm 0.025 $ &  $8.66 \pm 0.05$ &  $<0.45          $ & $  <0.10        $ & $6.97 \pm 0.70$ & $1.24\pm0.19 $   & $14.8   \pm    0.6 $ & $(4.79\pm 1.08)  \times10^{-3}$ &  $0.156 \pm 0.032 $  \\
  NGC 5750      & $2.19  \pm 0.20  $ &  $             $ &  $0.290\pm 0.098 $ & $  <2.01        $ & $1.89 \pm 0.24$ & $0.731\pm0.027$  & $17.9   \pm    0.2 $ & $(5.79\pm 0.26)  \times10^{-3}$ &  $0.198 \pm 0.011 $  \\
  NGC 5691      & $0.440 \pm 0.04  $ &  $             $ &  $0.211\pm 0.050 $ & $1.43 \pm  0.29 $ & $0.929\pm 0.11$ & $0.573\pm0.026$  & $21.2   \pm    0.4 $ & $(5.40\pm 0.50)  \times10^{-2}$ &  $0.685 \pm 0.065 $  \\
  C 013-010     & $8.26  \pm 0.76  $ &  $             $ &  $27.2 \pm 7.2   $ & $  <19.2        $ & $<36.3        $ & $19.1\pm0.47 $   & $23.1   \pm    0.3 $ & $(4.34\pm 0.37)  \times10^{-2}$ &  $27.1  \pm 1.77  $  \\
  NGC 3907B     & $4.67  \pm 0.043 $ &  $             $ &  $4.2  \pm 1.0   $ & $               $ & $6.36 \pm 1.66$ & $5.95\pm0.50 $   & $18.6   \pm    0.4 $ & $(1.17\pm 0.13)  \times10^{-2}$ &  $2.37  \pm 0.27  $  \\
  C 018-077     & $2.37  \pm 0.022 $ &  $9.01 \pm 0.03$ &  $11.4 \pm 2.2   $ & $  <18.1        $ & $16.6 \pm 3.71$ & $7.09\pm0.38 $   & $20.9   \pm    0.4 $ & $(2.19\pm 0.31)  \times10^{-2}$ &  $5.48  \pm 0.68  $  \\
  NGC 5478      & $4.78  \pm 0.44  $ &  $9.24 \pm 0.06$ &  $6.64 \pm 1.9   $ & $  31.1 \pm 4.9 $ & $17.1 \pm 2.80$ & $8.02\pm0.81 $   & $18.9   \pm    0.6 $ & $(1.14\pm 0.13)  \times10^{-2}$ &  $3.32  \pm 0.48  $  \\
  NGC 2861      & $3.43  \pm 0.32  $ &  $             $ &  $2.65 \pm 0.67  $ & $               $ & $3.18 \pm 0.34$ & $3.14\pm0.40 $   & $20.0   \pm    0.8 $ & $(1.36\pm 0.21)  \times10^{-2}$ &  $1.73  \pm 0.34  $  \\
  SDP 1         & $3.87  \pm 0.36  $ &  $             $ &  $               $ & $75   \pm  11   $ & $7.09 \pm 1.39$ & $15.9\pm1.5  $   & $24.5   \pm    1.3 $ & $(8.93\pm 3.14)  \times10^{-2}$ &  $36.3  \pm 10.4  $  \\
  SDP 4         & $2.69  \pm 0.25  $ &  $             $ &  $               $ & $1.6  \pm  0.20 $ & $<16.4        $ & $2.51\pm0.14 $   & $21.9   \pm    0.4 $ & $(2.79\pm 0.40)  \times10^{-2}$ &  $2.52  \pm 0.30  $  \\
  SDP 15        & $2.74  \pm 0.25  $ &  $             $ &  $<11.9          $ & $1.3  \pm 1.1   $ & $             $ & $4.18\pm0.55 $   & $23.2   \pm    1.2 $ & $(1.56\pm 0.25)  \times10^{-1}$ &  $10.8  \pm 1.30  $  \\
\hline     
\end{tabular}
\end{center}
}
\label{tab:derived}
\end{sidewaystable*}

\subsection{Dust-to-gas ratio and metallicity}
Figure~\ref{fig:mdust_lco_lhi} shows the dust mass from the SED models plotted against the CO luminosity and \hi\ mass, as well as the ratios as in Figure~\ref{fig:lfir_lco}. 
Errors on dust mass include a 10 per cent error due to the uncertainty of $\beta$ (as described above), added in quadrature to the random errors from the SED fit.
The \hi\ mass is estimated as 
\begin{equation}
M_\text{HI} = 2.36\times10^{-5}\,d_L^2\,S_\text{HI}\ \Msun
\end{equation}
\citep[\eg][]{Dunne2000}.
We have not attempted to estimate \Mhtwo\ due to uncertainties in the CO line ratios and the CO--\htwo\ conversion (which will be discussed below).

The galaxy with the lowest \Mdust/$L_\text{CO}$ in both CO lines is NGC~5713, which is atypical of the sample as it is a starburst galaxy from the SLUGS sample. 
Apart from this outlier, there is very little scatter in the dust/CO ratio, and a decrease in the ratio with  increasing luminosity.
\label{sec:outliers}
There are several outliers in the \Mdust/$M_\text{HI}$ ratios, in particular SDP~1 (with the highest \Mdust/$M_\text{HI}$) and NGC~5496 (with the lowest). Both of these were also outliers in the CO--\hi\ correlation, and in \hi--IR, as discussed in Section~\ref{sec:correlations}. SDP~1 is a spectroscopically identified AGN, has a disturbed visual morphology {(see Figure~\ref{fig:spectra})} and a high luminosity ($\Ltir=4.0\times10^{11}\Lsun$), hence it is atypical of the sample. It is also possible that the \hi\ self-absorption is underestimated for this source (in which case it should have a lower \Mdust/$M_\text{HI}$) -- the correction in Section~\ref{sec:hiselfabs} is based on morphology and inclination alone, but this is likely to be unreliable for mergers. 
The object with the second-highest \Mdust/$M_\text{HI}$ is NGC~5746, which is a dust-rich edge-on spiral, and it is possible that the self-absorption correction was underestimated for this galaxy also (NGC~5746 was not observed in CO so we cannot tell whether it is an outlier in \Mdust/$L_\text{CO}$ as well).
NGC~5496, which has the lowest \Mdust/$M_\text{HI}$, is another edge-on disk galaxy but is very blue and has low surface brightness (unlike NGC~5746 for example). Due to the high inclination, the \hi\ flux has been corrected by a factor of 1.32 for self-absorption; however without this correction this object remains a significant outlier, so it appears to have an intrinsically low dust/gas ratio. 
Figure~\ref{fig:mdust_lco_lhi} shows that it also has a low stellar mass and low metallicity (see also discussion below).

Overall, the scatter in \Mdust/$M_\text{HI}$ is greater than in \Mdust/$L_\text{CO}$, while the scatter in \Mdust/$L\subcott$ is lower than that in \Mdust/$L\subcoto$. 
This contrasts with the flux correlations (see $r_s$ in Figure~\ref{fig:chisq_sco}), which indicated that the long-wavelength emission from cold dust was better correlated with \hi\ and \coto. The discrepancy is likely to result {at least partially from the dependence of derived dust masses on temperature and $\beta$, but there are many other factors which affect this correlation. 
Scatter could be increased if the extent of \hi\ emission probed by the 14.3~arcmin Parkes beam is larger than the aperture within which the submm flux is measured ($\sim 1$~arcmin), so that the \hi\ measurements probe a larger gas reservoir than the dust measurements \citep[see also][]{Devereux1990}. 
It is also likely that there is a range of line ratios in the sample, and that \cott\ and \coto\ do not trace \htwo\ consistently due to a range of excitations.}

\begin{figure}
\begin{center}
\includegraphics[width=0.45\textwidth]{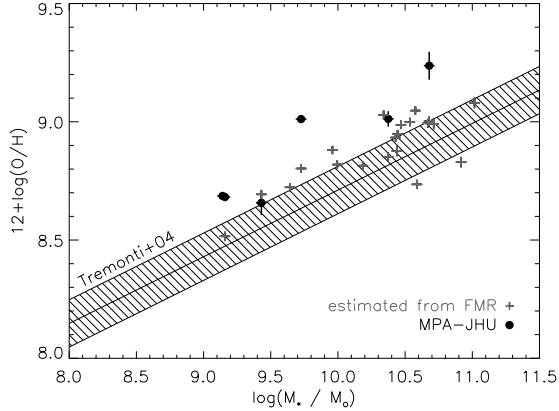} 
\end{center}
\caption{Metallicity estimates from the FMR, based on stellar mass and SFR as described in the text, compared with spectroscopic measurements from the \mbox{MPA-JHU} catalogue. The filled region shows the median $\pm1\sigma$ scatter of the mass--metallicity relation measured by \citet{Tremonti2004} from SDSS DR7.}
\label{fig:mass_mets}
\end{figure}

A range of metallicities in the sample could also influence these correlations, since both dust/gas and CO/\htwo\ ratios depend on metallicity. 
Hence the correlation between CO and \Mdust\ may be stronger because galaxies with higher metallicities have higher dust and CO fractions for the same gas mass.
We can investigate this with independent tracers of the gas-phase metallicity. Spectroscopic measurements from SDSS DR7 are available in the \mbox{MPA-JHU} catalogue \citep{Tremonti2004}, although only five galaxies in the current sample have metallicities in the catalogue since the others were not classified as star-forming. These values are printed in Figure~\ref{fig:mdust_lco_lhi} and in Table~\ref{tab:derived}. 
{The Figure also shows stellar masses for the galaxies, derived from $g-i$ photometry following the method proposed by \citet{Taylor2011}, as used in the GAMA survey.
We used photometry from the publicly-available GAMA DR1 \citep{Hill2011}, except for six low-surface-brightness galaxies (NGC~5496, 5584, 5690, 5705, 5746, and UGC~09215) where visual inspection revealed that the GAMA Kron apertures did not cover the full extent of the disk. 
For these six, we measured magnitudes in apertures conservatively defined from the SPIRE images. }
Stellar mass can be used as an approximate tracer of metallicity via the mass--metallicity relation \citep{Tremonti2004} or the fundamental metallicity relation \citep[FMR;][]{Mannucci2010}. The FMR combines SFR with stellar mass to reduce scatter in the relationship.
Figure~\ref{fig:mass_mets} shows the relationship between stellar mass and metallicities from \mbox{MPA-JHU} and derived from the FMR respectively. 
For the latter we have estimated SFR from the radio $L_{1.4}$ following \citet{Bell2003}.
The SFR is divided by a factor 1.6 to convert to the \citet{Chabrier2003} IMF used by \citet{Mannucci2010}.
Error bars allow for the uncertainty in stellar mass, {assuming photometric errors of $0.1$mag}, and the uncertainty in $L_{1.4}$.

Figure~\ref{fig:mass_mets} shows that spectroscopically measured metallicities in the sample are systematically high compared with those derived from the FMR for the same galaxies, and compared with the \citet{Tremonti2004} relation. Assuming that the galaxies with spectroscopic measurements are representative of the sample, this could indicate that galaxies in the sample are systematically metal-rich compared with both relations, which is likely considering the selection by dust emission. 
However, it is also possible that the spectroscopic metallicities are higher because they are measured from SDSS fibres covering the central $\sim3\arcsec$ of the galaxies only, and the metallicity in the disk is likely to be lower \citep{Vila-Costas1992}.
Overall, using the stellar masses to estimate the metallicity is subject to large uncertainties due to the complex interplay between stellar mass, SFR and metallicity \citep{Hunt2012,Lara-Lopez2013}, 
and such estimates are likely to be biased for this sample. Combined with the uncertain conversions from our data to \cooz\ (note the large range of line ratios in Table~\ref{tab:aperdata}), and uncertainty in the relationship between \cooz/\htwo\ and oxygen abundance \citep{Genzel2011}, this will lead to large uncertainties on any estimate of \Mhtwo.
Nevertheless, Figure~\ref{fig:mass_mets} indicates a large range of metallicities, which is likely to increase the scatter in the \Mdust/$M_\text{HI}$ ratios.
A broad metallicity range is supported by the diversity of optical colours and morphologies in the sample, which includes dusty spirals (NGC~2861, 3907B, 5478, 5690, 5740, CGCG~013-010, 018-077); blue low-surface-brightness disks (NGC~5496, 5584, 5691);  early-type galaxies (NGC~5719, 5750, SDP~4, 15); as well as the starburst NGC~5713.

\subsection{Slope of the CO--FIR relationship}
\label{sec:lums}
The slope of the relationship between SFR and molecular gas mass, or between their densities (the SK law), is of interest for understanding how star formation is regulated in galaxies.  
To measure the slope in the correlations, we fitted power laws to the data in Figure~\ref{fig:lfir_lco}. 
{We define the slope $n$ such that \mbox{$\Lfir \propto L_\text{CO}^{n}$}, and we fit $n_{32}=0.77\pm0.10$ ($\chi^2=36$) for \cott, and $n_{21}=0.80\pm0.27$ ($\chi^2=90$) for \coto.} 
Errors were estimated by increasing the observational errors to give a reduced $\chisq=1$ in the fit, following \citet{Tremaine2002}.
For reference, \citet{Yao2003} fitted a slope $n_{32}=1.4$ in their 60\mum-selected SLUGS sample (although they only measured CO in the central 15~arcsec of the galaxies), while \citet{Iono2009} found $n_{32}=0.93$ for a sample of local and distant IR-selected galaxies covering five orders of magnitude in luminosity. However, the index may vary between different luminosity regimes \citep[\eg][but see also \citealp{Ivison2011}]{Genzel2010,Wilson2012}, and most of the current sample are not LIRGs (see Figure~\ref{fig:lfir_lco}).
{It may also depend on the CO transition as a result of excitation conditions correlating with luminosity.}

{\citet{Gao2004a} fitted a slope of $n_{10}=1.25$ to the $L\subcooz$ -- \Lfir\ relation in a large sample of nearby galaxies spanning 2.5 orders of magnitude in luminosity, yet a much flatter slope of \mbox{$n_{10}=n_{32}=0.87$} was fitted by \citet{Mao2010} to both \cooz\ and \cott\ data spanning over four orders of magnitude.
Both of these samples, containing star-forming spirals as well as (U)LIRGs, showed consistent relationships over the full luminosity range.
An alternative scenario was presented by \citet{Wilson2012}, who analysed a sample of \hi-selected galaxies in NGLS and SINGS, in comparison to the LIRG sample of \citet{Iono2009}.} They split the local galaxies in NGLS into two luminosity ranges and found that galaxies with $8.3<\log_{10}\Lfir<9.5$ had systematically higher \Lfir/$L\subcott$ ratios than $9.5<\log_{10}\Lfir<10.7$ galaxies (\ie\ $n_{32}<1$; see Figure~5 of \citealt{Wilson2012}). This is consistent with the sub-linear correlation seen in the current data. On the other hand, they showed that the LIRGs from \citet{Iono2009} had higher ratios, and fitted a slope $n_{32}\sim1.2$ for all the galaxies with $\log_{10}\Lfir>9.5$.
The galaxies in the current sample fall in the range  $9.1<\log_{10}\Lfir<11.6$ (though only SDP~1 and CGCG~013-010 have $\log_{10}\Lfir>11$) and the mean (standard deviation) of $\log_{10}(\Lfir/L\subcott)$ is 1.7(0.3). This is consistent with galaxies in the same range from \citet{Wilson2012} and significantly lower than the mean for (U)LIRGs and SMGs. Low-redshift submm-selected galaxies therefore appear to be typical of the the normal, disk-mode star-forming sequence, or ``main sequence'' \citep{Daddi2010,Genzel2010}.
There is clearly some disagreement on the slope of the FIR--CO correlation, with different results being obtained in different samples and in different luminosity ranges. 
Metallicity is likely to play a role, since the CO/\htwo\ conversion is sensitive to metallicity \citep{Genzel2011}, and the estimates in Figure~\ref{fig:mass_mets} indicate a large range of metallicities in the sample.

{We define the slope $m$ in the $\Mdust\propto L_\text{CO}^m$ and $\Mdust\propto M_\text{HI}^m$ relations, and fit}
$m_{32}=0.60\pm0.08$ ($\chi^2=34$) 
for \cott, $m_{21}=0.47\pm0.20$ ($\chi^2=83$) 
for \coto, and $m_\hi=0.98\pm0.21$ ($\chi^2=111$) 
for \hi. Excluding two outliers (SDP~1 and NGC~5496, which may have unreliable self-absorption corrections) from the \hi\ correlation 
yields $m_\hi=0.97\pm0.15$ with $\chi^2=48$. 
Errors were again estimated by increasing the observational errors to give a reduced $\chisq=1$.
{A linear \Mdust--$M_\text{HI}$ correlation ($m_{\hi}=1$) is consistent with other studies in the literature \citep[\eg][]{Devereux1990,Dunne2000}, notwithstanding the scatter in these data. We find the mean ratio of \hi\ gas to total dust mass to be 200 (190 excluding the two outliers), which is similar to that in the SLUGS sample \citep{Dunne2000}.}
In comparison, \citet{Yao2003} fitted the \Mdust--$L\subcott$ correlation with a slope equivalent to $m_{32}=1.05\pm0.08$ to the SLUGS 60\mum\ sample.
Unlike our two-component fits, \citeauthor{Yao2003} used dust masses derived from the 850\mum\ flux assuming a single temperature fitted to 60, 100 and 850\mum\ photometry, so their temperatures may be systematically higher and dust masses lower compared with our analysis, although this should not lead them to over-estimate $m_{32}$ unless the bias is greater at low $L_\text{CO}$. 
\citet{Corbelli2012} fitted correlations between $L\subcott$ and luminosities in each of the PACS and SPIRE bands, finding roughly linear slopes and an evolution in the slope from short to long wavelengths. 
We see some tentative evidence for a similar wavelength dependence of the slope in the right-hand panel of Figure~\ref{fig:chisq_sco}, and this is qualitatively consistent with the difference between $n_{32}$ ($\sim$ slope of the short-wavelength luminosity correlation) and $m_{32}$ ($\sim$ slope of the long-wavelength luminosity correlation).

\begin{figure*}
\begin{center}
\begin{minipage}[t]{0.475\linewidth}
\includegraphics[width=\textwidth]{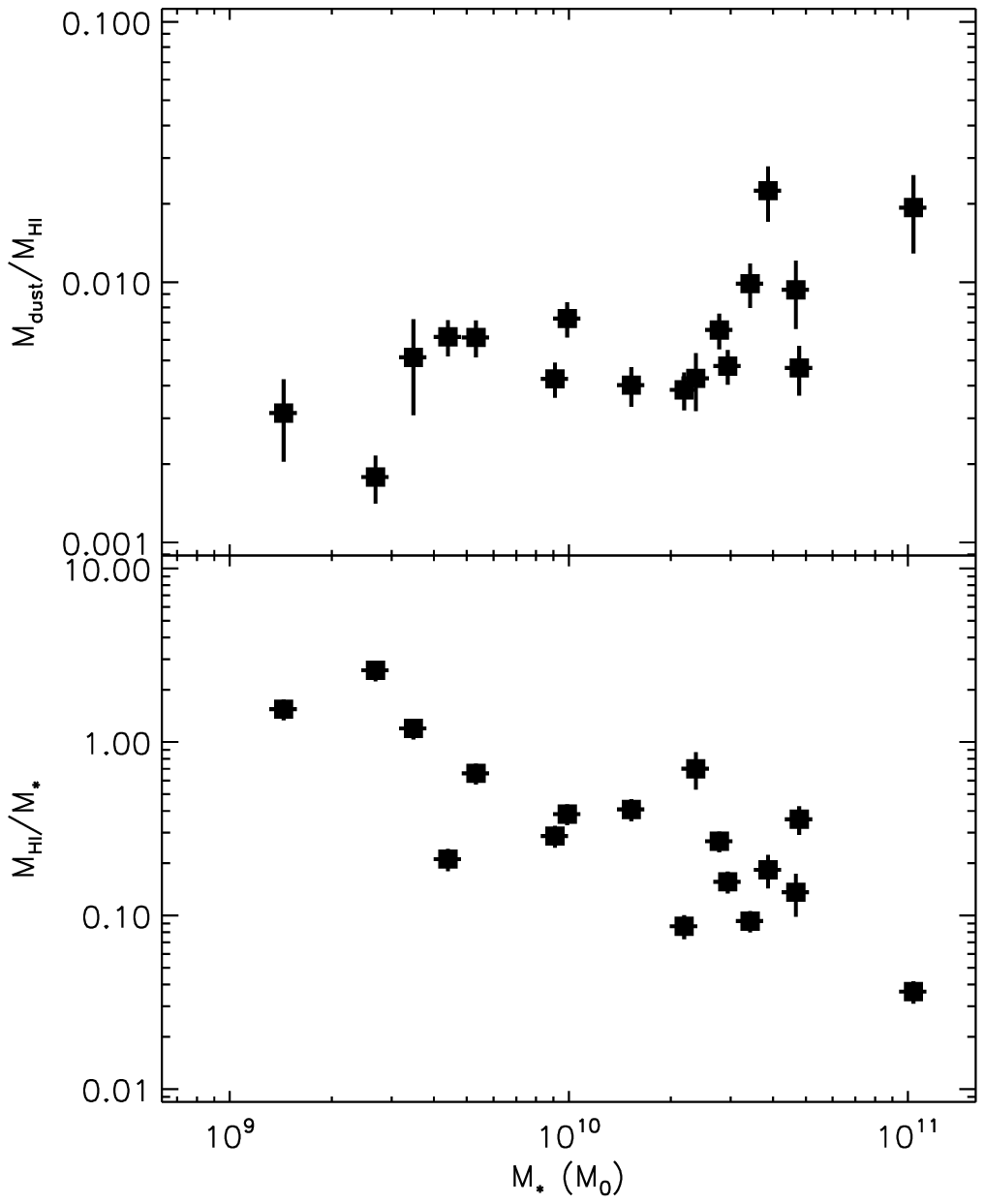}
\caption{Dust-to-\hi\ mass ratio and \hi-to-stellar mass ratio as a function of stellar mass.}
\label{fig:mdust_mhi_mstar}
\end{minipage} 
\hfill
\begin{minipage}[t]{0.475\linewidth}
\includegraphics[width=\textwidth]{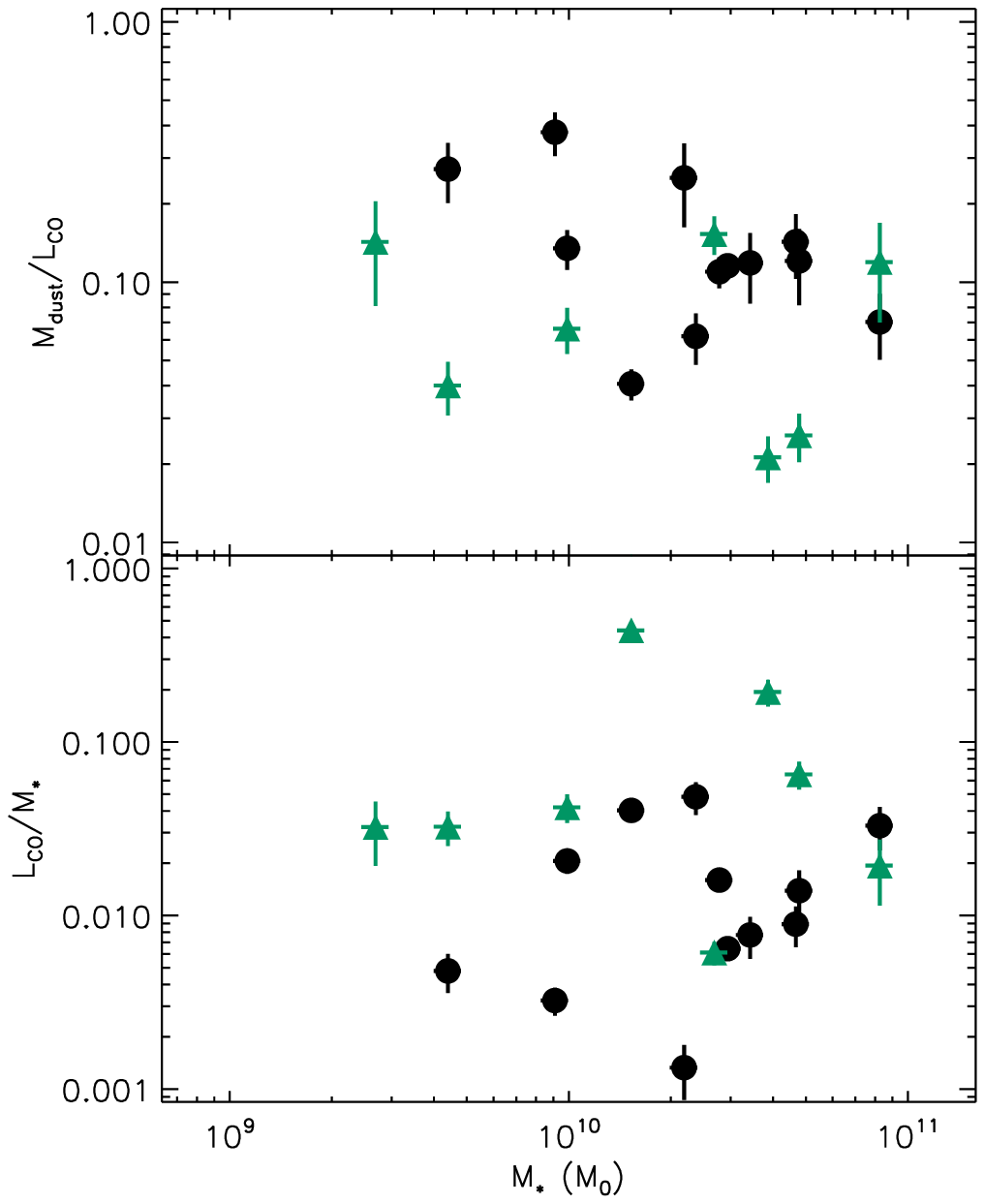}
\caption{Dust-to-$L_\text{CO}$ ratio and $L_\text{CO}$-to-stellar mass ratio as a function of stellar mass.}
\label{fig:mdust_lco_mstar}
\end{minipage}
\end{center}
\end{figure*}

\label{sec:disc_met}
The negative slope of $\Mdust/L_\text{CO}$ versus $L_\text{CO}$ {in Figure~\ref{fig:mdust_lco_lhi}} indicates an intrinsic variation across the sample. 
This variation could be in the dust/\htwo\ ratio, the CO/\htwo\ ratio, or the CO excitation. 
{To explore the possible dependence on metallicity, we can plot the ratios
against stellar mass, which is correlated with metallicity. As a result, more massive galaxies are expected to have higher dust/gas ratios, but they also have lower gas/stellar mass fractions, both from \cooz\ \citep{Saintonge2011} and \hi\ \citep{Cortese2011,Catinella2012}, and lower dust/stellar mass fractions \citep{Bourne2012,Cortese2012b}. Figure~\ref{fig:mdust_mhi_mstar} shows that the dust/\hi\ mass fraction increases and the \hi/stellar mass fraction decreases with stellar mass, as we expect \citep[and as shown by][]{Cortese2012b}. In Figure~\ref{fig:mdust_lco_mstar}, however, 
$\Mdust/L\subcott$ appears to weakly decrease with increasing \Mstar, while $\Mdust/L\subcoto$ is not significantly correlated with \Mstar, which does not follow the expected behaviour of the dust/\htwo\ ratio. Similarly, neither $L\subcott/\Mstar$ nor $L\subcoto/\Mstar$ show any anticorrelation with stellar mass, which we would expect in the $\Mhtwo/\Mstar$ ratio.
These correlations may therefore be confounded by metallicity variations, which could reduce both the dust/\htwo\ and CO/\htwo\ ratios at low stellar masses. Note for example the low CO content of low-metallicity dwarfs in comparison with expectations from the SK law \citep{Schruba2012} and in comparison with the [C{\sc ii}] molecular gas tracer \citep{Smith1997,Hunter2001,Israel2011,Madden2012}.

We might expect less scatter in Figure~\ref{fig:mdust_lco_mstar} using \cooz, which is a closer tracer of the total CO mass (and $\Mhtwo$, given $X_\text{CO}$), in which case the large scatter we see in the \cott\ and \coto\ ratios is a result of different excitation conditions across the sample. However, this explanation is difficult to reconcile with the fact that greater scatter is observed in the \coto\ than \cott\ correlations.
To test whether a correlation between $L_\text{CO}$ and excitation could explain the decreasing $\Mdust/L_\text{CO}$ in Figure~\ref{fig:mdust_lco_lhi}, we measure a slope of $-0.48\pm0.05$ in $\log(\Mdust/L\subcott)$ as a function of $\log(L\subcott)$. If \Mdust/L\subcooz\ is constant, then the $R_{31}$ line ratio must vary by a factor of $10\pm1$ over the range $7.2<\log(L\subcott)<9.5$ in order to explain the observed trend.
In conclusion, to translate the trends in $\Mdust/L\subcoto$ into $\Mdust/\Mhtwo$ (and hence total gas-to-dust), we need either global \cooz\ or gas-phase metallicity measurements.
}

\section{Conclusions}
\label{sec:conc}

We have measured the global fluxes in the \coto\ and \cott\ lines in a pilot sample of nearby galaxies selected in the submm from \hatlas. We exploit the wide areal coverage of \hatlas\ to blindly select at 500\mum, thus sampling low-redshift galaxies with the highest dust masses, a population that is relatively poorly studied in the local Universe. Galaxies in the sample are {a mixture of} dust-rich late-type spirals, starbursts, merging/interacting galaxies, low-surface-brightness disks, and dusty early-type galaxies.
We analyse the results in terms of the correlations between different tracers of dust and gas in the ISM, using $22-500\mum$ photometry from \hatlas, \IRAS, and \WISE, and \hi\ 21\,cm line fluxes from HIPASS and from the literature. Our conclusions are as follows:

\begin{enumerate}

\item We measure correlations between the CO and IR fluxes in all FIR and submm bands from 22 to 500\mum, and find that scatter in the correlations evolves with the IR wavelength used. The \cott\ line is better correlated with FIR bands close to the $\sim100\mum$ peak of the SED, consistent with it being a dense gas tracer and the FIR being correlated with SFR. The \coto\ line appears to be better correlated with the submm rather than FIR bands near the peak, although this needs to be confirmed with a larger sample. It is clear that the \coto--IR correlations have a different wavelength dependence to the \cott--IR correlations, consistent with \coto\ and \cott\ tracing different components of the molecular gas. 

\item In correlations between the \hi\ 21\,cm line and the FIR/submm, the scatter follows opposite trends with wavelength to the \cott. 
This indicates that the submm, which traces the cold dust mass, is also better correlated with the neutral atomic gas (which is robustly traced by the \hi\ line) than the shorter FIR wavelengths. These results suggest that the FIR luminosity ($\sim 100\mum$) traces the SFR due to the correlation with dense gas, while the submm is more closely correlated with the total gas mass due to the better correlation with \hi\ [and \coto]. We caution, however, that in these submm-selected galaxies, even the 100\mum\ fluxes contain significant emission from cold cirrus dust.

\item Differential correlation strengths as a function of wavelength are interpreted as evidence for dust at different temperatures occupying different phases of the ISM, with warm dust being more associated with dense molecular gas clouds, and cold dust inhabiting the diffuse neutral (\hi) ISM. 
This is consistent with cold dust being at least partially heated by radiation from older stellar populations rather than young OB stars, as shown by independent evidence \citep[\eg][]{Boquien2011,Bendo2011,Boselli2012}.

\item We observe a deviation from the trends described above when considering 22 and 60\mum\ fluxes. These show a poorer correlation with \cott\ than fluxes at longer wavelengths, while 22\mum\ fluxes are better correlated with \hi\ than fluxes at longer wavelengths. This can be explained by a contribution to this part of the IR SED by stochastically-heated very small grains in the diffuse phase of the ISM.

\item The slope ($n$) of the FIR--CO correlation appears to be sub-linear in this sample: we fit $\Lfir \propto L_\text{CO}^n$ with $n=0.77\pm0.10$ for \cott\ and $n=0.80\pm0.27$ for \coto. These values are lower than that measured in most other samples \citep{Yao2003,Iono2009,Mao2010}, although the relatively small number of data points and small luminosity range in this sample may account for the difference. 

\item We also find significantly sub-linear slopes of $m_{32}=0.60\pm0.08$ and $m_{21}=0.47\pm0.20$ in the $\Mdust\propto L\subcott^{m_{32}}$ and $\Mdust\propto L\subcott^{m_{21}}$ correlations respectively, in comparison to approximately linear slopes in the 60\mum\ SLUGS sample \citep{Yao2003} and the Virgo cluster spirals in HeVICS \citep{Corbelli2012}. {The explanation for these slopes could be a large range of metallicity across this sample, with higher-mass galaxies having higher metallicity, hence higher CO/\htwo\ and lower dust/CO ratios for the same dust/gas ratio. CO excitation may also play a role, since this is likely to correlate with luminosity, but a variation in $R_{31}$ by a factor of $10\pm1$ between $7.2<\log(L\subcott)<9.5$ would be required to fully explain the trend if the underlying $M_\text{dust}/M_\text{CO}$ ratio is constant.}

\item The FIR/CO luminosity ratios of this sample are in line with low/moderate luminosity SFGs in the local Universe, and not with (U)LIRGs and high-redshift SMGs, suggesting that the dustiest galaxies at low redshifts have modest star-formation efficiencies (= SFR/gas mass). 

This pilot study of a small sample has revealed some interesting trends in the correlation of various gas tracers with the FIR and submm, but many questions remain unanswered. Additional data for \coto\ and \cooz\ in particular would enable a better understanding of the correlations between total CO mass and the FIR/submm, and by comparing to $H_2$ gas masses estimated from the dust mass, it would be possible to investigate any variation of the CO $X$ factor in these dust- and gas-rich (and potentially high-metallicity) galaxies. Observations of [C\textsc{i}] emission with ALMA (currently available at intermediate redshifts in band 7) would also be valuable for such a comparison, and would improve our ability to interpret measurements of the more commonly observed CO lines.

\end{enumerate}

\section*{Acknowledgements}
The authors thank the referee for his/her helpful insights on the paper.
The \Herschel-ATLAS is a project with \Herschel, which is an ESA space observatory with science instruments provided by European-led Principal Investigator consortia and with important participation from NASA. The \hatlas\ website is \url{http://www.h-atlas.org/}.
The James Clerk Maxwell Telescope is operated by the Joint Astronomy Centre on behalf of the Science and Technology Facilities Council of the United Kingdom, the Netherlands Organisation for Scientific Research, and the National Research Council of Canada. This paper makes use of data obtained from the JCMT proposal M11AU04.
The Parkes telescope is part of the Australia Telescope which is funded by the Commonwealth of Australia for operation as a National Facility managed by CSIRO.
This publication makes use of data products from the Wide-field Infrared Survey Explorer, which is a joint project of the University of California, Los Angeles, and the Jet Propulsion Laboratory/California Institute of Technology, funded by the National Aeronautics and Space Administration.
This research has made use of the NASA/IPAC Extragalactic Database (NED), and the NASA/IPAC Infrared Science Archive, which are operated by the Jet Propulsion Laboratory, California Institute of Technology, under contract with the National Aeronautics and Space Administration.

\bibliographystyle{mn2e}
\bibliography{MyLibrary20120814}

\end{document}